\documentclass[12pt]{article}
\usepackage{amsmath,amsthm}
\usepackage{eufrak}
\usepackage[pdftex]{graphicx}

\newcommand{\ar}{\renewcommand{\arraystretch}{1.3}} 
\DeclareMathAlphabet{\bb}{U}{msb}{m}{n} \gdef\C{\bb C} \gdef\dZ{\bb
Z}    \gdef\R{\bb R}
\gdef\K{\bb K} \gdef\BH{\bb H}  

 \DeclareMathOperator{\spin}{{\bf
Spin}}

\DeclareMathOperator{\Sym}{Sym} 
 
 \DeclareMathOperator{\Mat}{Mat}
 
 \DeclareMathOperator{\SL}{SL}
\DeclareMathOperator{\SO}{SO}\DeclareMathOperator{\SU}{SU}
 \DeclareMathOperator{\GO}{O}\DeclareMathOperator{\GU}{U}

\newcommand{\cH}{{\cal H}}

\newcommand{\bsA}{{\bf\sf A}}
\newcommand{\sA}{{\sf A}}
\newcommand{\sB}{{\sf B}}

\newcommand{\sL}{{\sf L}}

\newcommand{\sH}{{\sf H}}

\newcommand{\sX}{{\sf X}}
\newcommand{\sY}{{\sf Y}}

\newcommand{\bsH}{{\boldsymbol{\sf H}}}

\newcommand{\bsZ}{{\boldsymbol{\sf Z}}}
\newcommand{\bsT}{{\boldsymbol{\sf T}}}

\newcommand{\bsX}{{\boldsymbol{\sf X}}}
\newcommand{\bsY}{{\boldsymbol{\sf Y}}}
\newcommand{\bsE}{{\boldsymbol{\sf E}}}
\newcommand{\bsC}{{\boldsymbol{\sf C}}}
\newcommand{\bsL}{{\boldsymbol{\sf L}}}
\newcommand{\bsK}{{\boldsymbol{\sf K}}}
\newcommand{\bsJ}{{\boldsymbol{\sf J}}}
\newcommand{\bsS}{{\boldsymbol{\sf S}}}
\newcommand{\bsQ}{{\boldsymbol{\sf Q}}}
\newcommand{\bsP}{{\boldsymbol{\sf P}}}
\newcommand{\bsB}{{\boldsymbol{\sf B}}}

\newcommand{\fK}{\mathfrak{K}}

\newcommand{\fW}{\mathfrak{W}}

\newcommand{\fg}{\mathfrak{g}}

\newcommand{\cl}{C\kern -0.2em \ell}

\newcommand{\ld}{\left[}
\newcommand{\rd}{\right]}
\newcommand{\lf}{\left\{}
\newcommand{\rf}{\right\}}

\newtheorem{thm}{Theorem}
\newtheorem{defn}{Definition}
\begin{document}
\title{The Periodic Table and the Group $\SO(4,4)$}
\author{V.~V. Varlamov\thanks{Siberian State Industrial University,
Kirova 42, Novokuznetsk 654007, Russia, e-mail:
varlamov@sibsiu.ru}}
\date{}
\maketitle
\begin{abstract}
The periodic system of chemical elements is represented within the framework of the weight diagram of the Lie algebra of the fourth rank of the rotation group of an eight-dimensional pseudo-Euclidean space. The hydrogen realization of the Cartan subalgebra and Weyl generators of the group algebra is studied. The root structure of the subalgebras of the group algebra of a conformal group in the framework of a twofold covering is analyzed. Based on the analysis, the Cartan-Weyl basis of the group algebra is determined. The root and weight diagrams are constructed. A mass formula associated with each node of the weight diagram is introduced. Spin is interpreted as the fourth generator of the Cartan subalgebra, whose two eigenvalues correspond to two three-dimensional projections of the weight diagram containing elements of the periodic system from hydrogen to moscovium (the first projection) and from helium to oganesson (the second projection). One of the main advantages of the proposed group-theoretic construction of the periodic system is the natural inclusion of antimatter in the general scheme.
\end{abstract}
\begin{flushright}
\begin{minipage}{16pc}{{\it
The Book of Nature is written in the language of mathematics
}
\vspace{0.2cm}\\
Galileo Galilei
}\end{minipage}
\end{flushright}

\section{Introduction\label{sec1}}
It is well known that science only reaches perfection when it begins to use mathematics (K. Marx). This fully applies to chemisty, where the periodic law is the most important generalization of this science, and the periodic table of Mendeleev is called by the authors of the article \cite{Schwarz} "Icon of Chemistry". Immediately after D.I. Mendeleev's discovery of the periodic system of chemical element, attempts began to be made to mathematically describe (mathematize) the periodic law (for the history of the issue, see \cite{Heik,Nova}). It is not surprising that the most suitable mathematical structure for describing the phenomenon of periodicity, i.e. repeatability (cyclicity), turned out to be the theory of groups.

Group-theoretic methods for studying the periodic system were proposed independently by several authors in early 70s of the last century. In 1971, the work of Rumer and Fet appeared \cite{RF71}, in which striking similarities were noted between the structure of the system of chemical elements and the structure of the energy spectrum of the hydrogen atom\footnote{Erwin Madelung was the first to use the "hydrogen" quantum numbers $n$, $l$, $m$, $s$ to number the elements of the periodic system. Madelung called the resulting classification of elements "empirical" because he could not connect it with the Bohr model. Apparently, it was precisely because of the lack of theoretical justification at that time (1936) that he published it as a reference material in \cite{Mad68}. When considering the history of the Madelung rule ($n+l,n$), many priority questions arise. As Ostrovsky  notes \cite{Ost01}, it is difficult to trace the origin of this rule, which looks like a kind of scientific folklore. So, in 1930 V. Karapetoff \cite{Kar30} used this rule to predict transuranic elements up to and including $Z=124$.}. This similarity is explained in \cite{RF71} within the framework of the Fock representation $F$ \cite{Fock35} for the spinor group $\spin(4)$ (the twofold covering of the group $\SO(4)$). However, the main disadvantage of the description in \cite{RF71} is the reducibility of the representation $F$, which did not allow us to consider the system as "elementary" in the sense of group dynamic. In 1972, Konopelchenko \cite{Kon72} extended the Fock representation $F$ to the representation $F^+$ of the conformal group thereby eliminating the above drawback. In the same 1972, Barut's article \cite{Bar72} appeared on the group structure of the periodic system within the conformal group $\SO(4,2)$. Barut introduced the quantum numbers\footnote{In order not to be biased against the quantum numbers $n$, $l$, $m_l$ and $m_s$ of the mechanical (planetary) Bohr model describing hydrogen-like systems, Barut intentionally introduced the symbols $\nu$, $\lambda$, $\mu_\lambda$ and $\mu_\sigma$, which primarily have a group-theoretic (not mechanical) meaning, athough their range of variation is the same like the "hydrogen" quantum numbers. Ostrovsky similarly distinguishes between ordinary ("hydrogen") quantum numbers and abstract $\SO(4,2)$-symbols, denoting the latter with a tilde sign: $\tilde{n}$, $\tilde{l}$, $\tilde{m}_l$, $\tilde{m}_s$ \cite{Ostr96}. This symbolism is intended to emphasize the holistic interpretation of the group-theoretic approach, in contrast to the mechanical reductionism of the Bohr model, in which quantum numbers correspond to the radial and orbital movements of the "constituent parts" of the atom, with the exception of the quantum number $m_s$, which, as is known, has no classical analogue, which once again indicates the palliative nature of the Bohr model.} $\nu$, $\lambda$, $\mu_\lambda$, $\mu_\sigma$ of the group $\SO(4,2)$ in order to consider chemical elements as different \textit{states} of \textit{a single quantum system}, which, in turn, is considered as a kind of superparticle. Following a suggestion made later by Wulfman \cite{Wulf78} in 1978, this pseudoparticle, whose spectrum is an atomic supermultiplet, will be designated by the name \textit{baruton}. Barut represents various states (or elements) by ket-vectors $\mid\alpha\rangle$, $\mid\beta\rangle$, $\mid\gamma\rangle$, $\ldots$, which form the basis of an infinite-dimensional Hilbert space. Within the framework of the group-theoretic description, various chemical elements are considered as \textit{structureless} particles, while it is assumed that atoms are \textit{non-composite}, and therefore their internal dynamics can be ignored\footnote{"Structureless" here should not be understood as the absence of any structure at all. First of all, this means the denial of the structure of the reductionist plan in the form of mechanical models (the Rutherford-Bohr planetary model, the quark model), introduced, as Heisenberg said, from the "repertoire of classical physics", i.e. models adequate at the level of macrophysics, but losing their meaning and obscuring the essence of the matter when transferred they are at the micro level. Within the framework of the group-theoretic approach, the structure of the holistic plan is implemented. Namely, the various states that are \textit{cyclic vectors} of $\K$-Hilbert spaces have the structure of the tensor product \cite{Var20,Var21b,Var21,Var22}. This structure defines the dynamic relationship between spin, charge and mass.}.

Simultaneously with these publications, the works of Octavio Novaro and co-authors \cite{NB72,BN73} appear, where the symmetry group of the periodic table proposes the group $G_{NB}=\SU(2)\otimes\SU(2)\otimes\SU(2)$ formed by three mutually commuting angular moments $\boldsymbol{J}_1$, $\boldsymbol{J}_2$, $\boldsymbol{J}_3$. The Novaro-Berrondo group $G_{NB}$ admits the following reduction:
\[
\SU(2)\otimes\SU(2)\otimes\SU(2)\supset\GO(4)\supset\SO(3).
\]
The irreducible representations of the group $G_{NB}$ have the form ($j_1,j_2,j_3$), where $j_1$, $j_2$, $j_3$ are the eigenvalues of the Casimir operators $\boldsymbol{J}^2_1$, $\boldsymbol{J}^2_2$, $\boldsymbol{J}^2_3$ and take integer and half-integer values. Physically valid representations of the group $G_{NB}$ have two types: ($j,j,0$) and $(j,0,j)$, $j=0,1,\frac{1}{2},\ldots$. This is a consequence of the Fock representation $F$$(j_1=j_2)$ for the group $\GO(4)$, which is part of the reduction chain for $G_{NB}$.

Already from the first works in this direction, an important difference appears between the two group-theoretic approaches. Ostrovsky was the first to point out this difference \cite{Ostr96}. Historically, the only quantum system studied (by methods of group theory) was the hydrogen atom, whose Hamiltonian was precisely known. When group theory began to be applied in atomic physics, this was a typical case. Following Ostrovsky's terminology \cite{Ostr96}, we will call it \textit{the atomic physics approach} (APA). However, when it comes to the periodic table, it is much more difficult to construct a Hamiltonian, let alone study its symmetry. \textit{The elementary particle approach} (EPA), the foundations of which are laid in the works of Rumer, Fet and Barut, is based on an analogy with groups of dynamic ("internal") symmetries of elementary particle physics, such as $\SU(2)$ (isotopic spin), $\SU(3)$ and $\SU(6)$. In this approach, chemical elements are considered as different states of some substance: Barut's "atomic matter"\footnote{Since within the framework of the EPA, various states form a \textit{single quantum system}, then, as a result, \textit{quantum transitions} between states (transmutation of elements) are possible. In this context, Barut's "atomic matter" should be understood as "primary matter" (\textit{prima materia}).} \cite{Bar72} or "a structureless particle with internal degrees of freedom" \cite{RF71}.

All further group-theoretic generalizations were associated with attempts to theoretically explain the so-called atomic \textit{magic numbers} describing the \textit{doubling of periods}: 2, 8, 8, 18, 18, 32, 32, $\ldots$. As noted by the author of the article \cite{Low69}, the lack of a theoretical explanation for the doubling of periods (which still exists, see \cite{Scerri, Tyss}) is equivalent to the lack of a theoretical undestanding of the periodic system of chemical elements as a whole\footnote{According to a widespread misconception, Bohr's planetary model explains the periodic table. However, Bohr deduced electronic configurations not from quantum theory, but based on the known chemical and spectroscopic properties of the elements. Moreover, the interpretation of the structure of the periodic table based on the sequence of filling of electronic atomic orbitals in accordance with their relative energies $\varepsilon_{nl}$ is very, very approximate. There is no universal sequence of orbital energies $\varepsilon_{nl}$, besides, such a sequence does not completely determine the order of settlement of atomic orbitals by electrons, since it is necessary to take into account configuration interactions (superposition of configurations in a multi-configuration approximation). The reason for the repetition of similar electronic configurations of atoms is unknown (for more details, see \cite{KK05}). As a result, Bohr's model can reproduce (approximate) Mendeleev's initial discovery only using mathematical approximations (within the framework of the one-electron Hartree approximation) -- it cannot predict (explain) the periodic table. At this point, the widespread notion of the reducibility of chemistry to physics is called into question \cite{Scerri2,Lombardi}. In this regard, a fairly broad discussion about the ontological status of chemistry has recently arisen in the journal Foundations of Chemistry.}. The doubling of periods means that the entire variety of chemical elements naturally decomposes into two sets with the sum of $n+l$ even or odd, where $n$ and $l$ are the main and orbital quantum numbers. As a result, elements from the same (even or odd) set are chemically more similar than elements from different sets \cite{San52,Neu70,Oda73}. Barut tried to explain the doubling of periods by reducing the representations $h$ of the conformal group $\SO(4,2)$ relative to the subgroup $\SO(3,2)$ (anti-de Sitter group) based on the following chain:
\[
\SO(4,2)\supset\SO(3,2)\supset\SO(3)\otimes\SO(2),
\]
according to which the representation $h$ decomposes into the sum $h=h_e\oplus h_o$, where $h_e$ corresponds to even $n+l$, $h_o$ corresponds to odd $n+l$. Ostrovsky, criticizing Barut's scheme, notes that according to this reduction, the subgroup $\GO(4)$ completely loses its meaning. It follows that the quantum number $n$, which is directly related to $\GO(4)$, does not appear in this scheme. However, this quantum number is essential for describing a periodic system\footnote{The doubling of periods proposed by Barut through two different representations of the anti-de Sitter group $\SO(3,2)$, Novaro \cite{Nov89} calls "a fatal flaw", since the resulting dimensions do not correspond to magic numbers. Novaro himself tried to explain the doubling of periods by distinguishing two types of representations $(j,j,0)$ and $(j,0,j)$ of the group $G_{NB}$ \cite{Nov89}.}. Moreover, the group $\GO(4)$ is associated with the first application of group theory in quantum mechanics. In a 1926 paper \cite{Pauli}, Pauli used Heisenberg matrix mechanics to obtain the spectrum of the hydrogen atom. In addition to the angular momentum $\bsL$, Pauli also introduced the quantum mechanical analogue $\bsA$ of the classical Laplace-Runge-Lenz vector. The invariance of the Hamiltonian with respect to these operators ($\bsL$ and $\bsA$) turned out to be sufficient to explain the complete degeneracy of the hydrogen spectrum. In addition, the corresponding algebra can be identified as the Lie algebra of the rotation group in four dimensions, isomorphic to the special orthogonal group $\SO(4)$, which was subsequently strictly established by Fock \cite{Fock} (and further by Bargman \cite{Bar36}).

In the works of Fet \cite{Fet79,Fet80}, the doubling of periods is interpreted by including the cyclic group $\dZ_2$ (i.e., the group of permutations of two elements) in the dynamic group:
\[
G_F=\GO(4,2)\otimes\SU(2)\otimes\dZ_2.
\]
Then, in 1981, Ostrovsky \cite{Ost81} introduces the group
\[
G_O=\GO(4,2)\otimes\SU(2)_S\otimes\SU(2)_T.
\]
Its subgroup $\GO(4)\otimes\SU(2)_S\otimes\SU(2)_T$ contains the symmetry $\GO(4)$, which leads to representations of dimension $n^2$. By extending the group $\GO(4)$ to $\GO(4)\otimes\SU(2)_S$, the dimensions of the representations are doubled to $2n^2$. The subscript $S$ here indicates the physical origin of the group $\SU(2)$ from the electron spin $m_s=\pm 1/2$. Ostrovsky called this "horizontal" doubling of period lengths \textit{spin doubling}. The vertical doubling of period lengths, known as the actual \textit{doubling of periods} in the periodic table, was formulated by Ostrovsky in group-theoretic form by introducing a second group $\SU(2)$, denoted by $\SU(2)_T$ and formally analogous to the isospin group. This leads to \textit{two copies} of the representations of the group $\GO(4,2)\otimes\SU(2)_S$, which are implemented in two different Hilbert spaces $\cH_+$ and $\cH_-$. Ostrovsky introduces three operators $T_+$, $T_-$ and $T_3$ of the algebra $\mathfrak{su}(2)_T$, where the operator $T_3$ acts as a Cartan generator distinguishing states from both subspaces $\cH_+$ and $\cH_-$, and the ladder operators $T_\pm$ act as shift operators between $\cH_+$ and $\cH_-$. A comparison with the Fet group $G_F$, where the doubling of periods is given by the cyclic group $\dZ_2$, shows that the construction $G_F$ does not lead to the use of ladder operators to connect two disjoint representations of the group $\GO(4,2)\otimes\SU(2)_S$, as is the case in the Ostrovsky's construction $G_O$. In his subsequent works \cite{Fet89,Fet}, Fet uses essentially the same group $\SO(4,2)\otimes\SU(2)\otimes\SU(2)^\prime$ as Ostrovsky's.

Historically, the graphical representation of the periodic law has been  very important. The graphical aspect is extremely important, since the first form in which the periodic law is represented is its well-known tabular representation. The book of Mazurs \cite{Masurs} (see also the book of van Spronson \cite{Spronson}) contains more than 700 different graphical representations of the periodic table. Mendeleev himself, as is known, preferred a three-dimensional spiral structure (similar to the design of Chancourtois). The original tables of Mendeleev and Meyer, in addition to a clear visual image, indicate the distribution of chemical elements according to periods, groups and subgroups, and also contain connections between homologous elements, which allowed Mendeleecv to predict new elements at that time (gallium and germanium). However, from the very first years after the appearance of the Mendeleev table (1869), attempts began to be made to go beyond the two-dimensional tabular representation. So, in 1882, Bailey's pyramidal model appeared, further improved by Thomsen (1895) and Bohr (1922). Moreover, screw and spiral models appeared earlier than the Mendeleev and Meyer tables (the "Telluric screw" by Chancourtois (1862) and the Hinrichs spiral (1867)). A spiral is a two-dimensional figure, while a screw has three dimensions. Both designs were used in order to give a clear idea of the periodic table of elements. The advantage of both spiral and helical (spiral) systems is that they form a single whole. The projection of the screw onto the plane is a spiral. The screw system provides an additional means of expressing analogies in properties and, importantly, differences. Since this advantage was initially underestimated, the screw was used less frequently than the spiral. It was only after the discovery of isotopes that some researchers came to the conclusion that the shape of the screw makes it possible to solve the problem of including isotopes in the periodic table. For various graphical representations of the periodic table, see \cite{Spronson,Masurs,Scerri}.

The main feature of the models listed above (tabular, pyramidal, spiral and helical) is the construction of a system of chemical elements depending on the increase in atomic weight. However, quantum mechanics and the first construction of the quantum mechanical model of the atom (Bohr's model) led to the understanding that the main structural characteristic of the periodic law is not a linear increase in atomic weight, but a structure determined by the order of quantum numbers. One of the first to realize this was the German mathematician Gerhard Haenzel. In the introduction to the article \cite{Haenzel} Haenzel writes: "Previous ideas about the periodic table of elements as tables with horizontal and vertical rows and columns (Meyer-Mendeleev) or with connecting lines between related elements\footnote{Here, apparently, Haenzel is referring to the pyramid models of the periodic table of Bailey, Thomsen and Bohr already mentioned above, in which related (homologous) elements were connected by lines.} no longer correspond to the current understanding of the systematics of the elements, since they do not agree with the results of wave mechanics and their conclusions, which fairly reflect the chemistry and physics of the periodic system. Based on the elementary geometric configuration of concentric regular polygons, a multilayer plane is obtained on which these polygons with an odd number of angles are distributed in such a way that each polygon is surrounded by the contour of the previous one. This plane, which thus originates from one of the first objects of geometry, turns out to be a suitable carrier of a system of chemical elements, since its structure and topology fully correspond to the theorems governing this system". In 1943, Wilhelm Finke in the article \cite{Finke} presents a system of Haenzel circles and rings in space. The three-dimensional Finke system, like a screw in relation to a spiral in systems constructed according to an increase in atomic weight, provides an additional means of expressing analogies in properties (Bailey-Thomsen-Bohr lines). However, a common disadvantage of the Haenzel and Finke systems is the artificial representation of the fourth quantum number $s$ (spin) in the form of two points on transversals.

This article is a continuation of a series of works \cite{Var1801,Var1802,Var1901,VPB22,Var2401,Var2402}, in which an attempt is made to construct a periodic system of chemical elements within the framework of a weight diagram of a group algebra of a certain symmetry group. We will begin by studying the structure of the group algebra $\mathfrak{so}(4,2)$ of the conformal group $\SO(4,2)$, a group, as time has shown, most suitable for studying the periodic system of chemical elements. With the exception of the Novaro-Berrondo group $G_{NB}$, this group is the main component of the Fet group $G_F$ and the Ostrovsky group $G_O$. For this reason, the group algebra $\mathfrak{so}(4,2)$ describes the basic structural properties of the periodic table. First of all, the Lie algebra $\mathfrak{so}(4,2)$ is a third-rank algebra, which makes it possible to graphically visualize the root and weight diagrams\footnote{For Lie algebras of rank $r>3$, the graphical implementation of root and weight diagrams becomes almost impossible. An alternative method for constructing such diagrams for Lie algebras of any rank is the Dynkin diagram method \cite{Dyn46}.}. The structure of the algebra $\mathfrak{so}(4,2)$ is determined by the allocation of the Cartan subalgebra and the subsequent construction of Weyl generators, which allows us to set the Cartan-Weyl basis of the algebra $\mathfrak{so}(4,2)$ and construct the corresponding root and weight diagrams. In paragraph 2, all theoretical constructions necessary for further study of the group algebra $\mathfrak{so}(4,2)$ are briefly considered: Cartan subalgebra, Weyl generators and Casimir invariants. In paragraph 3, the Barut representation is used for generators of the algebra $\mathfrak{so}(4,2)$, as the most convenient for subsequent reduction of $\mathfrak{so}(4,2)$ to its subalgebras $\mathfrak{so}(3,1)\simeq\mathfrak{sl}(2,\C)$, $\mathfrak{so}(4)$ and $\mathfrak{so}(2,2)$. In this case, a transition is used to a twofold covering of the conformal group $\SO(4,2)$, which is isomorphic to the group $\SU(2,2)$\footnote{From a purely algebraic point of view, it is more adequate to consider a conformal group with an inverse signature, i.e. $\SO(2,4)$, since in this case the twofold covering (spinor group) $\spin_+(2,4)\simeq\SU(2,2)$ has a quaternion division ring $\K\simeq\BH$ unlike $\spin_+(4,2)$ with a real ring $\K\simeq\R$, see \cite{Var04}.}. Unitary representations of the group $\SU(2,2)$, following the Thomas method \cite{Tomas}, were studied in \cite{Murai1,Murai2,Kihlberg,Yao1,Yao2,Yao3} mainly with respect to the maximal compact subgroup $K=\SU(2)\otimes\SU(2)\otimes\GU(1)$. The finite-dimensional representations of the group $\SU(2,2)$ are defined in the Yao basis \cite{Yao1} with respect to the subgroup $K$.

Despite the fact that the structure of the algebra $\mathfrak{so}(4,2)$ corresponding to type $A_3$ according to the Killing-Cartan classification is well studied (see \cite{Hum,Hall}), this study focuses on the physical application of general algebraic methods to the study of periodic system of chemical elements. In this regard, the starting point of the study is the hydrogen realization of the algebra $\mathfrak{so}(4,2)$, i.e. the Barut representation \cite{Bar72}, considered in section 3.1. Next, in paragraph 4, the Barut representation is formulated in the Yao basis \cite{Yao1} for the group $\SU(2,2)$, which is a twofold covering of the conformal group $\SO(4,2)$. In paragraph 5, we consider the subalgebra $\mathfrak{so}(3,1)\simeq\mathfrak{sl}(2,\C)$ corresponding to a physically important subgroup (the Lorentz group) $\SO(3,1)\subset\SO(4,2)$. For the algebra $\mathfrak{sl}(2,\C)$, the Cartan subalgebra, the Cartan-Weyl basis are determined, and root and weight diagrams are constructed. A mass formula is given that is directly related to each node of the weight diagram. A similar consideration for the subalgebras $\mathfrak{so}(4)$ and $\mathfrak{so}(2,2)$ is carried out in paragraphs 6 and 7. According to the results of the analysis of the root structure of the subalgebras ($\mathfrak{sl}(2,\C)$, $\mathfrak{so}(4)$ and $\mathfrak{so}(2,2)$) in paragraph 8, the root diagram of the Lie algebra $\mathfrak{so}(4,2)$ is constructed. The weight diagram ($\SO(4,2)$-tower) of the algebra $\mathfrak{so}(4,2)$ is defined in paragraph 9. It is shown that the weight diagrams of the Lie algebras of the second rank $\mathfrak{so}(4)$ and $\mathfrak{so}(2,2)$ are projections of $\SO(4,2)$-tower onto coordinate planes formed by Cartan generators of the algebra $\mathfrak{so}(4,2)$. A mass formula is given that is directly related to each node of the $\SO(4,2)$-tower. Since $\mathfrak{so}(4,2)$ is a third-rank algebra, the weight diagram ($\SO(4,2)$-tower) is defined in a three-dimensional weight space in which three generators act as coordinate axes forming the Cartan subalgebra $\fK\subset\mathfrak{so}(4,2)$. However, for a group-theoretic description of the structure of a periodic system defined by the four quantum numbers $(n,l,m,s)$, a third-rank algebra is not enough. A transition to the Lie algebra of the fourth rank is necessary. In paragraph 13 of this article, the group of rotations $\SO(4,4)$ of an eight-dimensional pseudo-Euclidean space $\R^{4,4}$ is considered. The Lie algebra $\mathfrak{so}(4,4)$ of the group $\SO(4,4)$ has the fourth rank. The fourth generator $\bsL_{78}$ (understood as a spin generator) of the Cartan subalgebra $\fK\subset\mathfrak{so}(4,4)$ commutes with all 15 generators of the subalgebra $\mathfrak{so}(4,2)$. As a consequence, the basis of the algebra of twofold covering $\spin_+(4,4)$ splits into two structurally identical bases, each of which is isomorphic to the Yao basis \cite{Yao1} for the group algebra of the group $\spin_+(4,2)\simeq\SU(2,2)$ covering the conformal group. This allows us to place two three-dimensional projections of the weight diagram in the algebra $\mathfrak{so}(4,4)$ using the Madelung basis. The first projection contains all the elements of the periodic table with a spin value $s=-1/2$, starting with hydrogen \textbf{H} ($Z=1$) to moscovium \textbf{Mc} ($Z=115$), the second projection contains all the elements with a value $s=+1/2$, starting from helium \textbf{He} ($Z=2$) to oganesson \textbf{Og} ($Z=118$).

\section{Cartan subalgebra and Weyl generators}
Let $\fg$ be a Lie algebra of dimension $n$ formed by generators $\bsX_i$ ($i=1\rightarrow n$) that satisfy permutation relations
\[
\ld\bsX_i,\bsX_j\rd=\sum^n_{k=1}f_{ijk}\bsX_k\equiv f_{ijk}\bsX_k,
\]
where $f_{ijk}$ are structural constants. $n$ generators $\bsX_i$ form \textit{basis} $\mathcal{B}$ of the Lie algebra $\fg$.
\begin{defn}(Maximal Abelian subalgebra) An Abelian subalgebra $\fK$ of the Lie algebra $\fg$ is called \textit{maximal} when there are no additional elements of the algebra $\fg$ that commute with all elements of the subalgebra $\fK$.
\end{defn}
The subalgebra $\fK$ is better known as the \textit{Cartan subalgebra} of the algebra $\fg$, and the number $m$ of elements of the Cartan subalgebra is called the \textit{rank} of the Lie algebra $\fg$.
\begin{defn}(Cartan subalgebra of the Lie algebra) Let $\fg$ be an $n$-dimensional Lie algebra, then the set of all mutually commuting basis elements $\lf\bsX_i=\bsH_i\rf$ ($i=1\rightarrow m$) of the algebra $\fg$ forms the basis of the maximal Abelian subalgebra $\fK\subset\fg$.
\end{defn}
\begin{defn} (Rank of the Lie algebra) The dimension $m$ of the Cartan subalgebra $\fK\subset\fg$ defines the \textit{rank} of the Lie algebra $\fg$.
\end{defn}
The elements $\bsH_i$ of the Cartan subalgebra $\fK$ are called \textit{Cartan generators} or \textit{Cartan elements}. Cartan generators satisfy permutation relations
\begin{equation}\label{Per0}
\ld\bsH_i,\bsH_j\rd=0,\quad\forall i,j=1,\ldots,m.
\end{equation}
This means that all $\bsH_i$ are simultaneously diagonalizable. Denoting their eigenvalues by $h_i$, we get
\[
\bsH_i\left|h_1,h_2,\ldots,h_m\right\rangle=h_i\left|h_1,h_2,\ldots,h_m\right\rangle,\quad\forall i=1\rightarrow m.
\]
The eigenvalues $h_i$ are called \textit{weights}. $h_i$ can be considered as components of the $m$-dimensional vector $\boldsymbol{h}$, which is called \textit{weight vector}. The weights of the Cartan subalgebra are used as quantum numbers to denote a given multiplet.

Next, from the remaining generators $\bsX_i$ of the algebra $\fg$ ($i=1\rightarrow n-m$), which are not elements of the Cartan subalgebra $\fK$, we form linear combinations, which, in turn, form the set of \textit{increasing} and \textit{decreasing} operators (\textit{ladder operators} $\bsE_\alpha$ -- \textit{Weyl generators} or \textit{Weyl elements}). Along with Cartan generators $\bsH_i$, they make up the \textit{Cartan-Weyl basis} $\lf\bsH_i,\bsE_\alpha\rf$ of the algebra $\fg$.

Thus, the Lie algebra $\fg$ can be decomposed into a direct sum consisting of the Cartan subalgebra $\fK$ ($m$ generators $\bsH_i$) and $n-m$ one-dimensional subalgebras $\fW_\alpha$ formed by Weyl generators $\bsE_\alpha$:
\[
\fg=\fK\bigoplus^{n-m}_{\alpha=1}\fW_\alpha=\fK\oplus\fW_1\oplus\fW_2\oplus\ldots\oplus\fW_{n-m}.
\]
Generators $\bsH_i$ and $\bsE_\alpha$ satisfy permutation relations
\begin{equation}\label{Per1}
\ld\bsH_i,\bsE_\alpha\rd=\alpha_i\bsE_\alpha,\quad\forall i=1\rightarrow m,\;\alpha=1\rightarrow n-m.
\end{equation}
The relations (\ref{Per1}) can be written as follows:
\[
\begin{bmatrix}
\ld\bsH_1,\bsE_\alpha\rd\\
\ld\bsH_2,\bsE_\alpha\rd\\
\vdots\\
\ld\bsH_m,\bsE_\alpha\rd
\end{bmatrix}=
\begin{bmatrix}
\alpha_1\\
\alpha_2\\
\vdots\\
\alpha_m
\end{bmatrix}\bsE_\alpha=\boldsymbol{\alpha}\bsE_\alpha.
\]
The various $\alpha_i$ are called the \textit{roots} of generator $\bsE_\alpha$. The set of roots $\alpha_i$ can be considered as a collection of components of the vector $\boldsymbol{\alpha}$, called the \textit{root vector}, which belongs to the $m$-dimensional \textit{root space}. The root vector for each Weyl generator $\bsE_\alpha$ can be depicted in a diagram called a \textit{Weyl diagram}, the dimension of which is equal to the rank $m$ of the Lie algebra $\fg$. It should be noted that by virtue of (\ref{Per0}) all Cartan generators $\bsH_i$ have roots $\alpha_i=0$ and, therefore, are located in the center of the Weyl diagram.

In general, the standard form of commutation relations for generators of the Lie algebra $\fg$ is written as follows:
\begin{equation}\label{Per2}
\begin{array}{lclcl}
\ld\bsH_i,\bsH_j\rd&=&0,&&\forall i,j=1,\ldots,m;\\
\ld\bsH_i,\bsE_\alpha\rd&=&\alpha_i\bsE_\alpha,&&\forall i=1\rightarrow m,\;\alpha=1\rightarrow n-m;\\
\ld\bsE_\alpha,\bsE_{-\alpha}\rd&=&\alpha^i\bsH_i;&&\\
\ld\bsE_\alpha,\bsE_\beta\rd&=&N^\gamma_{\alpha\beta}\bsE_\gamma,&&\beta\neq-\alpha.
\end{array}
\end{equation}
The quantities $N^\gamma_{\alpha\beta}$ can also be expressed in terms of root vectors, so we know the algebra $\fg$ if its roots are known. These roots have the property
\[
\sum_\alpha\alpha_i\alpha_j=\delta_{ij},
\]
where $\alpha$ can accept only $n-m$ values:
\begin{equation}\label{Roots}
\alpha=\pm 1,\,\pm 2,\,\ldots,\,\pm\frac{1}{2}(n-m).
\end{equation}

Next, \textit{Casimir invariants} $\bsC_\mu$ commute with all generators $\bsX_i$ of the algebra $\fg$:
\[
\ld\bsC_\mu,\bsX_i\rd=0,\quad\forall\mu=1\rightarrow m,\; i=1\rightarrow n.
\]
The number of Casimir invariants (operators) for a given Lie algebra $\fg$ is determined by the rank of this algebra.
\begin{thm} (Racah theorem) For every $n$-dimensional Lie algebra $\fg$ of rank $m$, there are a total of $m$ Casimir operators $\bsC_\mu$ ($\mu=1\rightarrow m$) that commute with generators $\bsX_i$ ($i=1\rightarrow n$) of the algebra $\fg$.
\end{thm}

As a consequence, all $\bsC_\mu$ also commute with Cartan generators $\bsH_i$:
\[
\ld\bsC_\mu,\bsH_i\rd=0,\quad\forall\mu,i=1\rightarrow m.
\]
Thus, it is possible to find a complete set of states that are simultaneously the eigenstates of all $\bsC_\mu$ and $\bsH_i$. Let's define a ket-vector of the following form:
\[
\left|c_1,c_2,\ldots,c_m;h_1,h_2,\ldots,h_m\right\rangle=\left|c_\mu;h_i\right\rangle,
\]
where $c_\mu$, $h_i$ are the eigenvalues of the operators $\bsC_\mu$, $\bsH_i$. Then
\[
\bsC_\mu\left|c_\mu;h_i\right\rangle=c_\mu\left|c_\mu;h_i\right\rangle,\quad
\bsH_i\left|c_\mu;h_i\right\rangle=h_i\left|c_\mu;h_i\right\rangle
\]
for all $\mu,i=1\rightarrow m$. We conclude that the Casimir invariants $\bsC_\mu$ and the generators $\bsH_i$ of the Cartan subalgebra $\fK$ allow us to label each state of the multiplet, while the ladder operators $\bsE_\alpha$ allow us to move between states inside the multiplet, as shown in the Weyl diagram. Thus, when the ladder operator $\bsE_\alpha$ acts on the ket-vector $\left|c_\mu;h_i\right\rangle$, it shifts the eigenvalue of the operators $\bsH_i$ by the value $\alpha_i$ in accordance with
\[
\bsE_\alpha\left|c_\mu;h_i\right\rangle\approx\left|c_\mu;h_i+\alpha_i\right\rangle.
\]

\section{The conformal group $\SO(4,2)$}
A special pseudo-orthogonal group in six dimensions, $\SO(4,2)$, corresponds to the rotation group of the six-dimensional pseudo-Euclidean space $\R^{4,2}$, or, equivalently, the set of $6\times 6$ orthogonal matrices leaving a quadratic form
\[
Q(\boldsymbol{r})=x^2_1+x^2_2+x^2_3+x^2_4-x^2_5-x^2_6=\boldsymbol{r}^T\boldsymbol{r}
\]
invariant, where $\boldsymbol{r}=\ld x_1,x_2,x_3,x_4,x_5,x_6\rd^T$.

The structure of the corresponding Lie algebra $\mathfrak{so}(4,2)$ is determined by the commutation properties of its generators $\bsL_{\alpha\beta}$. $\bsL_{\alpha\beta}$ form the basis of the algebra $\mathfrak{so}(4,2)$. The number of independent generators is easy to find: out of 36 possible combinations of indices $\alpha$ and $\beta$, six combinations disappear due to $\bsL_{\alpha\alpha}=0$, this reduces the number of generators to 30. Moreover, by virtue of $\bsL_{\alpha\beta}=-\bsL_{\beta\alpha}$, only 15 independent generators remain, the number of which can also be obtained using the formula $n(n-1)/2$. Thus,
\begin{equation}\label{LO}
\bsL\Leftrightarrow\begin{bmatrix}
0 & \bsL_{12} & \bsL_{13} & \bsL_{14} & \bsL_{15} & \bsL_{16}\\
  & 0     & \bsL_{23}  & \bsL_{24} & \bsL_{25} & \bsL_{26}\\
  &       & 0      & \bsL_{34} & \bsL_{35} & \bsL_{36}\\
  &       &        & 0     & \bsL_{45}& \bsL_{46}\\
  &       &        &       & 0       & \bsL_{56}\\
  &       &        &       &         & 0
\end{bmatrix}.
\end{equation}

The system of fifteen generators $\bsL_{\alpha\beta}$ of the algebra $\mathfrak{so}(4,2)$ satisfies the following permutation relations:
\begin{equation}\label{Commut}
\left[\bsL_{\alpha\beta},\bsL_{\gamma\delta}\right]=i\left(g_{\alpha\delta}\bsL_{\beta\gamma}+g_{\beta\gamma}\bsL_{\alpha\delta}
-g_{\alpha\gamma}\bsL_{\beta\delta}-g_{\beta\delta}\bsL_{\alpha\gamma}\right),
\end{equation}
where $\alpha,\beta,\gamma,\delta=1,\ldots,6$, $\alpha\neq\beta,\;\gamma\neq\delta$, at this point $g_{11}=g_{22}=g_{33}=g_{44}=1$, $g_{55}=g_{66}=-1$.

\subsection{Hydrogen implementation of the algebra $\mathfrak{so}(4,2)$}
In this section, we consider \textit{Barut representation} \cite{Bar72,ACP82} of the Lie algebra $\mathfrak{so}(4,2)$ of the conformal group $\SO(4,2)$. As noted in the introduction, the subgroup $\SO(4)$ is important in the full spectrum-generating group $\SO(4,2)$. The group $\SO(4)$ describes the degeneracy of the energy levels of the hydrogen atom \cite{Pauli,Fock35}. In the algebra $\mathfrak{so}(4,2)$, the subalgebra $\mathfrak{so}(4)$ is represented by the following generators:
\[
\sL_1=\bsL_{23},\quad\sL_2=\bsL_{31},\quad\sL_3=\bsL_{12},
\]
\begin{equation}\label{LA}
\sA_1=\bsL_{14},\quad\sA_2=\bsL_{24},\quad\sA_3=\bsL_{34}.
\end{equation}
Here $\sL_1$, $\sL_2$, $\sL_3$ are angular momentum generators, generators $\sA_1$, $\sA_2$, $\sA_3$ correspond to the Laplace-Runge-Lenz vector \cite{Pauli}. The commutation relations for (\ref{LA}) have the form
\[
\ld\sL_i,\sL_j\rd=i\varepsilon_{ijk}\sL_k,\quad\ld\sL_i,\sA_j\rd=i\varepsilon_{ijk}\sA_k,\quad
\ld\sA_i,\sA_j\rd=i\varepsilon_{ijk}\sL_k.
\]

Next, the full spectrum-generating algebra $\mathfrak{so}(4,2)$ should include the algebra $\mathfrak{so}(2,1)$, the generators $\Delta_i$ ($i=1\rightarrow 3$) of which act on the radial part of the hydrogen wave function $\psi(r,\theta,\phi)=R_{n,l}(r)Y^m_l(\theta,\phi)$, and since the generators $\sL_i$ act only on the angular part, then
\[
\ld\sL_i,\Delta_i\rd=0.
\]
Therefore, the generators $\Delta_i$ of the subalgebra $\mathfrak{so}(2,1)$ should not contain common indexes with the generators $\sL_i$ of angular momentum. Thus,
\[
\Delta_1=\bsL_{46},\quad\Delta_2=\bsL_{45},\quad\Delta_3=\bsL_{56}.
\]
Combining the generators $\sA_i$ with $\Delta_2$, we get the elements $\bsL_{15}$, $\bsL_{25}$, $\bsL_{35}$ of the fifth column in (\ref{LO}), which we denote by the symbol $\sB_i$:
\[
\ld\Delta_2,\sA_i\rd=i\sB_i.
\]
The three generators $\sB_i$ are components of the vector $\bsB$, which in a sense is conjugate to the Laplace-Runge-Lenz vector $\bsA$.

Similarly, the elements $\bsL_{16}$, $\bsL_{26}$, $\bsL_{36}$ of the sixth column in (\ref{LO}) are obtained by switching the vector $\bsA$ with $\Delta_1$:
\[
\ld\Delta_1,\sA_i\rd=i\Gamma_i.
\]
In turn, the three generators $\Gamma_i$ are components of the vector denoted by Barut $\boldsymbol{\Gamma}$.

Thus, all 15 generators of the algebra $\mathfrak{so}(4,2)$ can be represented in matrix form
\[
\bsL\Leftrightarrow\begin{bmatrix}
0 & \sL_3 & -\sL_2 & \sA_1 & \sB_1 & \Gamma_1\\
  & 0     & \sL_1  & \sA_2 & \sB_2 & \Gamma_2\\
  &       & 0      & \sA_3 & \sB_3 & \Gamma_3\\
  &       &        & 0     & \Delta_2& \Delta_1\\
  &       &        &       & 0       & \Delta_3\\
  &       &        &       &         & 0
\end{bmatrix}.
\]
\subsection{Cartan subalgebra}
Let's find the maximum subset of commuting generators of the algebra $\mathfrak{so}(4,2)$. As is known, two generators commute if they do not have common indexes. It is easy to see that among the generators of the algebra $\mathfrak{so}(4,2)$, this condition is satisfied by three generators $\bsL_{12}$, $\bsL_{34}$ and $\bsL_{56}$ (i.e. $\sL_3$, $\sA_3$ and $\Delta_3$, respectively):
\[
\ld\sL_3,\sA_3\rd=\ld\sL_3,\Delta_3\rd=\ld\sA_3,\Delta_3\rd=0.
\]
The triplet $\lf\sL_3,\sA_3,\Delta_3\rf$ forms the basis of \textit{maximal abelian subalgebra} $\fK\subset\mathfrak{so}(4,2)$ (\textit{Cartan subalgebra}). $\sL_3$, $\sA_3$ and $\Delta_3$ are called \textit{Cartan generators}. The dimension of the subalgebra $\fK$ defines the \textit{rank} of the Lie algebra $\mathfrak{so}(4,2)$. As a result, all root and weight diagrams for $\mathfrak{so}(4,2)$ will be three-dimensional.

\textit{Casimir invariants}. Due to Racah theorem and the fact that the algebra $\mathfrak{so}(4,2)$ has rank 3, it can be expected that the group $\SO(4,2)$ admits three independent \textit{Casimir invariants} $\bsC_\mu$ ($\mu=1\rightarrow 3$), which commute with all generators of the algebra $\mathfrak{so}(4,2)$. The most important Casimir operator of the group $\SO(4,2)$ is a quadratic combination of invariants of various subgroups:
\[
\bsC_2=\bsL^2+\bsA^2-\bsB^2-\boldsymbol{\Gamma}^2+\Delta^2_3-\Delta^2_1-\Delta^2_2.
\]
Here $\bsL^2+\bsA^2$ and $\Delta^2_3-\Delta^2_1-\Delta^2_2$ are known as Casimir operators of the groups $\SO(4)$ and $\SO(2,1)$, respectively. The other two Casimir operators of the group $\SO(4,2)$ are polynomials of the third and fourth degree with respect to the generators of the algebra $\mathfrak{so}(4,2)$:
\[
\bsC_3=\frac{1}{48}\varepsilon_{abcdef}\bsL^{ab}\bsL^{cd}\bsL^{ef},\quad
\bsC_4=\bsL_{ab}\bsL^{bc}\bsL_{cd}\bsL^{da}.
\]

\section{The group $\SU(2,2)$ and the Yao basis}
The twofold covering of the conformal group $\SO(4,2)$ is isomorphic to the group $\SU(2,2)$. $\SU(2,2)$ (a group of pseudounitary unimodular $4\times 4$ matrices) is defined as a transformation group of a four-dimensional complex space $\C^4$ (twistor space), leaving invariant the quadratic form $|Z_1|^2+|Z_2|^2-|Z_3|^2-|Z_4|^2$. In this case, \textit{twistors} (vectors of the space $\C^4$) are defined as reduced spinors for the conformal group (see Appendix B in \cite{Var17}).

The basis of the algebra $\mathfrak{su}(2,2)$ with respect to the maximal compact subgroup
\[
K=\SU(2)\otimes\SU(2)\otimes\GU(1)
\]
is related to the generators of the algebra $\mathfrak{so}(4,2)$ by the following relations \cite{Yao1,Yao2,Yao3}:
\begin{equation}\label{BY1}
\bsK_1=1/2\left(\bsL_{23}+\bsL_{14}\right),\quad\bsK_2=1/2\left(\bsL_{31}+\bsL_{24}\right),\quad
\bsK_3=1/2\left(\bsL_{12}+\bsL_{34}\right),
\end{equation}
\begin{equation}\label{BY2}
\bsJ_1=1/2\left(\bsL_{23}-\bsL_{14}\right),\quad\bsJ_2=1/2\left(\bsL_{31}-\bsL_{24}\right),\quad
\bsJ_3=1/2\left(\bsL_{12}-\bsL_{34}\right),
\end{equation}
\begin{equation}\label{BY3}
\bsT_1=1/2\left(-\bsL_{15}-\bsL_{26}\right),\quad\bsT_2=1/2\left(\bsL_{25}-\bsL_{16}\right),\quad
\bsT_0=1/2\left(-\bsL_{12}-\bsL_{56}\right).
\end{equation}
\begin{equation}\label{BY4}
\bsS_1=1/2\left(-\bsL_{15}+\bsL_{26}\right),\quad\bsS_2=1/2\left(-\bsL_{25}-\bsL_{16}\right),\quad
\bsS_0=1/2\left(\bsL_{12}-\bsL_{56}\right),
\end{equation}
\begin{equation}\label{BY5}
\bsP_1=1/2\left(-\bsL_{35}-\bsL_{46}\right),\quad\bsP_2=1/2\left(\bsL_{45}-\bsL_{36}\right),\quad
\bsP_0=1/2\left(-\bsL_{34}-\bsL_{56}\right),
\end{equation}
\begin{equation}\label{BY6}
\bsQ_1=1/2\left(\bsL_{35}-\bsL_{46}\right),\quad\bsQ_2=1/2\left(\bsL_{45}+\bsL_{36}\right),\quad
\bsQ_0=1/2\left(\bsL_{34}-\bsL_{56}\right).
\end{equation}
The Yao basis (\ref{BY1})--(\ref{BY6}) contains 18 generators, which creates a redundant generator system for the algebra $\mathfrak{so}(4,2)$, since the latter consists of 15 independent generators. The basis of the algebra $\mathfrak{so}(4,2)$ can be obtained from (\ref{BY1})--(\ref{BY6}) by excluding three generators using the relations
\begin{equation}\label{Em1}
\bsJ_3+\bsK_3=\bsS_0-\bsT_0=\bsL_{12}=\sL_3,
\end{equation}
\begin{equation}\label{Em2}
\bsJ_3-\bsK_3=\bsP_0-\bsQ_0=-\bsL_{34}=-\sA_3,
\end{equation}
\begin{equation}\label{Em3}
\bsP_0+\bsQ_0=\bsS_0+\bsT_0=-\bsL_{56}=-\Delta_3.
\end{equation}
It is easy to see that the six generators $\bsK_3$, $\bsJ_3$, $\bsP_0$, $\bsQ_0$, $\bsS_0$, $\bsT_0$ of the Yao basis by virtue of the relations (\ref{Em1})--(\ref{Em3}) emulate the Cartan subalgebra $\fK=\lf\sL_3,\sA_3,\Delta_3\rf$ of the Lie algebra $\mathfrak{so}(4,2)$.

Of the remaining 12 basis generators (\ref{BY1})--(\ref{BY6}), we form Weyl generators by means of the following linear combinations:
\[
\bsJ_\pm=\bsJ_1\pm i\bsJ_2,\quad\bsP_\pm=\bsP_1\pm i\bsP_2,\quad\bsS_\pm=\bsS_1\pm i\bsS_2,
\]
\begin{equation}\label{Envelope}
\bsK_\pm=\bsK_1\pm i\bsK_2,\quad\bsQ_\pm=\bsQ_1\pm i\bsQ_2,\quad\bsT_\pm=\bsT_1\pm i\bsT_2.
\end{equation}


\section{Subalgebra $\mathfrak{so}(3,1)$}
The Lorentz subgroup $\SO(3,1)$ within the conformal group $\SO(4,2)$ in the Barut representation can be formed by two groups of generators: $\sL_1$, $\sL_2$, $\sL_3$, $\sB_1$, $\sB_2$, $\sB_3$ or $\sL_1$, $\sL_2$, $\sL_3$, $\Gamma_1$, $\Gamma_2$, $\Gamma_3$. Let's take the first group. The commutation relations have the form
\begin{equation}\label{Com1}
\left.\begin{array}{lll} \ld\sL_1,\sL_2\rd=-i\sL_3, &
\ld\sL_2,\sL_3\rd=-i\sL_1, &
\ld\sL_3,\sL_1\rd=-i\sL_2,\\[0.1cm]
\ld\sB_1,\sB_2\rd=i\sL_3, & \ld\sB_2,\sB_3\rd=i\sL_1, &
\ld\sB_3,\sB_1\rd=i\sL_2,\\[0.1cm]
\ld\sL_1,\sB_1\rd=0, & \ld\sL_2,\sB_2\rd=0, &
\ld\sL_3,\sB_3\rd=0,\\[0.1cm]
\ld\sL_1,\sB_2\rd=-i\sB_3, & \ld\sL_1,\sB_3\rd=i\sB_2, & \\[0.1cm]
\ld\sL_2,\sB_3\rd=-i\sB_1, & \ld\sL_2,\sB_1\rd=i\sB_3, & \\[0.1cm]
\ld\sL_3,\sB_1\rd=-i\sB_2, & \ld\sL_3,\sB_2\rd=i\sB_1. &
\end{array}\right\}
\end{equation}
The generators $\bsL$ and $\bsB$ form the basis of the Lie algebra $\mathfrak{so}(3,1)\simeq\mathfrak{sl}(2,\C)$.

Let's introduce the following linear combinations:
\begin{equation}\label{Shell}
\bsX=\frac{1}{2}\left(\bsL+i\bsB\right),\quad\bsY=\frac{1}{2}\left(\bsL-i\bsB\right).
\end{equation}
Then
\[
\ld\sX_1,\sX_2\rd=-i\sX_3,\quad\ld\sX_2,\sX_3\rd=-i\sX_1,\quad\ld\sX_3,\sX_1\rd=-i\sX_2,
\]
\[
\ld\sY_1,\sY_2\rd=-i\sY_3,\quad\ld\sY_2,\sY_3\rd=-i\sY_1,\quad\ld\sY_3,\sY_1\rd=-i\sY_2,
\]
\[
\ld\bsX,\bsY\rd=0.
\]
It follows that the generators $\bsX$ and $\bsY$ form the bases of two independent algebras $\mathfrak{so}(3)$. Thus, the Lie algebra $\mathfrak{sl}(2,\C)$ of the group $\SL(2,\C)$ is isomorphic to the direct sum (the so-called Weyl's "unitary trick", see \cite[P.~28]{Knapp})
\begin{equation}\label{Sum2}
\mathfrak{sl}(2,\C)\simeq\mathfrak{su}(2)\oplus i\mathfrak{su}(2).
\end{equation}

Now define \textit{Cartan subalgebra} $\fK$ and the corresponding \textit{Weyl diagram} for the Lie algebra $\mathfrak{sl}(2,\C)$. To this end, it is necessary to move from the basis of the complex shell $\lf\sX_1,\sX_2,\sX_3,\sY_1,\sY_2,\sY_3\rf$ to the \textit{Cartan-Weyl basis}. The first step is to determine the maximum subset of mutually commuting generators of the algebra $\mathfrak{sl}(2,\C)$. Since in the direct sum (\ref{Sum2}) each subalgebra $\mathfrak{su}(2)$ is a Lie algebra of rank 1, it is natural to expect that the algebra $\mathfrak{sl}(2,\C)$ contains at most two commuting generators. From nine possible pairs of commuting generators $\lf\sX_i,\sY_j\rf$ ($i,j=1,2,3$), we choose a pair $\lf\sX_3,\sY_3\rf$ satisfying the condition $\ld\bsH_i,\bsH_j\rd=0$ (see formula (1)), i.e.
\begin{equation}\label{Ksl2}
\ld\sX_3,\sY_3\rd=0.
\end{equation}
The set $\lf\sX_3,\sY_3\rf$ forms a Cartan subalgebra $\fK\subset\mathfrak{sl}(2,\C)$. $\sX_3$ and $\sY_3$ are \textit{Cartan generators}, and the dimension of the subalgebra $\fK$ equal to 2 determines \textit{rank} of the Lie algebra $\mathfrak{sl}(2,\C)$.

Next, in order to determine \textit{Weyl generators} from the remaining generators $\sX_1$, $\sX_2$, $\sY_1$, $\sY_2$, we form the following linear combinations (\textit{increasing} and \textit{decreasing} operators):
\begin{equation}\label{GenW2}
\left.\begin{array}{cc}
\sX_+=\sX_1+i\sX_2, & \sX_-=\sX_1-i\sX_2,\\[0.1cm]
\sY_+=\sY_1+i\sY_2, & \sY_-=\sY_1-i\sY_2.
\end{array}\right\}
\end{equation}
Four Weyl generators (\ref{GenW2}), along with two Cartan generators $\sX_3$ and $\sY_3$, make up the \textit{Cartan-Weyl basis} of the algebra $\mathfrak{sl}(2,\C)$:
\[
\lf\sX_3,\sY_3,\sX_+,\sX_-,\sY_+,\sY_-\rf.
\]
Weyl generators $\bsE_\alpha=\lf\sX_\pm,\sY_\pm\rf$ and generators $\bsH_i=\lf\sX_3,\sY_3\rf$ of the Cartan subalgebra $\fK\subset\mathfrak{sl}(2,\C)$ satisfy permutation relations (2) and (3), where $\forall i,j=1,2;\;\alpha=1\rightarrow 4$:
\begin{equation}\label{Per2s}
\ld\sX_3,\sX_+\rd=-\sX_+,\quad\ld\sX_3,\sX_-\rd=\sX_-,\quad\ld\sX_+,\sX_-\rd=-2\sX_3,
\end{equation}
\begin{equation}\label{Per3s}
\ld\sY_3,\sY_+\rd=-\sY_+,\quad\ld\sY_3,\sY_-\rd=\sY_-,\quad\ld\sY_+,\sY_-\rd=-2\sY_3.
\end{equation}
In this case, the commutator $\ld\bsE_\alpha,\bsE_{-\alpha}\rd=\ld\sX_+,\sX_-\rd$ gives the roots of $\pm 2$. Thus, in accordance with formula (4), we have four different roots: $\alpha=\pm 1,\,\pm 2$.

Since $\mathfrak{sl}(2,\C)$ is a Lie algebra of rank 2, it follows from the Racah theorem (see paragraph 1) that there are two independent \textit{Casimir invariants} $\bsC_\mu$ that commute with all generators of the algebra $\mathfrak{sl}(2,\C)$, including two Cartan elements $\bsH_i$:
\begin{equation}\label{Csl4}
\ld\bsC_\mu,\bsH_i\rd=0,\quad\forall\mu=1\rightarrow 2;\;i=1\rightarrow 2.
\end{equation}
The Casimir invariants for the algebra $\mathfrak{sl}(2,\C)$ have the form
\begin{eqnarray}
\bsC_1&\equiv&\bsX^2=\sX^2_1+\sX^2_2+\sX^2_3=\frac{1}{4}\left(\bsA^2-\bsB^2+2i\bsA\bsB\right),\nonumber\\
\bsC_2&\equiv&\bsY^2=\sY^2_1+\sY^2_2+\sY^2_3=\frac{1}{4}\left(\bsA^2-\bsB^2-2i\bsA\bsB\right).\nonumber
\end{eqnarray}
Within the framework of the complex shell of the algebra $\mathfrak{sl}(2,\C)$, these invariants, better known as \textit{Laplace-Beltrami operators}, lead to Fuchs class differential equations for hyperspheric functions (see \cite{Var06}). The Laplace-Beltrami operators contain the Casimir operators $\bsA^2-\bsB^2$ and $\bsA\bsB$ of the Lorentz group as parts of a complex-valued function.

By virtue of (\ref{Ksl2}) and (\ref{Csl4}), a complete set of states is defined in the eigenspace $\sH_E$ of the energy operator $H$, which are simultaneously eigenstates of the operators $\bsX^2$, $\bsY^2$, $\sX_3$ and $\sY_3$. Let's make a ket-vector $\left|l,\dot{l};m,\dot{m}\right\rangle$. It should be noted that $l$ and $\dot{l}$ are not quantum numbers, but only \textit{set} them, real quantum numbers, i.e. the eigenvalues of the Casimir operators $\bsX^2$ and $\bsY^2$ are $l(l+1)$ and $\dot{l}(\dot{l}+1)$ according to the following relations:
\begin{eqnarray}
\bsX^2\left|l,\dot{l};m,\dot{m}\right\rangle&=&l(l+1)\left|l,\dot{l};m,\dot{m}\right\rangle,
\nonumber\\
\bsY^2\left|l,\dot{l};m,\dot{m}\right\rangle&=&\dot{l}(\dot{l}+1)\left|l,\dot{l};m,\dot{m}\right\rangle,
\nonumber
\end{eqnarray}
where $l,\dot{l}\in\lf 0,\frac{1}{2},1,\frac{3}{2},\ldots\rf$. Each subspace $\sH_E$ has dimension $(2l+1)(2\dot{l}+1)$, which implies that
\begin{eqnarray}
\sX_3\left|l,\dot{l};m,\dot{m}\right\rangle&=&m\left|l,\dot{l};m,\dot{m}\right\rangle,\label{X3_2}\\
\sY_3\left|l,\dot{l};m,\dot{m}\right\rangle&=&\dot{m}\left|l,\dot{l};m,\dot{m}\right\rangle\label{Y3_2}
\end{eqnarray}
with $m\in\lf-l,-l+1,\ldots,l-1,l\rf$ and $\dot{m}\in\lf-\dot{l},-\dot{l}+1,\ldots,\dot{l}-1,\dot{l}\rf$. The eigenvalues $m$ and $\dot{m}$ are the \textit{weights} of the Cartan generators $\sX_3$ and $\sY_3$.

The construction of the \textit{Weyl diagram} of the group $\SL(2,\C)$ is based on the Cartan subalgebra $\fK=\lf\sX_3,\sY_3\rf$, where the generators $\sX_3$ and $\sY_3$ form the basis of a two-dimensional orthogonal coordinate system. In these diagrams, the weights $m$ and $\dot{m}$ are used as coordinates to construct each state of the $\SL(2,\C)$-multiplet in the plane $(\sX_3,\sY_3)$, i.e. they form components of a two-dimensional \textit{weight vector} $\boldsymbol{h}=(m,\dot{m})$, which exits the origin of the coordinate system to the state $\left|l,\dot{l};m,\dot{m}\right\rangle$. The Weyl generators $\bsE_\alpha=\lf\sX_\pm,\sY_\pm\rf$ allow us to move between the states of the $\SL(2,\C)$-multiplet by shifting the eigenvalues $m$ and $\dot{m}$ of any ket-vector $\left|l,\dot{l};m,\dot{m}\right\rangle$ by the value that is given by the roots $\alpha_1$ and $\alpha_2$ of this Weyl generator relative to the Cartan generators $\sX_3$ and $\sY_3$:
\[
\bsE_\alpha\left|l,\dot{l};m,\dot{m}\right\rangle\;\longrightarrow\;
\left|l,\dot{l};m+\alpha_1,\dot{m}+\alpha_2\right\rangle.
\]
As an example, consider the action of the generator $\sX_+$ on the state $\left|l,\dot{l};m,\dot{m}\right\rangle$. By virtue of $\ld\bsH_i,\bsE_\alpha\rd=\alpha_i\bsE_\alpha$, (\ref{Per2s}) and (\ref{X3_2}), we get
\begin{eqnarray}
\sX_3\sX_+\left|l,\dot{l};m,\dot{m}\right\rangle&=&\left(\ld\sX_3,\sX_+\rd+\sX_+\sX_3\right)
\left|l,\dot{l};m,\dot{m}\right\rangle\nonumber\\
&=&\left(\sX_++ m\sX_+\right)\left|l,\dot{l};m,\dot{m}\right\rangle\nonumber\\
&=&\left(m+1\right)\sX_+\left|l,\dot{l};m,\dot{m}\right\rangle.\nonumber
\end{eqnarray}
Thus, $\sX_+$ increases the eigenvalue $m$ by $+1$, which is equal to the root of the generator $\sX_+$ relative to the Cartan generator $\sX_3$ according to (\ref{Per2s}). Similarly, using (\ref{Y3_2}), we get
\begin{eqnarray}
\sY_3\sX_+\left|l,\dot{l};m,\dot{m}\right\rangle&=&\left(\ld\sY_3,\sX_+\rd+\sX_+\sY_3\right)
\left|l,\dot{l};m,\dot{m}\right\rangle\nonumber\\
&=&\left(0+\dot{m}\sX_+\right)\left|l,\dot{l};m,\dot{m}\right\rangle\nonumber\\
&=&\dot{m}\sX_+\left|l,\dot{l};m,\dot{m}\right\rangle,\nonumber
\end{eqnarray}
where the generator $\sX_+$ leaves the eigenvalue $\dot{m}$ unchanged. Therefore,
\[
\sX_+\left|l,\dot{l};m,\dot{m}\right\rangle\;\longrightarrow\;\left|l,\dot{l};m+1,\dot{m}\right\rangle.
\]
The actions of the other three Weyl generators are defined similarly:
\[
\sX_-\left|l,\dot{l};m,\dot{m}\right\rangle\;\longrightarrow\;\left|l,\dot{l};m-1,\dot{m}\right\rangle,
\]
\[
\sY_+\left|l,\dot{l};m,\dot{m}\right\rangle\;\longrightarrow\;\left|l,\dot{l};m,\dot{m}+1\right\rangle,
\]
\[
\sY_-\left|l,\dot{l};m,\dot{m}\right\rangle\;\longrightarrow\;\left|l,\dot{l};m,\dot{m}-1\right\rangle.
\]
Let's graphically show the actions of the generators $\bsE_\alpha$ on the \textit{root diagram}. To this end, take the roots $\alpha_1$ and $\alpha_2$ of each Weyl element $\bsE_\alpha$ as components of the two-dimensional \textit{root vector} $\boldsymbol{\alpha}=(\alpha_1,\alpha_2)$ and place them in the two-dimensional \textit{weight space} formed by the plane $(\sX_3,\sY_3)$. This gives the root diagram of the Lie algebra $\mathfrak{sl}(2,\C)$, as shown in Figure \ref{pic1}, where, for simplicity, we have designated the various root vectors $\boldsymbol{\alpha}$ with the corresponding symbol of the Weyl generator $\bsE_\alpha$.
\begin{figure}[h]
\unitlength=1mm
\begin{center}
\begin{picture}(20,30)
\put(10,-8){\vector(0,1){34}}
\put(9.15,6.5){$\bullet$}
\put(-5,7.5){\vector(1,0){33}}
\put(25,9){$\sX_3$}
\put(15,9){$\sX_+$}
\put(12,4){$\scriptscriptstyle+\frac{1}{2}$}
\put(20,4){$\scriptscriptstyle+1$}
\put(0,4){$\scriptscriptstyle-\frac{1}{2}$}
\put(-6,4){$\scriptscriptstyle-1$}
\put(2.5,9){$\sX_-$}
\put(5,25){$\sY_3$}
\put(5,13){$\scriptscriptstyle+\frac{1}{2}$}
\put(5,1){$\scriptscriptstyle-\frac{1}{2}$}
\put(5,19){$\scriptscriptstyle+1$}
\put(5,-6){$\scriptscriptstyle-1$}
\put(11,16){$\sY_+$}
\put(11,-3.5){$\sY_-$}
\thicklines
\put(10,7.5){\vector(1,0){14}}
\put(10,7.5){\vector(0,1){14}}
\put(10,7.5){\vector(-1,0){14}}
\put(10,7.5){\vector(0,-1){14}}
\end{picture}
\end{center}
\vspace{0.3cm}
\caption{The root diagram of the Lie algebra $\mathfrak{sl}(2,\C)$. The action of each Weyl generator is shown in the ($\sX_3,\sY_3$)-plane.\label{pic1}}
\end{figure}

Obviously, the generators $\sX_-$ and $\sX_+$ (roots $\alpha_1=-1$, $\alpha_2=+1$) allow us to move one step \textit{left} and \textit{right}, respectively, while moving \textit{up} and \textit{down} are set by the generators $\sY_+$ and $\sY_-$. Thus, the states of the $\SL(2,\C)$-multiplet are translated into each other by repeated action of these ladder operators. With the generators of the algebra $\mathfrak{sl}(2,\C)$, located in Figure 1, it immediately becomes obvious that they correspond to two different manifolds: the generators $\sX_3$, $\sX_+$ and $\sX_-$ (forming the first subalgebra $\mathfrak{su}(2)$ in (\ref{Sum2})) correspond to the manifold $(1,0)$, whereas the generators $\sY_3$, $\sY_+$ and $\sY_-$ (the second subalgebra $i\mathfrak{su}(2)$), as can be seen, form the manifold $(0,1)$ on the weight diagram. The Weyl diagrams for the first three $\SL(2,\C)$ multiplets are shown in Figure \ref{pic2}.
\begin{figure}[h]
\unitlength=1.5mm
\begin{center}
\begin{picture}(30,30)(37,-10)
\put(50,0){$\overset{(0,0)}{\bullet}$}
\put(40,0.5){\vector(1,0){25}}\put(63,1.5){$\sX_3$}\put(30,0.5){$a)$}
\put(55,-1){$\scriptscriptstyle+\frac{1}{2}$}
\put(57,0.5){\line(0,1){0.5}}
\put(45,-1){$\scriptscriptstyle-\frac{1}{2}$}
\put(47,0.5){\line(0,1){0.5}}
\put(42,0.5){\line(0,1){0.5}}
\put(60.5,-1){$\scriptscriptstyle+1$}
\put(62,0.5){\line(0,1){0.5}}
\put(40.5,-1){$\scriptscriptstyle-1$}
\put(52,-12){\vector(0,1){26}}\put(48.5,13){$\sY_3$}
\put(49.25,5.25){$\scriptscriptstyle+\frac{1}{2}$}
\put(52,5.5){\line(1,0){0.5}}
\put(49.25,-5.75){$\scriptscriptstyle-\frac{1}{2}$}
\put(52,-5.5){\line(1,0){0.5}}
\put(49.25,10.25){$\scriptscriptstyle+1$}
\put(52,10.5){\line(1,0){0.5}}
\put(49.25,-10.75){$\scriptscriptstyle-1$}
\put(52,-10.5){\line(1,0){0.5}}
\put(55,5){$\overset{(\frac{1}{2},0)}{\bullet}$}
\put(55,-6){$\overset{(0,\frac{1}{2})}{\bullet}$}
\put(45,5){$\overset{(0,\frac{1}{2})}{\bullet}$}
\put(45,-6){$\overset{(\frac{1}{2},0)}{\bullet}$}
\end{picture}
\end{center}
\end{figure}

\begin{figure}[h]
\unitlength=1.5mm
\begin{center}
\begin{picture}(30,30)(37,-10)
\put(50,0){$\overset{(0,0)}{\bullet}$}
\put(38,0.5){\vector(1,0){28}}\put(64,1.5){$\sX_3$}\put(30,0.5){$b)$}
\put(55,-1){$\scriptscriptstyle+\frac{1}{2}$}
\put(57,0.5){\line(0,1){0.5}}
\put(45,-1){$\scriptscriptstyle-\frac{1}{2}$}
\put(47,0.5){\line(0,1){0.5}}
\put(42,0.5){\line(0,1){0.5}}
\put(60.5,-1){$\scriptscriptstyle+1$}
\put(62,0.5){\line(0,1){0.5}}
\put(40.5,-1){$\scriptscriptstyle-1$}
\put(52,-14){\vector(0,1){30}}\put(48.5,15){$\sY_3$}
\put(49.25,5.25){$\scriptscriptstyle+\frac{1}{2}$}
\put(52,5.5){\line(1,0){0.5}}
\put(49.25,-5.75){$\scriptscriptstyle-\frac{1}{2}$}
\put(52,-5.5){\line(1,0){0.5}}
\put(49.25,10.25){$\scriptscriptstyle+1$}
\put(52,10.5){\line(1,0){0.5}}
\put(49.25,-10.75){$\scriptscriptstyle-1$}
\put(52,-10.5){\line(1,0){0.5}}
\put(55,5){$\overset{(\frac{1}{2},0)}{\bullet}$}
\put(55,-6){$\overset{(0,\frac{1}{2})}{\bullet}$}
\put(45,5){$\overset{(0,\frac{1}{2})}{\bullet}$}
\put(45,-6){$\overset{(\frac{1}{2},0)}{\bullet}$}
\put(49.5,10){$\overset{(\frac{1}{2},\frac{1}{2})}{\bullet}$}
\put(49.5,-11){$\overset{(\frac{1}{2},\frac{1}{2})}{\bullet}$}
\put(60,10){$\overset{(1,0)}{\bullet}$}
\put(60,-11){$\overset{(0,1)}{\bullet}$}
\put(40,10){$\overset{(0,1)}{\bullet}$}
\put(40,-11){$\overset{(1,0)}{\bullet}$}
\end{picture}
\end{center}
\end{figure}
\vspace{0.8cm}
\begin{figure}[h]
\unitlength=1.5mm
\begin{center}
\begin{picture}(30,30)(37,-10)
\put(50,0){$\overset{(0,0)}{\bullet}$}
\put(33,0.5){\vector(1,0){39}}\put(70,1.5){$\sX_3$}\put(30,0.5){$c)$}
\put(55,-1){$\scriptscriptstyle+\frac{1}{2}$}
\put(57,0.5){\line(0,1){0.5}}
\put(45,-1){$\scriptscriptstyle-\frac{1}{2}$}
\put(47,0.5){\line(0,1){0.5}}
\put(42,0.5){\line(0,1){0.5}}
\put(60.5,-1){$\scriptscriptstyle+1$}
\put(62,0.5){\line(0,1){0.5}}
\put(65.25,-1){$\scriptscriptstyle+\frac{3}{2}$}
\put(40.5,-1){$\scriptscriptstyle-1$}
\put(52,-20){\vector(0,1){40}}\put(48.5,19){$\sY_3$}
\put(49.25,5.25){$\scriptscriptstyle+\frac{1}{2}$}
\put(52,5.5){\line(1,0){0.5}}
\put(49.25,-5.75){$\scriptscriptstyle-\frac{1}{2}$}
\put(52,-5.5){\line(1,0){0.5}}
\put(49.25,10.25){$\scriptscriptstyle+1$}
\put(52,10.5){\line(1,0){0.5}}
\put(49.25,-10.75){$\scriptscriptstyle-1$}
\put(52,-10.5){\line(1,0){0.5}}
\put(55,5){$\overset{(\frac{1}{2},0)}{\bullet}$}
\put(55,-6){$\overset{(\frac{1}{2},0)}{\bullet}$}
\put(45,5){$\overset{(0,\frac{1}{2})}{\bullet}$}
\put(45,-6){$\overset{(\frac{1}{2},0)}{\bullet}$}
\put(49.5,10){$\overset{(\frac{1}{2},\frac{1}{2})}{\bullet}$}
\put(49.5,-11){$\overset{(\frac{1}{2},\frac{1}{2})}{\bullet}$}
\put(60,10){$\overset{(1,0)}{\bullet}$}
\put(60,-11){$\overset{(0,1)}{\bullet}$}
\put(40,10){$\overset{(0,1)}{\bullet}$}
\put(40,-11){$\overset{(1,0)}{\bullet}$}
\put(35,15){$\overset{(0,\frac{3}{2})}{\bullet}$}
\put(35,-16){$\overset{(\frac{3}{2},0)}{\bullet}$}
\put(45,15){$\overset{(\frac{1}{2},1)}{\bullet}$}
\put(45,-16){$\overset{(1,\frac{1}{2})}{\bullet}$}
\put(55,15){$\overset{(1,\frac{1}{2})}{\bullet}$}
\put(55,-16){$\overset{(\frac{1}{2},1)}{\bullet}$}
\put(65,15){$\overset{(\frac{3}{2},0)}{\bullet}$}
\put(65,-16){$\overset{(0,\frac{3}{2})}{\bullet}$}
\end{picture}
\end{center}
\vspace{0.7cm}
\caption{The first three weight diagrams (Weyl diagrams) of the Lie algebra $\mathfrak{sl}(2,\C)$: a) $(\tfrac{1}{2},\tfrac{1}{2})$-multiplet; b) $(1,1)$-multiplet; c) $(\tfrac{3}{2},\tfrac{3}{2})$- multiplet.\label{pic2}}
\end{figure}
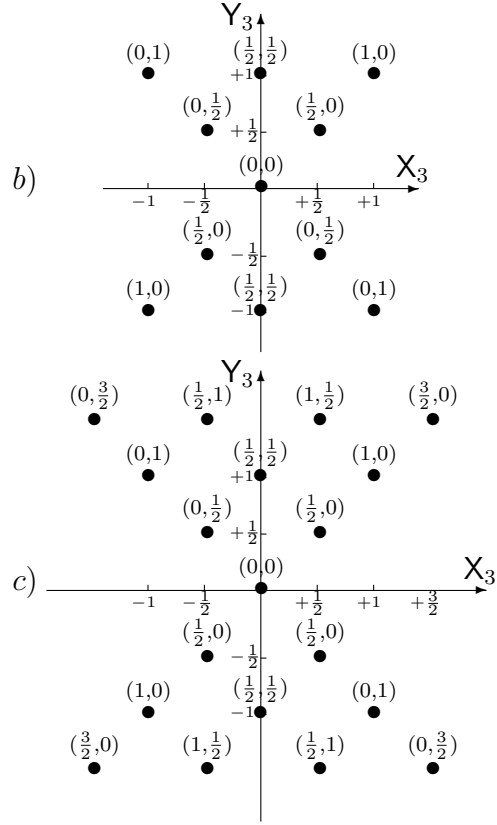

\begin{figure}[h]
\unitlength=1.5mm
\begin{center}
\begin{picture}(100,85)(6,-35)
\put(50,0){$\overset{(0,0)}{\bullet}$}\put(47,5.5){\line(1,0){10}}\put(52.25,2.75){\line(0,1){7.25}}
\put(47,-5.5){\line(1,0){10}}\put(52.25,-7.15){\line(0,1){7.25}}
\put(55,5){$\overset{(\frac{1}{2},0)}{\bullet}$}
\put(55,-6){$\overset{(0,\frac{1}{2})}{\bullet}$}
\put(45,5){$\overset{(0,\frac{1}{2})}{\bullet}$}
\put(45,-6){$\overset{(\frac{1}{2},0)}{\bullet}$}
\put(40,10){$\overset{(0,1)}{\bullet}$}\put(42,10.5){\line(1,0){10}}\put(47.25,7.75){\line(0,1){7.25}}
\put(40,-11){$\overset{(1,0)}{\bullet}$}\put(42,-10.5){\line(1,0){10}}\put(47.25,-12.75){\line(0,1){7.25}}
\put(50,10){$\overset{(\frac{1}{2},\frac{1}{2})}{\bullet}$}
\put(50,-11){$\overset{(\frac{1}{2},\frac{1}{2})}{\bullet}$}
\put(52,10.5){\line(1,0){10}}\put(57.25,7.75){\line(0,1){7.25}}
\put(52,-10.5){\line(1,0){10}}\put(57.25,-12.75){\line(0,1){7.25}}
\put(60,10){$\overset{(1,0)}{\bullet}$}
\put(60,-11){$\overset{(0,1)}{\bullet}$}
\put(35,15){$\overset{(0,\frac{3}{2})}{\bullet}$}\put(37,15.5){\line(1,0){10}}\put(42.25,12.75){\line(0,1){7.25}}
\put(35,-16){$\overset{(\frac{3}{2},0)}{\bullet}$}\put(37,-15.5){\line(1,0){10}}\put(42.25,-17.75){\line(0,1){7.25}}
\put(45,15){$\overset{(\frac{1}{2},1)}{\bullet}$}\put(47,15.5){\line(1,0){10}}\put(52.25,12.75){\line(0,1){7.25}}
\put(45,-16){$\overset{(\frac{1}{2},1)}{\bullet}$}\put(47,-15.5){\line(1,0){10}}\put(52.25,-17.75){\line(0,1){7.25}}
\put(55,15){$\overset{(1,\frac{1}{2})}{\bullet}$}\put(57,15.5){\line(1,0){10}}\put(62.25,12.75){\line(0,1){7.25}}
\put(55,-16){$\overset{(1,\frac{1}{2})}{\bullet}$}\put(57,-15.5){\line(1,0){10}}\put(62.25,-17.75){\line(0,1){7.25}}
\put(65,15){$\overset{(\frac{3}{2},0)}{\bullet}$}
\put(65,-16){$\overset{(0,\frac{3}{2})}{\bullet}$}
\put(30,20){$\overset{(0,2)}{\bullet}$}\put(32,20.5){\line(1,0){10}}\put(37.25,17.75){\line(0,1){7.25}}
\put(30,-21){$\overset{(2,0)}{\bullet}$}\put(32,-20.5){\line(1,0){10}}\put(37.25,-22.75){\line(0,1){7.25}}
\put(40,20){$\overset{(\frac{1}{2},\frac{3}{2})}{\bullet}$}
\put(40,-21){$\overset{(\frac{3}{2},\frac{1}{2})}{\bullet}$}
\put(42,20.5){\line(1,0){10}}\put(47.25,17.75){\line(0,1){7.25}}
\put(42,-20.5){\line(1,0){10}}\put(47.25,-22.75){\line(0,1){7.25}}
\put(50,20){$\overset{(1,1)}{\bullet}$}\put(52,20.5){\line(1,0){10}}\put(57.25,17.75){\line(0,1){7.25}}
\put(50,-21){$\overset{(1,1)}{\bullet}$}\put(52,-20.5){\line(1,0){10}}\put(57.25,-22.75){\line(0,1){7.25}}
\put(60,20){$\overset{(\frac{3}{2},\frac{1}{2})}{\bullet}$}
\put(60,-21){$\overset{(\frac{1}{2},\frac{3}{2})}{\bullet}$}
\put(62,20.5){\line(1,0){10}}\put(67.25,17.75){\line(0,1){7.25}}
\put(62,-20.5){\line(1,0){10}}\put(67.25,-22.75){\line(0,1){7.25}}
\put(70,20){$\overset{(2,0)}{\bullet}$}
\put(70,-21){$\overset{(0,2)}{\bullet}$}
\put(25,25){$\overset{(0,\frac{5}{2})}{\bullet}$}\put(27,25.5){\line(1,0){10}}\put(32.25,22.75){\line(0,1){7.25}}
\put(25,-26){$\overset{(\frac{5}{2},0)}{\bullet}$}\put(27,-25.5){\line(1,0){10}}\put(32.25,-27.75){\line(0,1){7.25}}
\put(35,25){$\overset{(\frac{1}{2},2)}{\bullet}$}\put(37,25.5){\line(1,0){10}}\put(42.25,22.75){\line(0,1){7.25}}
\put(35,-26){$\overset{(2,\frac{1}{2})}{\bullet}$}\put(37,-25.5){\line(1,0){10}}\put(42.25,-27.75){\line(0,1){7.25}}
\put(45,25){$\overset{(1,\frac{3}{2})}{\bullet}$}\put(47,25.5){\line(1,0){10}}\put(52.25,22.75){\line(0,1){7.25}}
\put(45,-26){$\overset{(\frac{3}{2},1)}{\bullet}$}\put(47,-25.5){\line(1,0){10}}\put(52.25,-27.75){\line(0,1){7.25}}
\put(55,25){$\overset{(\frac{3}{2},1)}{\bullet}$}\put(57,25.5){\line(1,0){10}}\put(62.25,22.75){\line(0,1){7.25}}
\put(55,-26){$\overset{(1,\frac{3}{2})}{\bullet}$}\put(57,-25.5){\line(1,0){10}}\put(62.25,-27.75){\line(0,1){7.25}}
\put(65,25){$\overset{(2,\frac{1}{2})}{\bullet}$}\put(67,25.5){\line(1,0){10}}\put(72.25,22.75){\line(0,1){7.25}}
\put(65,-26){$\overset{(\frac{1}{2},2)}{\bullet}$}\put(67,-25.5){\line(1,0){10}}\put(72.25,-27.75){\line(0,1){7.25}}
\put(75,25){$\overset{(\frac{5}{2},0)}{\bullet}$}
\put(75,-26){$\overset{(0,\frac{5}{2})}{\bullet}$}
\put(20,30){$\overset{(0,3)}{\bullet}$}\put(22,30.5){\line(1,0){10}}\put(27.25,27.75){\line(0,1){7.25}}
\put(20,-31){$\overset{(3,0)}{\bullet}$}\put(22,-30.5){\line(1,0){10}}\put(27.25,-32.75){\line(0,1){7.25}}
\put(30,30){$\overset{(\frac{1}{2},\frac{5}{2})}{\bullet}$}
\put(30,-31){$\overset{(\frac{5}{2},\frac{1}{2})}{\bullet}$}
\put(32,30.5){\line(1,0){10}}\put(37.25,27.75){\line(0,1){7.25}}
\put(32,-30.5){\line(1,0){10}}\put(37.25,-32.75){\line(0,1){7.25}}
\put(40,30){$\overset{(1,2)}{\bullet}$}\put(42,30.5){\line(1,0){10}}\put(47.25,27.75){\line(0,1){7.25}}
\put(40,-31){$\overset{(2,1)}{\bullet}$}\put(42,-30.5){\line(1,0){10}}\put(47.25,-32.75){\line(0,1){7.25}}
\put(50,30){$\overset{(\frac{3}{2},\frac{3}{2})}{\bullet}$}
\put(50,-31){$\overset{(\frac{3}{2},\frac{3}{2})}{\bullet}$}
\put(52,30.5){\line(1,0){10}}\put(57.25,27.75){\line(0,1){7.25}}
\put(52,-30.5){\line(1,0){10}}\put(57.25,-32.75){\line(0,1){7.25}}
\put(60,30){$\overset{(2,1)}{\bullet}$}\put(62,30.5){\line(1,0){10}}\put(67.25,27.75){\line(0,1){7.25}}
\put(60,-31){$\overset{(1,2)}{\bullet}$}\put(62,-30.5){\line(1,0){10}}\put(67.25,-32.75){\line(0,1){7.25}}
\put(70,30){$\overset{(\frac{5}{2},\frac{1}{2})}{\bullet}$}
\put(70,-31){$\overset{(\frac{1}{2},\frac{5}{2})}{\bullet}$}
\put(72,30.5){\line(1,0){10}}\put(77.25,27.75){\line(0,1){7.25}}
\put(72,-30.5){\line(1,0){10}}\put(77.25,-32.75){\line(0,1){7.25}}
\put(80,30){$\overset{(3,0)}{\bullet}$}
\put(80,-31){$\overset{(0,3)}{\bullet}$}
\put(15,35){$\overset{(0,\frac{7}{2})}{\bullet}$}\put(17,35.5){\line(1,0){10}}\put(22.25,32.75){\line(0,1){7.25}}
\put(15,-36){$\overset{(\frac{7}{2},0)}{\bullet}$}\put(17,-35.5){\line(1,0){10}}\put(22.25,-37.75){\line(0,1){7.25}}
\put(25,35){$\overset{(\frac{1}{2},3)}{\bullet}$}\put(27,35.5){\line(1,0){10}}\put(32.25,32.75){\line(0,1){7.25}}
\put(25,-36){$\overset{(3,\frac{1}{2})}{\bullet}$}\put(27,-35.5){\line(1,0){10}}\put(32.25,-37.75){\line(0,1){7.25}}
\put(35,35){$\overset{(1,\frac{5}{2})}{\bullet}$}\put(37,35.5){\line(1,0){10}}\put(42.25,32.75){\line(0,1){7.25}}
\put(35,-36){$\overset{(\frac{5}{2},1)}{\bullet}$}\put(37,-35.5){\line(1,0){10}}\put(42.25,-37.75){\line(0,1){7.25}}
\put(45,35){$\overset{(\frac{3}{2},2)}{\bullet}$}\put(47,35.5){\line(1,0){10}}\put(52.25,32.75){\line(0,1){7.25}}
\put(45,-36){$\overset{(2,\frac{3}{2})}{\bullet}$}\put(47,-35.5){\line(1,0){10}}\put(52.25,-37.75){\line(0,1){7.25}}
\put(55,35){$\overset{(2,\frac{3}{2})}{\bullet}$}\put(57,35.5){\line(1,0){10}}\put(62.25,32.75){\line(0,1){7.25}}
\put(55,-36){$\overset{(\frac{3}{2},2)}{\bullet}$}\put(57,-35.5){\line(1,0){10}}\put(62.25,-37.75){\line(0,1){7.25}}
\put(65,35){$\overset{(\frac{5}{2},1)}{\bullet}$}\put(67,35.5){\line(1,0){10}}\put(72.25,32.75){\line(0,1){7.25}}
\put(65,-36){$\overset{(1,\frac{5}{2})}{\bullet}$}\put(67,-35.5){\line(1,0){10}}\put(72.25,-37.75){\line(0,1){7.25}}
\put(75,35){$\overset{(3,\frac{1}{2})}{\bullet}$}\put(77,35.5){\line(1,0){10}}\put(82.25,32.75){\line(0,1){7.25}}
\put(75,-36){$\overset{(\frac{1}{2},3)}{\bullet}$}\put(77,-35.5){\line(1,0){10}}\put(82.25,-37.75){\line(0,1){7.25}}
\put(85,35){$\overset{(\frac{7}{2},0)}{\bullet}$}
\put(85,-36){$\overset{(0,\frac{7}{2})}{\bullet}$}
\put(10,40){$\overset{(0,4)}{\bullet}$}\put(12,40.5){\line(1,0){10}}
\put(10,-41){$\overset{(4,0)}{\bullet}$}\put(12,-40.5){\line(1,0){10}}
\put(20,40){$\overset{(\frac{1}{2},\frac{7}{2})}{\bullet}$}\put(22,40.5){\line(1,0){10}}
\put(20,-41){$\overset{(\frac{7}{2},\frac{1}{2})}{\bullet}$}\put(22,-40.5){\line(1,0){10}}
\put(30,40){$\overset{(1,3)}{\bullet}$}\put(32,40.5){\line(1,0){10}}
\put(30,-41){$\overset{(3,1)}{\bullet}$}\put(32,-40.5){\line(1,0){10}}
\put(40,40){$\overset{(\frac{3}{2},\frac{5}{2})}{\bullet}$}\put(42,40.5){\line(1,0){10}}
\put(40,-41){$\overset{(\frac{5}{2},\frac{3}{2})}{\bullet}$}\put(42,-40.5){\line(1,0){10}}
\put(50,40){$\overset{(2,2)}{\bullet}$}\put(52,40.5){\line(1,0){10}}
\put(50,-41){$\overset{(2,2)}{\bullet}$}\put(52,-40.5){\line(1,0){10}}
\put(60,40){$\overset{(\frac{5}{2},\frac{3}{2})}{\bullet}$}\put(62,40.5){\line(1,0){10}}
\put(60,-41){$\overset{(\frac{3}{2},\frac{5}{2})}{\bullet}$}\put(62,-40.5){\line(1,0){10}}
\put(70,40){$\overset{(3,1)}{\bullet}$}\put(72,40.5){\line(1,0){10}}
\put(70,-41){$\overset{(1,3)}{\bullet}$}\put(72,-40.5){\line(1,0){10}}
\put(80,40){$\overset{(\frac{7}{2},\frac{1}{2})}{\bullet}$}\put(82,40.5){\line(1,0){10}}
\put(80,-41){$\overset{(\frac{1}{2},\frac{7}{2})}{\bullet}$}\put(82,-40.5){\line(1,0){10}}
\put(90,40){$\overset{(4,0)}{\bullet}$}
\put(90,-41){$\overset{(0,4)}{\bullet}$}
\put(11.5,45){$\vdots$}
\put(21.5,45){$\vdots$}
\put(31.5,45){$\vdots$}
\put(41.5,45){$\vdots$}
\put(52,42){\vector(0,1){10}}\put(48,50){$\sY_3$}
\put(61.5,45){$\vdots$}
\put(71.5,45){$\vdots$}
\put(81.5,45){$\vdots$}
\put(91.5,45){$\vdots$}
\put(11.5,-45){$\vdots$}
\put(21.5,-45){$\vdots$}
\put(31.5,-45){$\vdots$}
\put(41.5,-45){$\vdots$}
\put(51.5,-45){$\vdots$}
\put(61.5,-45){$\vdots$}
\put(71.5,-45){$\vdots$}
\put(81.5,-45){$\vdots$}
\put(91.5,-45){$\vdots$}
\put(10,0.5){\line(1,0){42}}\put(50,0.5){\vector(1,0){42}}\put(90,2){$\sX_3$}
\put(16.5,32){$\vdots$}
\put(16.5,29){$\vdots$}
\put(16.5,26){$\vdots$}
\put(16.5,23){$\vdots$}
\put(16.5,20){$\vdots$}
\put(16.5,17){$\vdots$}
\put(16.5,14){$\vdots$}
\put(16.5,11){$\vdots$}
\put(16.5,9){$\vdots$}
\put(16.5,6){$\vdots$}
\put(16.5,3){$\vdots$}
\put(16.5,1.5){$\cdot$}
\put(16.5,0){$\cdot$}
\put(16.5,-32){$\vdots$}
\put(16.5,-29){$\vdots$}
\put(16.5,-26){$\vdots$}
\put(16.5,-23){$\vdots$}
\put(16.5,-20){$\vdots$}
\put(16.5,-17){$\vdots$}
\put(16.5,-14){$\vdots$}
\put(16.5,-11){$\vdots$}
\put(16.5,-9){$\vdots$}
\put(16.5,-6){$\cdot$}
\put(14.5,-3){$-\frac{7}{2}$}
\put(21.5,27){$\vdots$}
\put(21.5,24){$\vdots$}
\put(21.5,21){$\vdots$}
\put(21.5,18){$\vdots$}
\put(21.5,15){$\vdots$}
\put(21.5,13){$\vdots$}
\put(21.5,9){$\vdots$}
\put(21.5,6){$\vdots$}
\put(21.5,3){$\vdots$}
\put(21.5,1.5){$\cdot$}
\put(21.5,0){$\cdot$}
\put(21.5,-27){$\vdots$}
\put(21.5,-24){$\vdots$}
\put(21.5,-21){$\vdots$}
\put(21.5,-18){$\vdots$}
\put(21.5,-15){$\vdots$}
\put(21.5,-13){$\vdots$}
\put(21.5,-9){$\vdots$}
\put(21.5,-6){$\vdots$}
\put(19.5,-3){$-3$}
\put(26.5,22){$\vdots$}
\put(26.5,19){$\vdots$}
\put(26.5,16){$\vdots$}
\put(26.5,13){$\vdots$}
\put(26.5,10){$\vdots$}
\put(26.5,7){$\vdots$}
\put(26.5,4){$\vdots$}
\put(26.5,1){$\vdots$}
\put(26.5,-22){$\vdots$}
\put(26.5,-19){$\vdots$}
\put(26.5,-16){$\vdots$}
\put(26.5,-13){$\vdots$}
\put(26.5,-10){$\vdots$}
\put(26.5,-7){$\vdots$}
\put(26.5,-4){$\vdots$}
\put(26.5,-5){$\cdot$}
\put(24.5,-3){$-\frac{5}{2}$}
\put(31.5,17){$\vdots$}
\put(31.5,14){$\vdots$}
\put(31.5,11){$\vdots$}
\put(31.5,8){$\vdots$}
\put(31.5,5){$\vdots$}
\put(31.5,2){$\vdots$}
\put(31.5,0.5){$\cdot$}
\put(31.5,-17){$\vdots$}
\put(31.5,-14){$\vdots$}
\put(31.5,-11){$\vdots$}
\put(31.5,-8){$\vdots$}
\put(31.5,-5){$\vdots$}
\put(29.5,-3){$-2$}
\put(36.5,12){$\vdots$}
\put(36.5,9){$\vdots$}
\put(36.5,6){$\vdots$}
\put(36.5,3){$\vdots$}
\put(36.5,1.5){$\cdot$}
\put(36.5,0){$\cdot$}
\put(36.5,-12){$\vdots$}
\put(36.5,-9){$\vdots$}
\put(36.5,-6){$\vdots$}
\put(34.5,-3){$-\frac{3}{2}$}
\put(41.5,7){$\vdots$}
\put(41.5,4){$\vdots$}
\put(41.5,1){$\vdots$}
\put(39.5,-3){$-1$}
\put(46.5,2){$\vdots$}
\put(46.5,0.5){$\cdot$}
\put(41.5,-7){$\vdots$}
\put(41.5,-4.5){$\cdot$}
\put(44.5,-2){${\scriptstyle-\frac{1}{2}}$}
\put(51,-3){$0$}
\put(56.5,2){$\vdots$}
\put(56.5,0.5){$\cdot$}
\put(56.5,-2){${\scriptstyle\frac{1}{2}}$}
\put(61.5,7){$\vdots$}
\put(61.5,4){$\vdots$}
\put(61.5,1){$\vdots$}
\put(61.5,-7){$\vdots$}
\put(61.5,-4.5){$\cdot$}
\put(61.5,-3){$1$}
\put(66.5,12){$\vdots$}
\put(66.5,9){$\vdots$}
\put(66.5,6){$\vdots$}
\put(66.5,3){$\vdots$}
\put(66.5,1.5){$\cdot$}
\put(66.5,0){$\cdot$}
\put(66.5,-12){$\vdots$}
\put(66.5,-9){$\vdots$}
\put(66.5,-6){$\vdots$}
\put(66.5,-3){$\frac{3}{2}$}
\put(71.5,17){$\vdots$}
\put(71.5,14){$\vdots$}
\put(71.5,11){$\vdots$}
\put(71.5,8){$\vdots$}
\put(71.5,5){$\vdots$}
\put(71.5,2){$\vdots$}
\put(71.5,0.5){$\cdot$}
\put(71.5,-17){$\vdots$}
\put(71.5,-14){$\vdots$}
\put(71.5,-11){$\vdots$}
\put(71.5,-8){$\vdots$}
\put(71.5,-5){$\vdots$}
\put(71.5,-3){$2$}
\put(76.5,22){$\vdots$}
\put(76.5,19){$\vdots$}
\put(76.5,16){$\vdots$}
\put(76.5,13){$\vdots$}
\put(76.5,10){$\vdots$}
\put(76.5,7){$\vdots$}
\put(76.5,4){$\vdots$}
\put(76.5,1){$\vdots$}
\put(76.5,-22){$\vdots$}
\put(76.5,-19){$\vdots$}
\put(76.5,-16){$\vdots$}
\put(76.5,-13){$\vdots$}
\put(76.5,-10){$\vdots$}
\put(76.5,-7){$\vdots$}
\put(76.5,-5){$\cdot$}
\put(76.5,-3){$\frac{5}{2}$}
\put(81.5,27){$\vdots$}
\put(81.5,24){$\vdots$}
\put(81.5,21){$\vdots$}
\put(81.5,18){$\vdots$}
\put(81.5,15){$\vdots$}
\put(81.5,13){$\vdots$}
\put(81.5,9){$\vdots$}
\put(81.5,6){$\vdots$}
\put(81.5,3){$\vdots$}
\put(81.5,1.5){$\cdot$}
\put(81.5,0){$\cdot$}
\put(81.5,-27){$\vdots$}
\put(81.5,-24){$\vdots$}
\put(81.5,-21){$\vdots$}
\put(81.5,-18){$\vdots$}
\put(81.5,-15){$\vdots$}
\put(81.5,-13){$\vdots$}
\put(81.5,-9){$\vdots$}
\put(81.5,-6){$\vdots$}
\put(81.5,-3){$3$}
\put(86.5,32){$\vdots$}
\put(86.5,29){$\vdots$}
\put(86.5,26){$\vdots$}
\put(86.5,23){$\vdots$}
\put(86.5,20){$\vdots$}
\put(86.5,17){$\vdots$}
\put(86.5,14){$\vdots$}
\put(86.5,11){$\vdots$}
\put(86.5,9){$\vdots$}
\put(86.5,6){$\vdots$}
\put(86.5,3){$\vdots$}
\put(86.5,1.5){$\cdot$}
\put(86.5,0){$\cdot$}
\put(86.5,-32){$\vdots$}
\put(86.5,-29){$\vdots$}
\put(86.5,-26){$\vdots$}
\put(86.5,-23){$\vdots$}
\put(86.5,-20){$\vdots$}
\put(86.5,-17){$\vdots$}
\put(86.5,-14){$\vdots$}
\put(86.5,-11){$\vdots$}
\put(86.5,-9){$\vdots$}
\put(86.5,-6){$\cdot$}
\put(86.5,-3){$\frac{7}{2}$}
\put(53.8,1.7){$\cdot$}\put(54.3,2.2){$\cdot$}\put(54.8,2.7){$\cdot$}\put(55.3,3.3){$\cdot$}\put(55.8,3.8){$\cdot$}
\put(56.3,4.3){$\cdot$}
\put(52.8,-0.7){$\cdot$}\put(53.3,-1.2){$\cdot$}\put(53.8,-1.7){$\cdot$}\put(54.3,-2.3){$\cdot$}\put(54.8,-2.8){$\cdot$}
\put(55.3,-3.3){$\cdot$}
\put(58.8,6.8){$\cdot$}\put(59.3,7.3){$\cdot$}\put(59.8,7.8){$\cdot$}\put(60.3,8.3){$\cdot$}\put(60.8,8.8){$\cdot$}
\put(61.3,9.3){$\cdot$}
\put(57.8,-5.8){$\cdot$}\put(58.3,-6.3){$\cdot$}\put(58.8,-6.8){$\cdot$}\put(59.3,-7.3){$\cdot$}\put(59.8,-7.8){$\cdot$}
\put(60.3,-8.3){$\cdot$}
\put(63.8,11.8){$\cdot$}\put(64.3,12.3){$\cdot$}\put(64.8,12.8){$\cdot$}\put(65.3,13.3){$\cdot$}\put(65.8,13.8){$\cdot$}
\put(66.3,14.3){$\cdot$}
\put(62.8,-10.8){$\cdot$}\put(63.3,-11.3){$\cdot$}\put(63.8,-11.8){$\cdot$}\put(64.3,-12.3){$\cdot$}\put(64.8,-12.8){$\cdot$}
\put(65.3,-13.3){$\cdot$}
\put(68.8,16.8){$\cdot$}\put(69.3,17.3){$\cdot$}\put(69.8,17.8){$\cdot$}\put(70.3,18.3){$\cdot$}\put(70.8,18.8){$\cdot$}
\put(71.3,19.3){$\cdot$}
\put(67.8,-15.8){$\cdot$}\put(68.3,-16.3){$\cdot$}\put(68.8,-16.8){$\cdot$}\put(69.3,-17.3){$\cdot$}\put(69.8,-17.8){$\cdot$}
\put(70.3,-18.3){$\cdot$}
\put(33.8,21.8){$\cdot$}\put(34.3,22.3){$\cdot$}\put(34.8,22.8){$\cdot$}\put(35.3,23.3){$\cdot$}\put(35.8,23.8){$\cdot$}
\put(36.3,24.3){$\cdot$}
\put(32.3,-21.8){$\cdot$}\put(32.8,-22.3){$\cdot$}\put(33.3,-22.8){$\cdot$}\put(33.8,-23.3){$\cdot$}\put(34.3,-23.8){$\cdot$}
\put(34.8,-24.3){$\cdot$}
\put(38.8,26.8){$\cdot$}\put(39.3,27.3){$\cdot$}\put(39.8,27.8){$\cdot$}\put(40.3,28.3){$\cdot$}\put(40.8,28.8){$\cdot$}
\put(41.3,29.3){$\cdot$}
\put(37.3,-26.8){$\cdot$}\put(37.8,-27.3){$\cdot$}\put(38.3,-27.8){$\cdot$}\put(38.8,-28.3){$\cdot$}\put(39.3,-28.8){$\cdot$}
\put(39.8,-29.3){$\cdot$}
\put(43.8,31.8){$\cdot$}\put(44.3,32.3){$\cdot$}\put(44.8,32.8){$\cdot$}\put(45.3,33.3){$\cdot$}\put(45.8,33.8){$\cdot$}
\put(46.3,34.3){$\cdot$}
\put(42.3,-31.8){$\cdot$}\put(42.8,-32.3){$\cdot$}\put(43.3,-32.8){$\cdot$}\put(43.8,-33.3){$\cdot$}\put(44.3,-33.8){$\cdot$}
\put(44.8,-34.3){$\cdot$}
\put(48.8,36.8){$\cdot$}\put(49.3,37.3){$\cdot$}\put(49.8,37.8){$\cdot$}\put(50.3,38.3){$\cdot$}\put(50.8,38.8){$\cdot$}
\put(51.3,39.3){$\cdot$}
\put(47.3,-36.8){$\cdot$}\put(47.8,-37.3){$\cdot$}\put(48.3,-37.8){$\cdot$}\put(48.8,-38.3){$\cdot$}\put(49.3,-38.8){$\cdot$}
\put(49.8,-39.3){$\cdot$}
\put(47.3,4.4){$\cdot$}\put(47.8,3.9){$\cdot$}\put(48.3,3.4){$\cdot$}\put(48.8,2.9){$\cdot$}\put(49.3,2.4){$\cdot$}
\put(49.8,1.9){$\cdot$}
\put(50.8,-0.7){$\cdot$}\put(50.3,-1.2){$\cdot$}\put(49.8,-1.7){$\cdot$}\put(49.3,-2.2){$\cdot$}\put(48.8,-2.7){$\cdot$}
\put(48.3,-3.2){$\cdot$}
\put(42.3,9.4){$\cdot$}\put(42.8,8.9){$\cdot$}\put(43.3,8.4){$\cdot$}\put(43.8,7.9){$\cdot$}\put(44.3,7.4){$\cdot$}
\put(44.8,6.9){$\cdot$}
\put(45.8,-5.8){$\cdot$}\put(45.3,-6.3){$\cdot$}\put(44.8,-6.8){$\cdot$}\put(44.3,-7.3){$\cdot$}\put(43.8,-7.8){$\cdot$}
\put(43.3,-8.3){$\cdot$}
\put(37.3,14.4){$\cdot$}\put(37.8,13.9){$\cdot$}\put(38.3,13.4){$\cdot$}\put(38.8,12.9){$\cdot$}\put(39.3,12.4){$\cdot$}
\put(39.8,11.9){$\cdot$}
\put(40.8,-10.8){$\cdot$}\put(40.3,-11.3){$\cdot$}\put(39.8,-11.8){$\cdot$}\put(39.3,-12.3){$\cdot$}\put(38.8,-12.8){$\cdot$}
\put(38.3,-13.3){$\cdot$}
\put(32.3,19.4){$\cdot$}\put(32.8,18.9){$\cdot$}\put(33.3,18.4){$\cdot$}\put(33.8,17.9){$\cdot$}\put(34.3,17.4){$\cdot$}
\put(34.8,16.9){$\cdot$}
\put(35.8,-15.8){$\cdot$}\put(35.3,-16.3){$\cdot$}\put(34.8,-16.8){$\cdot$}\put(34.3,-17.3){$\cdot$}\put(33.8,-17.8){$\cdot$}
\put(33.3,-18.3){$\cdot$}
\put(67.3,24.4){$\cdot$}\put(67.8,23.9){$\cdot$}\put(68.3,23.4){$\cdot$}\put(68.8,22.9){$\cdot$}\put(69.3,22.4){$\cdot$}
\put(69.8,21.9){$\cdot$}
\put(71.3,-21.8){$\cdot$}\put(70.8,-22.3){$\cdot$}\put(70.3,-22.8){$\cdot$}\put(69.8,-23.3){$\cdot$}\put(69.3,-23.8){$\cdot$}
\put(68.8,-24.3){$\cdot$}
\put(62.3,29.4){$\cdot$}\put(62.8,28.9){$\cdot$}\put(63.3,28.4){$\cdot$}\put(63.8,27.9){$\cdot$}\put(64.3,27.4){$\cdot$}
\put(64.8,26.9){$\cdot$}
\put(66.3,-26.8){$\cdot$}\put(65.8,-27.3){$\cdot$}\put(65.3,-27.8){$\cdot$}\put(64.8,-28.3){$\cdot$}\put(64.3,-28.8){$\cdot$}
\put(63.8,-29.3){$\cdot$}
\put(57.3,34.4){$\cdot$}\put(57.8,33.9){$\cdot$}\put(58.3,33.4){$\cdot$}\put(58.8,32.9){$\cdot$}\put(59.3,32.4){$\cdot$}
\put(59.8,31.9){$\cdot$}
\put(61.3,-31.8){$\cdot$}\put(60.8,-32.3){$\cdot$}\put(60.3,-32.8){$\cdot$}\put(59.8,-33.3){$\cdot$}\put(59.3,-33.8){$\cdot$}
\put(58.8,-34.3){$\cdot$}
\put(52.3,39.4){$\cdot$}\put(52.8,38.9){$\cdot$}\put(53.3,38.4){$\cdot$}\put(53.8,37.9){$\cdot$}\put(54.3,37.4){$\cdot$}
\put(54.8,36.9){$\cdot$}
\put(56.3,-36.8){$\cdot$}\put(55.8,-37.3){$\cdot$}\put(55.3,-37.8){$\cdot$}\put(54.8,-38.3){$\cdot$}\put(54.3,-38.8){$\cdot$}
\put(53.8,-39.3){$\cdot$}
\end{picture}
\end{center}
\vspace{0.9cm}
\caption{Extended Weyl diagram of the algebra $\mathfrak{sl}(2,\C)$. A state is associated with each node $(l,\dot{l})$ of the diagram, the mass of which is determined by the formula (\ref{Mass2}).\label{pic3}}
\end{figure}
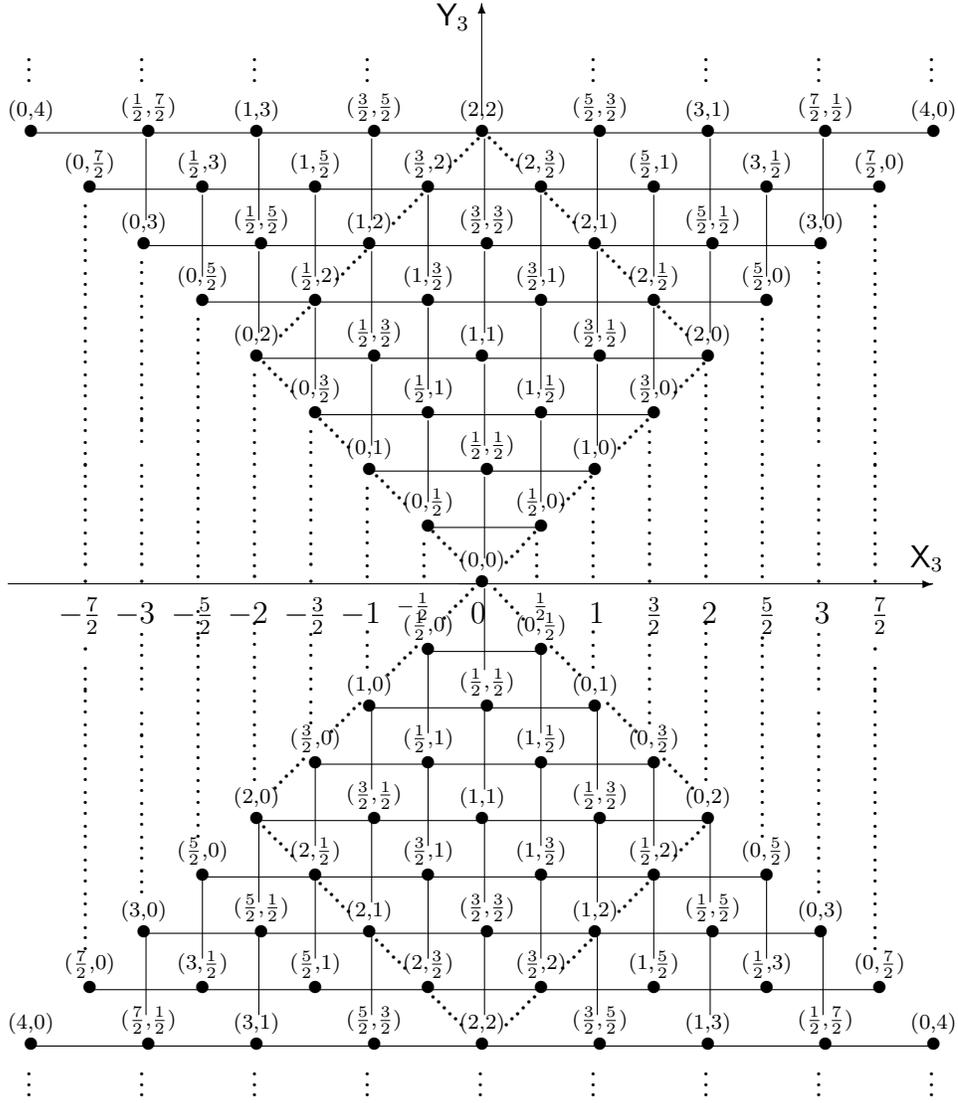

An extended Weyl diagram for $(l,\dot{l})$-manifolds is shown in Figure 3. The mass formula \cite{Var15d} is associated with each node of the weight diagram:
\begin{equation}\label{Mass2}
m=2m_e\left(l+\frac{1}{2}\right)\left(\dot{l}+\frac{1}{2}\right),
\end{equation}
where $m_e$ is the rest mass of the electron. The formula (\ref{Mass2}) describes the mass spectrum of elementary particles with an accuracy of 0.41\% (see \cite{Var1701,Var17a,Var23,Var23a}). Solutions of relativistic wave equations \cite{Var03a,Var07b,Var16b} for arbitrary spin chains ($(l,\dot{l})$-manifolds of the weight diagram in Figure \ref{pic3}) are defined as series of hyperspheric functions on the Lorentz group \cite{Var06}.

\section{Subalgebra $\mathfrak{so}(4)$}
For each generator of the subalgebra $\fK\subset\mathfrak{so}(4,2)$ there are six generators that do not share indexes with it. For example, for $\bsL_{56}$ ($\boldsymbol{\Delta}_3$), these generators are $\bsL_{12}$, $\bsL_{23}$, $\bsL_{31}$, $\bsL_{14}$, $\bsL_{24}$ and $\bsL_{34}$. Due to the permutation relations for the Lie algebra $\mathfrak{so}(4,2)$ (see formula (6)), all these generators commute with $\bsL_{56}$ and, therefore, can be positioned in the horizontal plane of the root diagram of the algebra $\mathfrak{so}(4,2)$. They correspond to the components $\sL_i$ and $\sA_i$ of the angular momentum vector $\bsL$ and the Laplace-Runge-Lenz vector $\bsA$.

Introducing linear combinations
\begin{equation}\label{BY1_2}
\bsK_1=1/2\left(\bsL_{23}+\bsL_{14}\right),\quad\bsK_2=1/2\left(\bsL_{31}+\bsL_{24}\right),\quad
\bsK_3=1/2\left(\bsL_{12}+\bsL_{34}\right),
\end{equation}
\begin{equation}\label{BY2_2}
\bsJ_1=1/2\left(\bsL_{23}-\bsL_{14}\right),\quad\bsJ_2=1/2\left(\bsL_{31}-\bsL_{24}\right),\quad
\bsJ_3=1/2\left(\bsL_{12}-\bsL_{34}\right)
\end{equation}
(see generators (8) and (9) of the Yao basis (\ref{BY1})--(\ref{BY6})), we obtain
\[
\ld\bsK,\bsJ\rd=0,
\]
and also
\[
\ld\bsK_i,\bsK_j\rd=i\varepsilon_{ijk}\bsK_k,\quad\ld\bsJ_i,\bsJ_j\rd=i\varepsilon_{ijk}\bsJ_k.
\]
It follows that the generators $\bsK$ and $\bsJ$ form the bases of two independent algebras $\mathfrak{so}(3)$. Thus, the Lie algebra $\mathfrak{so}(4)$ of the group $\SO(4)$ is isomorphic to the direct sum
\begin{equation}\label{Sum1}
\mathfrak{so}(4)\simeq\mathfrak{su}(2)\oplus \mathfrak{su}(2).
\end{equation}
Accordingly, this means that the group $\SO(4)$ is isomorphic to the direct product $\SO(3)\otimes\SO(3)$.

It is easy to see that the generators $\bsK_3$ and $\bsJ_3$ form the Cartan subalgebra of the Lie algebra $\mathfrak{so}(4)$, $\fK_{\mathfrak{so}(4)}=\lf\bsK_3,\bsJ_3\rf$. From the remaining generators in (\ref{BY1_2})--(\ref{BY2_2}) we will make up four Weyl generators
\[
\begin{array}{cc}
\bsK_+=\bsK_1+i\bsK_2, & \bsK_-=\bsK_1-i\bsK_2,\\[0.1cm]
\bsJ_+=\bsJ_1+i\bsJ_2, & \bsJ_-=\bsJ_1-i\bsJ_2.
\end{array}
\]
Then the Cartan-Weyl basis of the subalgebra $\mathfrak{so}(4)$ will take the form
\begin{equation}\label{CWso4}
\lf\bsK_3,\bsJ_3,\bsK_+,\bsK_-,\bsJ_+,\bsJ_-\rf.
\end{equation}
The basis generators (\ref{CWso4}) satisfy the relations
\begin{equation}\label{JKCom}
\ar\left.
\begin{array}{lll}
\left[\bsK_3,\bsK_+\right]=\bsK_+,&\quad\left[\bsK_3,\bsK_-\right]=-\bsK_-,&\quad\left[\bsK_+,\bsK_-\right]=2\bsK_3,\\
\left[\bsJ_3,\bsJ_+\right]=\bsJ_+,&\quad\left[\bsJ_3,\bsJ_-\right]=-\bsJ_-,&\quad\left[\bsJ_+,\bsJ_-\right]=2\bsJ_3,\\
&\left[\bsK_i,\bsJ_j\right]=0\quad(i,j=+,-,3).&
\end{array}\right\}
\end{equation}
It follows from the relations (\ref{JKCom}) that the root structure of the generators $\bsK_\pm$ and $\bsJ_\pm$ is similar to the root structure of the Weyl generators $\sX_\pm$ and $\sY_\pm$ of the subalgebra $\mathfrak{sl}(2,\C)$. The root diagram of the subalgebra $\mathfrak{so}(4)$ is shown in Figure \ref{pic4}.

\begin{figure}[h]
\unitlength=1mm
\begin{center}
\begin{picture}(20,30)
\put(10,-8){\vector(0,1){34}}
\put(9.15,6.5){$\bullet$}
\put(-5,7.5){\vector(1,0){33}}
\put(25,9){$\bsK_3$}
\put(15,9){$\bsK_+$}
\put(12,4){$\scriptscriptstyle+\frac{1}{2}$}
\put(20,4){$\scriptscriptstyle+1$}
\put(0,4){$\scriptscriptstyle-\frac{1}{2}$}
\put(-6,4){$\scriptscriptstyle-1$}
\put(2.5,9){$\bsK_-$}
\put(5,25){$\bsJ_3$}
\put(5,13){$\scriptscriptstyle+\frac{1}{2}$}
\put(5,1){$\scriptscriptstyle-\frac{1}{2}$}
\put(5,19){$\scriptscriptstyle+1$}
\put(5,-6){$\scriptscriptstyle-1$}
\put(11,16){$\bsJ_+$}
\put(11,-3.5){$\bsJ_-$}
\thicklines
\put(10,7.5){\vector(1,0){14}}
\put(10,7.5){\vector(0,1){14}}
\put(10,7.5){\vector(-1,0){14}}
\put(10,7.5){\vector(0,-1){14}}
\end{picture}
\end{center}
\vspace{0.3cm}
\caption{The root diagram of the Lie algebra $\mathfrak{so}(4)$. The action of each Weyl generator is shown in the ($\bsK_3,\bsJ_3$)-plane.\label{pic4}}
\end{figure}
The corresponding weight diagram is similar to the diagram in Figure 3 with the replacement of the Cartan generators $\sX_3$ and $\sY_3$ by $\bsK_3$ and $\bsJ_3$.

\section{Subalgebra $\mathfrak{so}(2,2)$}
For the following generator $\bsL_{34}$ ($\sA_3$), which is part of the subalgebra $\fK\subset\mathfrak{so}(4,2)$, we have six generators $\bsL_{12}$, $\bsL_{15}$, $\bsL_{25}$, $\bsL_{16}$, $\bsL_{26}$, $\bsL_{56}$, which do not have common indexes with $\bsL_{34}$. These generators correspond to the "hydrogen operators" $\sL_3$, $\sB_1$, $\sB_2$, $\Gamma_1$, $\Gamma_2$, $\Delta_3$ and thus can be positioned in the plane ($\sL_3$,$\Delta_3$). These six generators form linear combinations of the Yao basis:
\[
\bsT_1=1/2\left(-\bsL_{15}-\bsL_{26}\right),\quad\bsT_2=1/2\left(\bsL_{25}-\bsL_{16}\right),\quad
\bsT_0=1/2\left(-\bsL_{12}-\bsL_{56}\right).
\]
\[
\bsS_1=1/2\left(-\bsL_{15}+\bsL_{26}\right),\quad\bsS_2=1/2\left(-\bsL_{25}-\bsL_{16}\right),\quad
\bsS_0=1/2\left(\bsL_{12}-\bsL_{56}\right)
\]
(see formulas (10) and (11) in the basis (\ref{BY1})--(\ref{BY6})). It is easy to see that the components of the generators $\bsT$ and $\bsS$, mutually commute,
\begin{equation}\label{ST}
\ld\bsT,\bsS\rd=0.
\end{equation}
In this case, the components of $\bsT$ form the Lie algebra $\mathfrak{so}(2,1)$:
\[
\ld\bsT_1,\bsT_2\rd=i\bsT_0,\quad\ld\bsT_2,\bsT_0\rd=-i\bsT_1,\quad\ld\bsT_0,\bsT_1\rd=-i\bsT_2.
\]
Similarly for $\bsS$:
\[
\ld\bsS_1,\bsS_2\rd=i\bsS_0,\quad\ld\bsS_2,\bsS_0\rd=-i\bsS_1,\quad\ld\bsS_0,\bsS_1\rd=-i\bsS_2.
\]
By virtue of (\ref{ST}), both algebras are completely separated. By analogy with the algebra $\mathfrak{so}(4)$ (plane ($\sL_3$,$\sA_3$)), which allows decomposition into a direct sum of two algebras $\mathfrak{su}(2)$, in this case we get the Lie algebra $\mathfrak{so}(2,2)$, which is locally isomorphic to the direct sum of two algebras $\mathfrak{so}(2,1)$:
\begin{equation}\label{sum3}
\mathfrak{so}(2,2)\simeq\mathfrak{so}(2,1)\oplus\mathfrak{so}(2,1).
\end{equation}
By virtue of this analogy, the root diagram for the algebra $\mathfrak{so}(2,2)$ will be similar to the diagram for $\mathfrak{so}(4)$ in Figure 4. Further, two commuting generators $\bsT_0$ and $\bsS_0$ can be taken as a basis $\lf\bsT_0,\bsS_0\rf$ for the Cartan subalgebra $\fK\subset\mathfrak{so}(2,2)$. From the remaining generators $\bsT_1$, $\bsT_2$, $\bsS_1$ and $\bsS_2$, we will form the following linear combinations:
\[
\begin{array}{cc}
\bsT_+=\bsT_1+i\bsT_2, & \bsT_-=\bsT_1-i\bsT_2,\\[0.1cm]
\bsS_+=\bsS_1+i\bsS_2, & \bsS_-=\bsS_1-i\bsS_2,
\end{array}
\]
which form a linearly independent set of \textit{Weyl generators} of the algebra $\mathfrak{so}(2,2)$. Thus, the Cartan-Weyl basis of the subalgebra $\mathfrak{so}(2,2)$ has the form
\begin{equation}\label{CWso22}
\lf\bsT_0,\bsS_0,\bsT_+,\bsT_-,\bsS_+,\bsS_-\rf.
\end{equation}
The basis generators (\ref{CWso22}) satisfy the relations
\[
\left[\bsT_0,\bsT_+\right]=-\bsT_+,\quad\left[\bsT_0,\bsT_-\right]=\bsT_-,\quad\left[\bsT_+,\bsT_-\right]=-2\bsT_0,
\]
\[
\left[\bsS_0,\bsS_+\right]=-\bsS_+,\quad\left[\bsS_0,\bsS_-\right]=\bsS_-,\quad\left[\bsS_+,\bsS_-\right]=-2\bsS_0,
\]
\[
\left[\bsT_i,\bsS_j\right]=0\quad(i,j=+,-,0).
\]

The roots of the Weyl generators $\bsT_\pm$, $\bsS_\pm$ form components of a 2-dimensional root vector $\boldsymbol{\alpha}=(\alpha_1,\alpha_2)$, which can be positioned in a 2-dimensional weight space formed by the plane $(\bsT_0,\bsS_0)$. For this implementation of the $\mathfrak{so}(2,2)$ subalgebra, the root vectors have the form
\[
\boldsymbol{\alpha}(\bsT_+)=(1,0),\quad\boldsymbol{\alpha}(\bsT_-)=(-1,0),
\]
\[
\boldsymbol{\alpha}(\bsS_+)=(0,1),\quad\boldsymbol{\alpha}(\bsS_-)=(0,-1).
\]
It is easy to see that the root and weight diagrams of the algebra $\mathfrak{so}(2,2)$ are similar to the corresponding diagrams of the algebras $\mathfrak{sl}(2,\C)$ and $\mathfrak{so}(4)$ (see Figures \ref{pic1}--\ref{pic4}).

In turn, the following elements do not have common indexes with $\bsL_{12}$ ($\sL_3$): $\bsL_{34}$, $\bsL_{35}$, $\bsL_{36}$, $\bsL_{45}$, $\bsL_{46}$ and $\bsL_{56}$, corresponding to the generators $\sA_3$, $\sB_3$, $\Gamma_3$, $\Delta_2$, $\Delta_1$ and $\Delta_3$. Since all these generators commute with $\sL_3$, they can all be positioned in the ($\sA_3,\Delta_3$)-plane. These generators form linear combinations
\[
\bsP_1=1/2\left(-\bsL_{35}-\bsL_{46}\right),\quad\bsP_2=1/2\left(\bsL_{45}-\bsL_{36}\right),\quad
\bsP_0=1/2\left(-\bsL_{34}-\bsL_{56}\right),
\]
\[
\bsQ_1=1/2\left(\bsL_{35}-\bsL_{46}\right),\quad\bsQ_2=1/2\left(\bsL_{45}+\bsL_{36}\right),\quad
\bsQ_0=1/2\left(\bsL_{34}-\bsL_{56}\right).
\]
(see generators (12) and (13) of the Yao basis (\ref{BY1})--(\ref{BY6})). It is easy to verify that
\[
\ld\bsP_1,\bsP_2\rd=i\bsP_0,\quad\ld\bsP_2,\bsP_0\rd=-i\bsP_1,\quad\ld\bsP_0,\bsP_1\rd=-i\bsP_2,
\]
\[
\ld\bsQ_1,\bsQ_2\rd=i\bsQ_0,\quad\ld\bsQ_2,\bsQ_0\rd=-i\bsQ_1,\quad\ld\bsQ_0,\bsQ_1\rd=-i\bsQ_2,
\]
\[
\left[\bsP_i,\bsQ_j\right]=0\quad(i,j=0,1,2).
\]
It follows that the generators $\bsP_i$ and $\bsQ_i$ mutually commute and form two independent Lie algebras $\mathfrak{so}(2,1)$. As in the previous case (plane ($\sL_3,\Delta_3$)), we have the isomorphism $\mathfrak{so}(2,2)\simeq\mathfrak{so}(2,1)\oplus\mathfrak{so}(2,1)$. In turn, for the plane ($\sA_3,\Delta_3$) we have two Cartan generators $\bsP_0$ and $\bsQ_0$ forming the basis $\lf\bsP_0,\bsQ_0\rf$ of the subalgebra $\fK\subset\mathfrak{so}(2,2)$.

Next, from the remaining generators $\bsP_1$, $\bsP_2$, $\bsQ_1$ and $\bsQ_2$, we form the following linear combinations:
\[
\begin{array}{cc}
\bsP_+=\bsP_1+i\bsP_2, & \bsP_-=\bsP_1-i\bsP_2,\\[0.1cm]
\bsQ_+=\bsQ_1+i\bsQ_2, & \bsQ_-=\bsQ_1-i\bsQ_2,
\end{array}
\]
defining Weyl generators for a given implementation of the algebra $\mathfrak{so}(2,2)$ (plane ($\sA_3,\Delta_3$)). The Cartan-Weyl basis in this case has the form
\begin{equation}\label{CWso22_1}
\lf\bsP_0,\bsQ_0,\bsP_+,\bsP_-,\bsQ_+,\bsQ_-\rf.
\end{equation}
As in the previous case, the basis generators (\ref{CWso22_1}) satisfy the relations
\[
\left[\bsP_0,\bsP_+\right]=-\bsP_+,\quad\left[\bsP_0,\bsP_-\right]=\bsP_-,\quad\left[\bsP_+,\bsT_-\right]=-2\bsP_0,
\]
\[
\left[\bsQ_0,\bsQ_+\right]=-\bsQ_+,\quad\left[\bsQ_0,\bsQ_-\right]=\bsQ_-,\quad\left[\bsQ_+,\bsQ_-\right]=-2\bsQ_0,
\]
\[
\left[\bsP_i,\bsQ_j\right]=0\quad(i,j=+,-,0).
\]
From where for root vectors in the plane $(\bsP_0,\bsQ_0)$ we will get
\[
\boldsymbol{\alpha}(\bsP_+)=(1,0),\quad\boldsymbol{\alpha}(\bsP_-)=(-1,0),
\]
\[
\boldsymbol{\alpha}(\bsQ_+)=(0,1),\quad\boldsymbol{\alpha}(\bsQ_-)=(0,-1).
\]
The corresponding root structure is identical to the previous implementation of the algebra $\mathfrak{so}(2,2)$ with the replacement of generators $\bsT_0$, $\bsS_0$, $\bsT_\pm$, $\bsS_\pm$ by $\bsP_0$, $\bsQ_0$, $\bsP_\pm$, $\bsQ_\pm$.

\section{The root diagram of the algebra $\mathfrak{so}(4,2)$}
The above analysis of the root structure of the subalgebras $\mathfrak{so}(4)$ and $\mathfrak{so}(2,2)$ of the algebra $\mathfrak{so}(4,2)$ in each of the orthogonal planes ($\sL_3,\sA_3$), ($\sL_3,\Delta_3$) and ($\sA_3,\Delta_3$) now allows us to put all the results together and build a \textit{root diagram} for the algebra $\mathfrak{so}(4,2)$.

The total number of generators of the subalgebras $\mathfrak{so}(4)$ and $\mathfrak{so}(2,2)$ is 18. However, the generators $\bsK_3$, $\bsJ_3$, $\bsT_0$, $\bsS_0$, $\bsP_0$ and $\bsQ_0$ are not they are linearly independent because they satisfy the following relations:
\begin{equation}\label{Yao2}
\bsJ_3-\bsK_3=\bsP_0-\bsQ_0,\quad\bsJ_3+\bsK_3=\bsS_0-\bsT_0,\quad\bsP_0+\bsQ_0=\bsS_0+\bsT_0.
\end{equation}
Therefore, the set of 18 generators defines a redundant system from which one can obtain the basis of the algebra $\mathfrak{so}(4,2)$ by excluding three generators $\sL_3$, $\sA_3$, $\Delta_3$ using (\ref{Yao2}). This reduces the number of generators to 15, as it should be for algebra $\mathfrak{so}(4,2)$.

As noted above, the three commuting generators $\sL_3$, $\sA_3$ and $\Delta_3$ are \textit{Cartan generators} of the subalgebra $\fK\subset\mathfrak{so}(4,2)$, $\fK=\lf\sL_3,\sA_3,\Delta_3\rf$. They form the basis of a three-dimensional orthogonal system and are located at the origin of the root diagram. Along with the remaining 12 \textit{Weyl generators} $\bsK_\pm$, $\bsJ_\pm$, $\bsT_\pm$, $\bsS_\pm$, $\bsP_\pm$, $\bsQ_\pm$, they form \textit{Cartan-Weyl basis} for algebra $\mathfrak{so}(4,2)$:
\[
\lf\sL_3,\sA_3,\Delta_3,\bsK_+,\bsK_-,\bsJ_+,\bsJ_-,\bsT_+,\bsT_-,\bsS_+,\bsS_-,\bsP_+,\bsP_-,\bsQ_+,\bsQ_-\rf.
\]
Let the common characters be $\bsH_i$ $(i=1\rightarrow 3)$ and $\bsE_\alpha$ $(\alpha=1\rightarrow 12)$ for various Cartan and Weyl generators of the algebra $\mathfrak{so}(4,2)$ are denoted. In the order of the various generators $\bsE_\alpha$ in the root diagram \textit{roots} $\alpha_i$ of each $\bsE_i$ should be determined relative to the three Cartan generators according to the general definition
\[
\ld\bsH_i,\bsE_\alpha\rd=\alpha_i\bsE_\alpha,\quad\forall i=1,2;\;\alpha=1\rightarrow 4.
\]
In the previous paragraphs, the roots were defined relative to $\bsK_3$, $\bsJ_3$, $\bsT_0$, $\bsS_0$ and $\bsP_0$, $\bsQ_0$, and not relative to $\sL_3$, $\sA_3$ and $\Delta_3$. Therefore, the new \textit{root vectors} have the form
\[
\boldsymbol{\alpha}(\bsK_+)=(+1,+1,0),\quad\boldsymbol{\alpha}(\bsT_+)=(+1,0,+1),
\quad\boldsymbol{\alpha}(\bsP_+)=(0,+1,+1),
\]
\[
\boldsymbol{\alpha}(\bsK_-)=(-1,-1,0),\quad\boldsymbol{\alpha}(\bsT_-)=(-1,0,-1),
\quad\boldsymbol{\alpha}(\bsP_-)=(0,-1,-1),
\]
\[
\boldsymbol{\alpha}(\bsJ_+)=(-1,+1,0),\quad\boldsymbol{\alpha}(\bsS_+)=(-1,0,+1),
\quad\boldsymbol{\alpha}(\bsQ_+)=(0,-1,+1),
\]
\[
\boldsymbol{\alpha}(\bsJ_-)=(+1,-1,0),\quad\boldsymbol{\alpha}(\bsS_-)=(+1,0,-1),
\quad\boldsymbol{\alpha}(\bsQ_-)=(0,+1,-1),
\]
\[
\boldsymbol{\alpha}(\sL_3)=(0,0,0),\quad\boldsymbol{\alpha}(\sA_3)=(0,0,0),
\quad\boldsymbol{\alpha}(\Delta_3)=(0,0,0).
\]
Obviously, the four generators $\bsK_+$, $\bsK_-$, $\bsJ_+$, $\bsJ_-$ lead to the root diagram of the algebra $\mathfrak{so}(4)$ with vertex points ($\pm 1,\pm 1$) in ($\sL_3,\sA_3$)-plane (respectively ($\bsK_3,\bsJ_3$)-plane). Similarly, the generators $\bsT_+$, $\bsT_-$, $\bsS_+$, $\bsS_-$ and $\bsP_+$, $\bsP_-$, $\bsQ_+$, $\bsQ_-$ lead to similar root diagrams of algebras $\mathfrak{so}(2,2)$ in the planes ($\sL_3,\Delta_3$) (resp. ($\bsT_0,\bsS_0$)) and ($\sA_3,\Delta_3$) (resp. ($\bsP_0,\bsQ_0$)).

\begin{figure}[ht]
\unitlength=1mm
\begin{center}
\begin{picture}(100,75)(50,60)
\put(100,65){\vector(0,1){70}}
\put(99,99){$\bullet$}
\put(101,97){$0$}
\put(65,100){\vector(1,0){70}}
\put(100,100){\vector(-4,-3){30}}
\put(100,100){\line(4,3){30}}
\put(132,101){$\bsY\rightarrow\sA_3$}
\put(120,101){$\bsP_+$}
\put(120,96){$\bsJ_+$}
\put(111,97){$\scriptstyle+\frac{1}{2}$}
\put(126,97){$\scriptstyle+1$}
\put(83,97){$\scriptstyle-\frac{1}{2}$}
\put(70,97){$\scriptstyle-1$}
\put(77,101){$\bsP_-$}
\put(77,96){$\bsJ_-$}
\put(101,133){$\bsZ\rightarrow\boldsymbol{\Delta}_3$}
\put(100,114){$\scriptstyle+\frac{1}{2}$}
\put(100,85){$\scriptstyle-\frac{1}{2}$}
\put(101,127){$\scriptstyle+1$}
\put(101,71){$\scriptstyle-1$}
\put(101,120){$\bsS_+$}
\put(94,120){$\bsQ_+$}
\put(101,77){$\bsS_-$}
\put(94,77){$\bsQ_-$}
\put(71,74){$\bsX\rightarrow\sL_3$}
\put(76,79){$\scriptstyle+1$}
\put(87,88){$\scriptstyle+\frac{1}{2}$}
\put(122,114){$\scriptstyle-1$}
\put(110,106){$\scriptstyle-\frac{1}{2}$}
\put(83,84){$\bsK_+$}
\put(79,89){$\bsT_+$}
\put(117,109){$\bsK_-$}
\put(112,113){$\bsT_-$}
\thicklines
\put(100,100){\vector(1,0){30}}
\put(100,100){\vector(0,1){30}}
\put(100,100){\vector(-1,0){30}}
\put(100,100){\vector(0,-1){30}}
\put(100,100){\vector(-4,-3){25}}
\put(100,100){\vector(4,3){25}}
\end{picture}
\end{center}
\vspace{-1cm}
\caption{The root diagram of the Lie algebra $\mathfrak{so}(4,2)$.}
\label{pic5}
\end{figure}
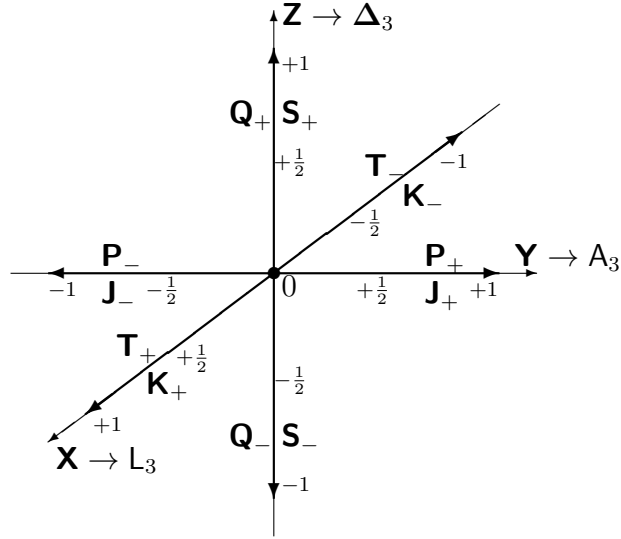

\begin{figure}[ht] %
\centering
\includegraphics[width=11cm]{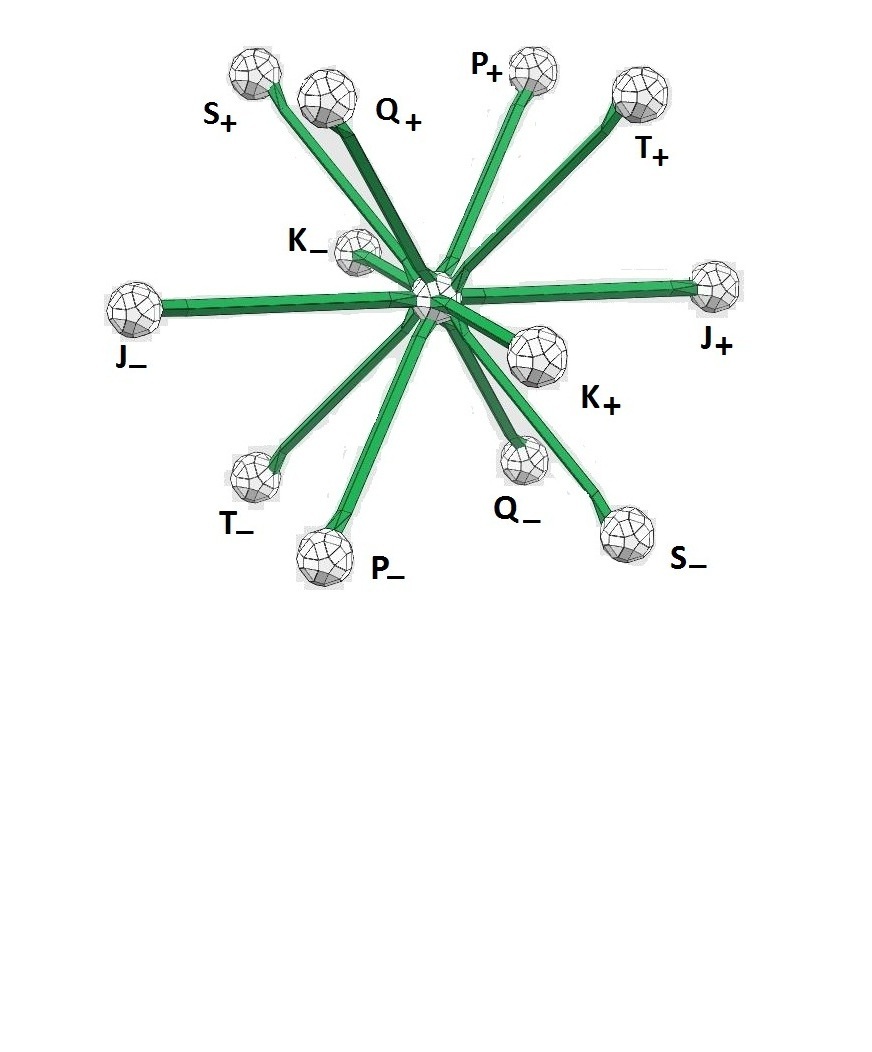}\\
\vspace{-6cm}
\caption{The root diagram of the Lie algebra $\mathfrak{so}(4,2)$, constructed in the Zometool system.}
\label{pic6}
\end{figure}

Thus, the representation of 15 generators on the root diagram of the algebra $\mathfrak{so}(4,2)$ consists of three Cartan generators, and three squares in three perpendicular planes formed by Weyl generators. In Figure \ref{pic5}, for clarity, the vertices of these three squares are aligned with the axes $\bsX=\bsK_3-\bsT_0+\bsP_0\rightarrow\sL_3$, $\bsY=-\bsJ_3-\bsP_0+\bsS_0\rightarrow\sA_3$, $\bsZ=-\bsS_0-\bsQ_0+\bsK_3\rightarrow\Delta_3$.
When rotated by $45^\circ$ relative to each other, these vertices (Weyl generators) form a \textit{cuboctahedron} (see \cite{Hall,Tyss}). The root diagram of the algebra $\mathfrak{so}(4,2)$ in the form of a cuboctahedron is shown in Figure \ref{pic6}.
\begin{figure}[ht] %
\centering
\vspace{-0.5cm}
\includegraphics[width=16cm]{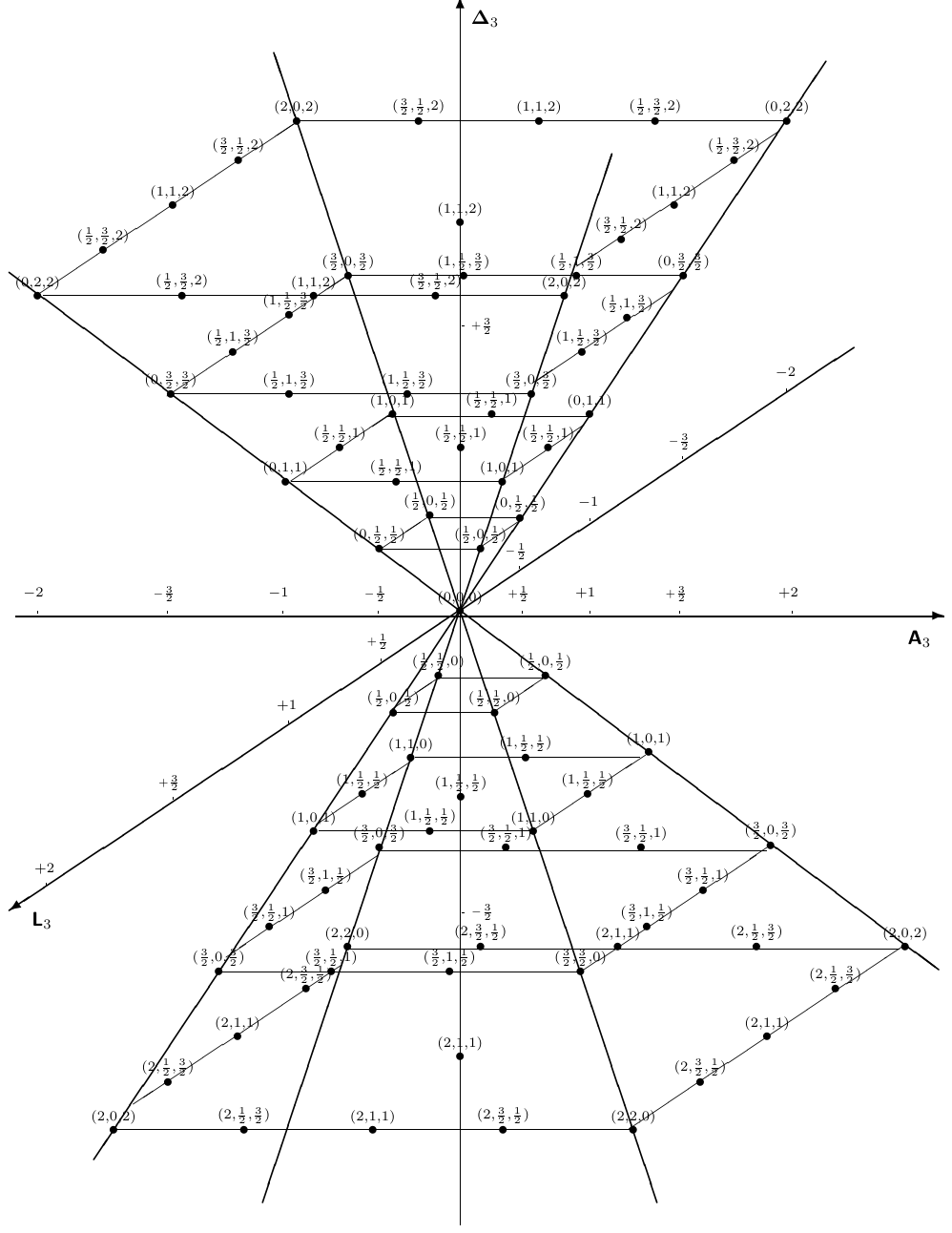}\\
\vspace{-6cm}
\caption{The weight diagram of the Lie algebra $\mathfrak{so}(4,2)$.}
\label{pic7}
\end{figure}
\section{The weight diagram of the Lie algebra $\mathfrak{so}(4,2)$}
Let's proceed to the construction of the \textit{weight diagram} of the algebra $\mathfrak{so}(4,2)$. The weight space is defined by three Cartan generators $\sL_3$, $\sA_3$ and $\Delta_3$, which serve as the basis of a three-dimensional orthogonal coordinate system. The horizontal plane formed by the generators $\sL_3$ and $\sA_3$ includes various $\SO(4)$-manifolds. In paragraph 6, these manifolds are defined with respect to the generators $\bsK_3$ and $\bsJ_3$, which in this diagram correspond to diagonal directions by virtue of their definition following from (\ref{BY1})--(\ref{BY2}):
\[
\bsK_3=\frac{1}{2}\left(\sL_3+\sA_3\right),\quad\bsJ_3=\frac{1}{2}\left(\sL_3-\sA_3\right).
\]
The vertical direction formed by the eigenvalues of the operator $\Delta_3$ adds a radial ladder operator to the variety.

The graphical representation of the weight diagram shown in Figure \ref{pic7} resembles a four-sided pyramid turned upside down and reflected from the plane ($\sL_3,\sA_3$).
We will call this construction $\SO(4,2)$-\textit{tower}. Each given floor of the $\SO(4,2)$-tower is characterized by the main quantum number $n$. The horizontal stripes (floors) correspond to various $l$-subshells, and the points are individual $m$-components (finite-dimensional representations of the group $\SO(4,2)$). The projection of the four-sided pyramid shown in Figure \ref{pic7} onto the plane ($\sL_3,\sA_3$) is the cone of representations of the group $\SO(4)$ (weight diagram of the algebra $\mathfrak{so}(4)$). In turn, projections of $\SO(4,2)$-tower on the plane ($\sL_3,\Delta_3$) and ($\sA_3,\Delta_3$) lead to weight diagrams for subalgebras of $\mathfrak{so}(2,2)$. Finite-dimensional representations of $\boldsymbol{\tau}_{l,\dot{l},n}$ of the group $\SO(4,2)$ are realized in symmetric spaces $\Sym_{(k,r,p)}$ of dimension
\[
\dim\Sym_{(k,r,p)}=(k+1)(r+1)(p+1).
\]

The following mass formula is associated with each node of the weight diagram in Figure \ref{pic7}:
\begin{equation}\label{Mass3}
m=2m_{\textrm{H}}\left(l+\frac{1}{2}\right)\left(\dot{l}+\frac{1}{2}\right)\left(\nu+\frac{1}{2}\right),
\end{equation}
where $m_{\textrm{H}}$ is the relative atomic mass of hydrogen. The output of the formula (\ref{Mass3}) is similar to the output of the mass formula (\ref{Mass2}), see \cite{Var23,Var23a}, which is a special case of (\ref{Mass3}) for $\nu=0$ and $m_{\textrm{H}}\rightarrow m_e$, i.e., when reducing the conformal group $\SO(4,2)$ to its Lorentz subgroup $\SO(3,1)$.

\section{Haenzel Polygonfl\"{a}che}
In order to find an explicit correspondence between individual $m$-components (points in Figure \ref{pic7}) and chemical elements, we use the geometric interpretation of the periodic table proposed by Haenzel in 1943 \cite{Haenzel}.

\begin{figure}[ht] %
\centering
\includegraphics[width=16cm]{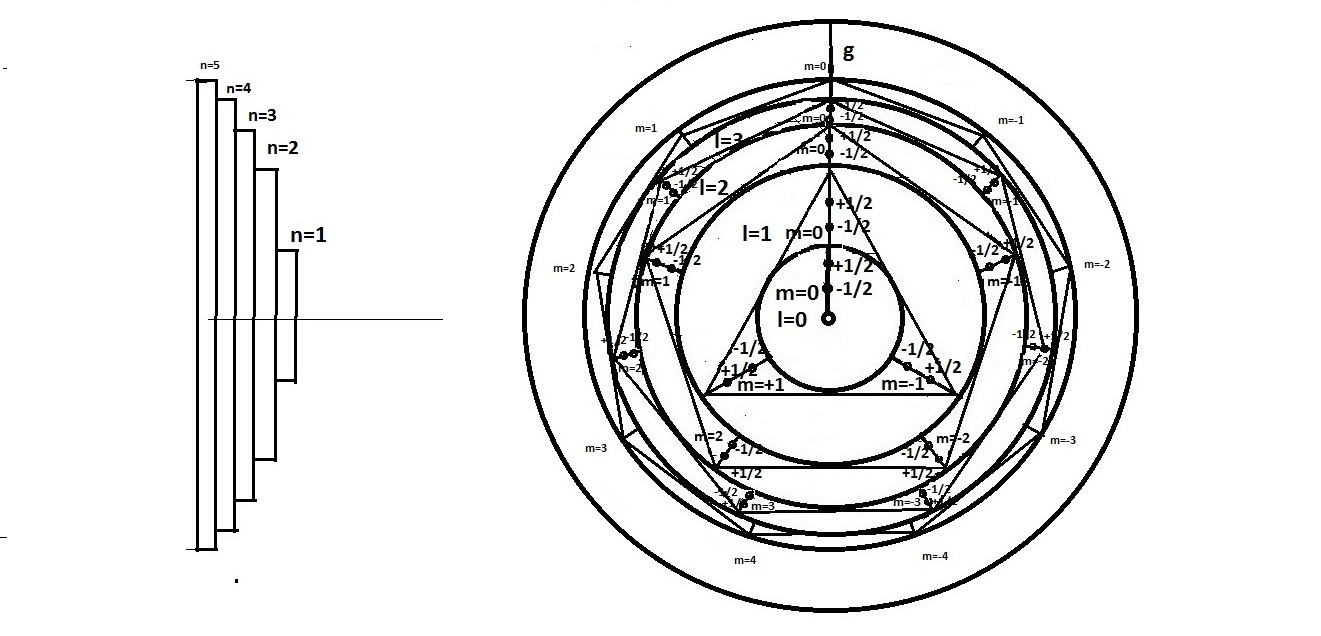}\\
\caption{Haenzel circles and rings.}
\label{pic8}
\end{figure}

The main idea of the article \cite{Haenzel} is the arrangement of the chemical elements of the periodic table according to the quantum numbers $n$, $l$, $m$ and $s$, which Haenzel designates $n_1$, $n_2$, $n_3$ and $n_4$\footnote{Next, we will adhere to the standard notation $n$, $l$, $m$, $s$.}. As is known, the ranges of variation of quantum numbers are as follows:
\begin{eqnarray}
n&=&1,\;2,\;3,\;\ldots,\;\nu,\;\ldots,\nonumber\\
l&=&0,\;1,\;2,\;\ldots,\;n-1,\nonumber\\
m&=&0,\;\pm 1,\;\pm 2,\;\ldots,\;\pm (n-1),\nonumber\\
s&=&-1/2,\;+1/2.\nonumber
\end{eqnarray}
The general scheme of the structure of Haenzel sheets and rings is shown in Figure \ref{pic8}.
In Figure \ref{pic8}, a row of superimposed flat sheets (circles) is numbered from bottom to top: $n=1,2,3\ldots,\nu,\ldots$. All sheets contain a single circle, the area of which is indicated by $l=0$, and a semi-direct $g$ with the direction $m=0$. The unit circle of the lowest sheet $n=1$ is cut along its edge and attached to the sheet $n=2$ located above it, so that you can exit the area $n=1$, $l=0$ only by climbing onto the sheet $n=2$. On this sheet, the unit circle $n=2$, $l=0$ is inscribed in an equilateral triangle (with an angle lying on $g$). The area of the ring between the unit circle and the circle described around the triangle is denoted by $l=1$. Three transversals (transverse lines) to the corners are marked in it: $m=0,\pm 1$. The outer edge of the ring $l=1$ is attached to the plane $n=3$. It contains a single circle $l=0$, the first circular ring $l=1$ with triangular transversals $m=0,\pm 1$ and the second circular ring $l=2$ with pentagonal transversals $m=0,\pm 1,\pm 2$. The outer edge of the ring $l=2$ is connected to the plane $n=4$, etc. In general, the sheet $n=\nu$ contains circular rings $l=0,1,2,\ldots,\nu-1$ and the transversals contained therein $m=0,\pm 1,\pm 2,\ldots,\pm(\nu-1)$.

According to Figure \ref{pic8}, it is assumed that on each transversal in separate circular rings there are two symmetrically located points that differ from each other by the fourth quantum number $s=\pm 1/2$ (spin). All these points are called "eigenvalue points" of Polygonfl\"{a}che, and their fours are the eigenvalues of the quantum system. The system of eigenvalues is distributed on the surface as follows: on the bottom sheet $n=1$ there are two points of eigenvalues $n=1$, $l=0$, $m=0$, $s=\pm 1/2$. There are 8 dots on the second sheet: $n=2$, $l=0$, $m=0$, $s=\pm 1/2$; $l=1$, $m=0,\pm 1$, $s=\pm 1/2$. There are 18 dots on the third sheet. In general, the sheet $n=\nu$ contains $2\nu^2$ points of eigenvalues. At the same time, \textit{a given sheet $n$ contains rings from $l=0$ to $l=n-1$, and in each ring with the number $l$ there are $2l+1$ transversals, so that the plane $n$ contains $\sum^{n-1}_{l=0}(2l+1)$ transversals each}.

\begin{figure}[ht] %
\centering
\vspace{-0.5cm}
\includegraphics[width=16cm]{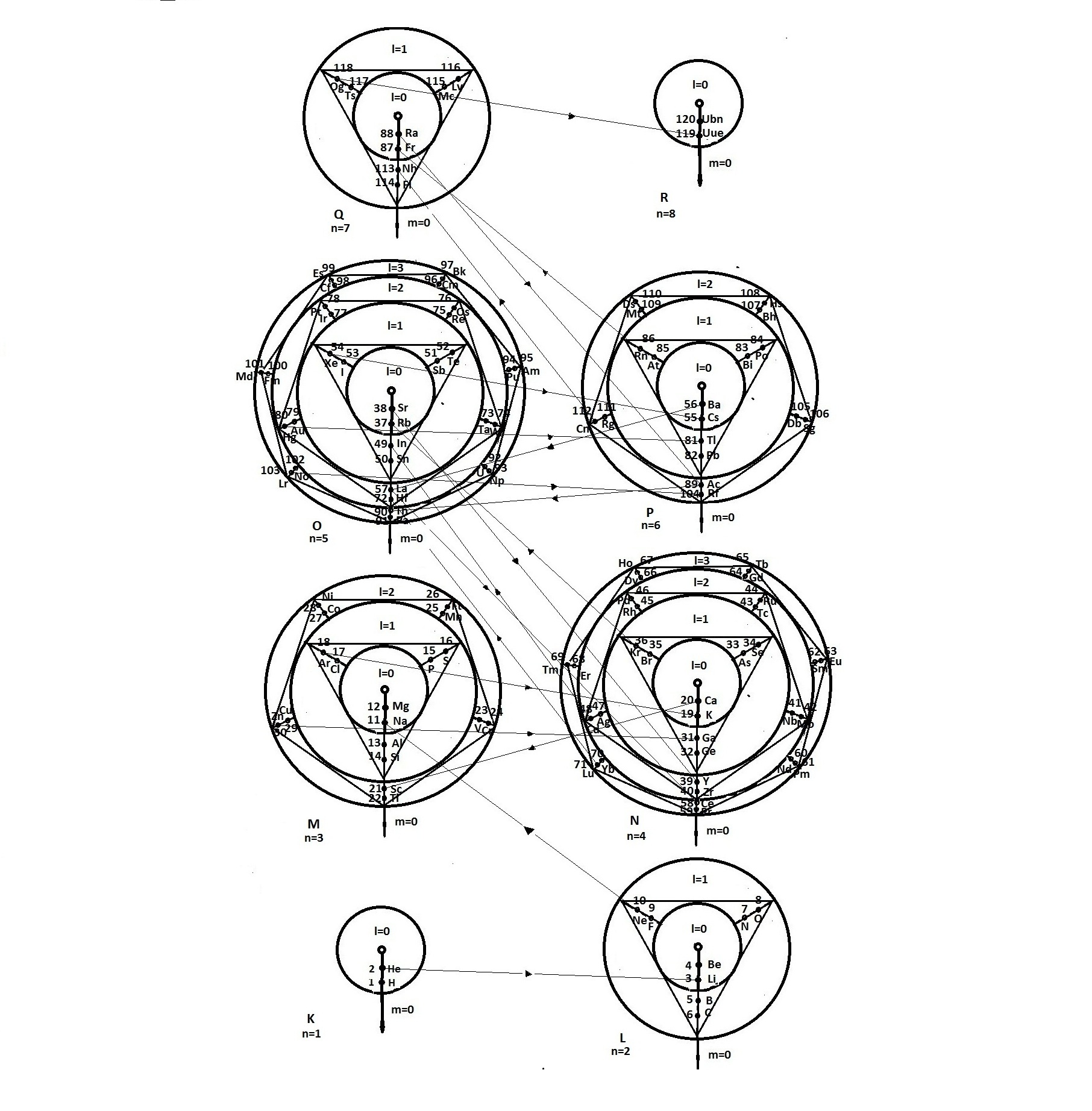}\\
\caption{Polygonfl\"{a}che and the periodic table.}
\label{pic9}
\end{figure}

Figure \ref{pic9} shows the application of Polygonfl\"{a}che to the periodic table of chemical elements.

In the lower sheet $n=1$ there are one- and two-electron systems (\textbf{H} and \textbf{He}), above in $n=2$ there is an eight-row sequence from lithium to neon, on the next sheet there is an eighteen-row sequence, etc. At the time of Haenzel's writing of the article \cite{Haenzel} (1942), only 94 elements were known. Taking into account modern knowledge of the periodic table, Figure \ref{pic9} shows the following extension of Polygonfl\"{a}che: 1) 24 new elements have been added, ranging from americium \textbf{Am} ($Z=95$) to oganesson \textbf{Og} ($Z=118$), which led to the expansion of sheet $n=5$ to ring $l=3$ and sheet $n=7$ to ring $l=2$; 2) added sheet $n=8$, $l=0$ containing two hypothetical elements \textbf{Uue} ($Z=119$) and \textbf{Ubn} ($Z=120$), since these elements logically complete the $f$-shell.

\section{The three-dimensional Finke system}
\begin{figure}[ht] %
\centering
\vspace{-0.5cm}
\includegraphics[width=16cm]{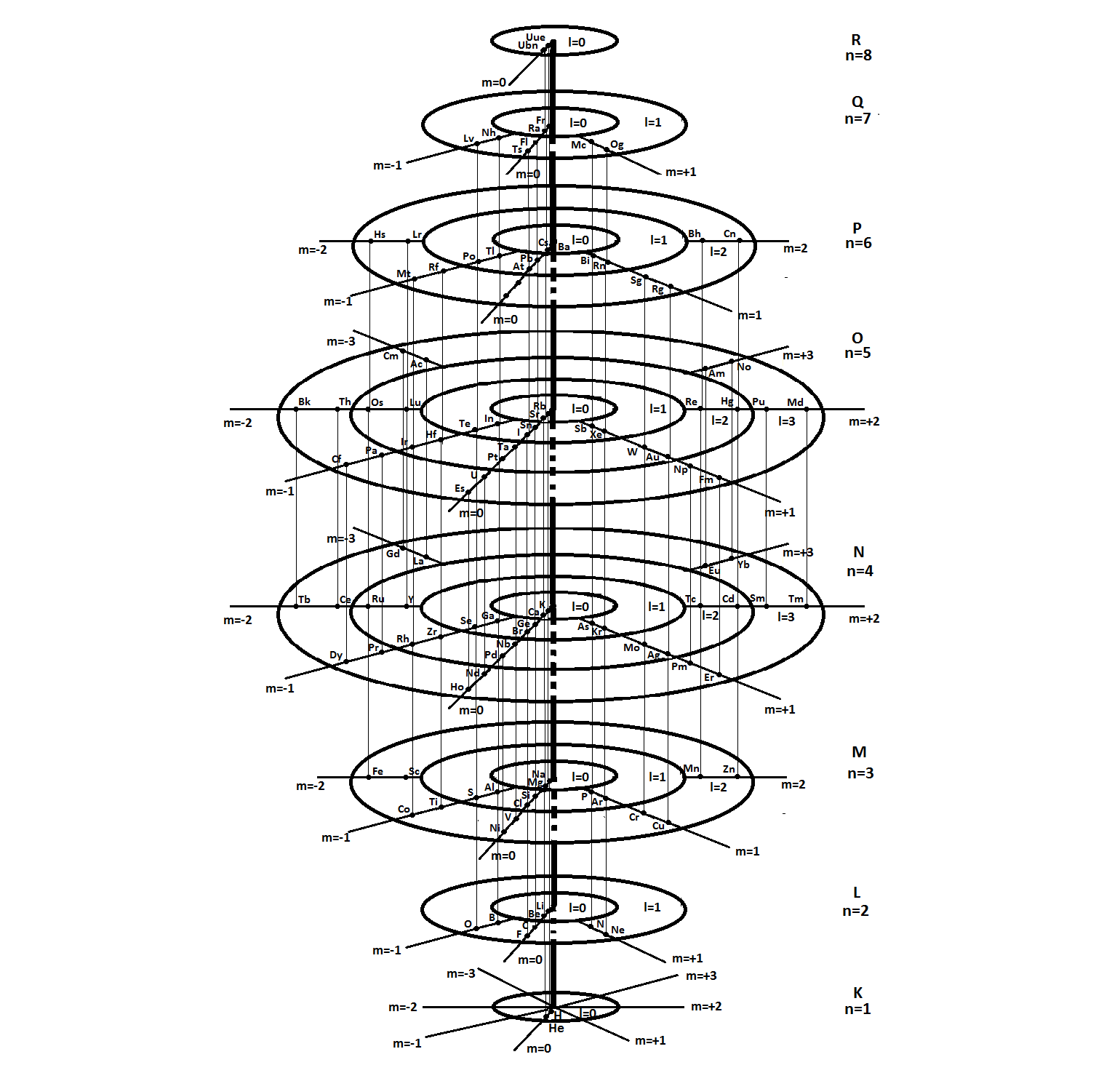}\\
\caption{The three-dimensional Finke system.}
\label{pic10}
\end{figure}

In the same year (1943), the Zeitschrift f\"{u}r Physik published an article by Wilhelm Finke \cite{Finke}, devoted to remarks on Haenzel Polygonfl\"{a}che. First, Finke brings Polygonal\"{a}che in line with Madelung's rule. Secondly, it pushes the Haenzel circles in space, thereby forming a three-dimensional representation of the periodic table\footnote{It should be noted that the three-dimensionality of Polygonfl\"{a}che was already implied by Haenzel, based on the structure of the multilayer surface shown in Figure \ref{pic8}.} (see Figure \ref{pic10}).
Among the main advantages of the three-dimensional representation, it should be noted the appearance of a graphical connection between homologous elements (Bailey-Thomsen-Bohr lines). However, this loses the order of filling the quantum levels (Haenzel circles and rings) of the periodic system, which in Polygonfl\"{a}che is carried out using directional lines (see Figure \ref{pic9}). In this regard, both systems (Haenzel and Finke) should be considered complementary. As in the case of Polygonfl\"{a}che, elements unknown at the time of 1943 were added to the Finke system in Figure \ref{pic10}, i.e. the circles $O$ and $Q$ were expanded, and the circle $R$ with hypothetical elements was added.

\section{A general note on the Haenzel and Finke systems}
In both systems, the first three quantum numbers $n$, $l$ and $m$ have a clear geometric representation, originally borrowed by Bohr in his planetary model of the atom from celestial mechanics (radial and azimuthal motions). In contrast, the fourth quantum number $s$ (spin) on a plane (Polygonal\"{a}che) or in space (Finke system) has an obviously artificial representation in the form of two points on transversals. All this suggests that the concept of spin does not fit into a three-dimensional representation. Therefore, all attempts to give a mechanical interpretation of spin within the framework of three-dimensional space were unsuccessful (for example, the original Uhlenbeck-Goudsmith idea of a spinning top). \textit{Spin is an object of the four-dimensional world}, or, perhaps more correctly, -- \textit{the effect of the fourth dimension}\footnote{In 1971, V.A. Fock, in his work \cite{Fock71}, posed the main question for the doctrine of the principle of periodicity and the theory of the periodic system: "Do the properties of atoms and their constituent parts fit into the framework of purely spatial representations, or do we need to somehow expand the concepts of space and spatial symmetry in order to accommodate the inherent properties of atoms and their components?" \cite[P.~108]{Fock71}. Obviously, by spatial representations and symmetries, Fock means the usual three-dimensional Euclidean space in which the Bohr model is implemented. At the end of the article \cite{Fock71} Fock himself answers the question he posed: "Purely spatial degrees of freedom of an electron are not enough to describe the properties of the electron shell of an atom and it is necessary to go beyond purely spatial concepts in order to express the laws that underlie the periodic table. The electron's new degree of freedom, its spin, makes it possible to describe properties of physical systems that are alien to classical concepts. This internal degree of freedom of the electron is essential for the formulation of the properties of multielectronic systems, and thus for the theoretical justification of the periodic table" \cite[P.~116]{Fock71}.}. Thus, we come to the need to define a four-dimensional representation for an adequate description of spin, which includes all four degrees of freedom ($n,l,m,s$) in a single way.

\section{Group $\SO(4,4)$}
A special pseudo-orthogonal group in eight dimensions, $\SO(4,4)$, corresponds to the rotation group of the eight-dimensional pseudo-Euclidean space $\R^{4,4}$, or, equivalently, the set of $8\times 8$ orthogonal matrices leaving a quadratic form
\[
Q(\boldsymbol{r})=x^2_1+x^2_2+x^2_3+x^2_4-x^2_5-x^2_6-x^2_7-x^2_8=\boldsymbol{r}^T\boldsymbol{r},
\]
where $\boldsymbol{r}=\ld x_1, x_2,x_3,x_4,x_5,x_6,x_7, x_8\rd^T$, invariant.

The structure of the corresponding Lie algebra $\mathfrak{so}(4,4)$ is determined by the commutation properties of its generators $\bsL_{\alpha\beta}$. $\bsL_{\alpha\beta}$ form the basis of the algebra $\mathfrak{so}(4,4)$. The number of independent generators is easy to find: out of 64 possible combinations of the indices $\alpha$ and $\beta$, eight combinations disappear by virtue of $\bsL_{\alpha\alpha}=0$, this reduces the number of generators to 56. Moreover, by virtue of $\bsL_{\alpha\beta}=-\bsL_{\beta\alpha}$ only 28 independent generators remain, the number of which can also be obtained using the formula $n(n-1)/2$, where $n=p+q$ is the dimension of the space $\R^{p,q}$. Thus,
\begin{equation}\label{LO2}
\bsL\Leftrightarrow\begin{bmatrix}
0 & \bsL_{12} & \bsL_{13} & \bsL_{14} & \bsL_{15} & \bsL_{16} & \bsL_{17} & \bsL_{18}\\
  & 0     & \bsL_{23}  & \bsL_{24} & \bsL_{25} & \bsL_{26} & \bsL_{27} & \bsL_{28}\\
  &       & 0      & \bsL_{34} & \bsL_{35} & \bsL_{36} & \bsL_{37} & \bsL_{38}\\
  &       &        & 0     & \bsL_{45}& \bsL_{46} & \bsL_{47} & \bsL_{48}\\
  &       &        &       & 0       & \bsL_{56} & \bsL_{57} & \bsL_{58}\\
  &       &        &       &         & 0         & \bsL_{67} & \bsL_{68}\\
  &       &        &       &         &           & 0         & \bsL_{78}\\
  &       &        &       &         &           &           & 0
\end{bmatrix}.
\end{equation}

The system of 28 generators $\bsL_{\alpha\beta}$ of the algebra $\mathfrak{so}(4,4)$ satisfies the following permutation relations:
\begin{equation}\label{Commut2}
\left[\bsL_{\alpha\beta},\bsL_{\gamma\delta}\right]=i\left(g_{\alpha\delta}\bsL_{\beta\gamma}+g_{\beta\gamma}\bsL_{\alpha\delta}
-g_{\alpha\gamma}\bsL_{\beta\delta}-g_{\beta\delta}\bsL_{\alpha\gamma}\right),
\end{equation}
where $\alpha,\beta,\gamma,\delta=1,\ldots,8$, $\alpha\neq\beta,\;\gamma\neq\delta$, while $g_{11}=g_{22}=g_{33}=g_{44}=1$, $g_{55}=g_{66}=g_{77}=g_{88}=-1$. Thus, we have a 28-dimensional Lie algebra $\mathfrak{so}(4,4)$.

Let's find the maximum subset of commuting generators of the algebra $\mathfrak{so}(4,4)$. As is known, two generators commute if they do not have common indexes. It is easy to see that among the generators of the algebra $\mathfrak{so}(4,4)$, four generators satisfy this condition
\begin{equation}\label{Cartan}
\bsL_{12},\;\bsL_{34},\;\bsL_{56},\;\bsL_{78}.
\end{equation}
The four generators $\lf\bsL_{12},\bsL_{34},\bsL_{56},\bsL_{78}\rf$ form the basis of \textit{maximal abelian subalgebra} $\fK\subset\mathfrak{so}(4,4)$ (\textit{Cartan subalgebra}). The generators (\ref{Cartan}) are called \textit{Cartan generators}. The dimension of the subalgebra $\fK$ defines the \textit{rank} of the Lie algebra, hence $\mathfrak{so}(4,4)$ is a Lie algebra of the fourth rank. Thus, all root and weight diagrams for $\mathfrak{so}(4,2)$ will be four-dimensional\footnote{As a consequence, this excludes explicit visualization of these diagrams for the algebra $\mathfrak{so}(4,4)$.}.

Let's make up the basis of the algebra $\mathfrak{so}(4,4)$ with respect to \textit{maximal compact subgroup}
\[
K=\SU(2)\otimes\SU(2)\otimes\GU(1)
\]
of the group $\SO(4,4)$ by means of the following linear combinations:
\begin{equation}\label{B1}
{}^1\bsK_1=1/2\left(\bsL_{23}+\bsL_{14}\right),\quad{}^1\bsK_2=1/2\left(\bsL_{31}+\bsL_{24}\right),\quad
{}^1\bsK_3=1/2\left(\bsL_{12}+\bsL_{34}\right),
\end{equation}
\begin{equation}\label{B2}
{}^1\bsJ_1=1/2\left(\bsL_{23}-\bsL_{14}\right),\quad{}^1\bsJ_2=1/2\left(\bsL_{31}-\bsL_{24}\right),\quad
{}^1\bsJ_3=1/2\left(\bsL_{12}-\bsL_{34}\right),
\end{equation}
\begin{equation}\label{B3}
{}^1\bsT_1=1/2\left(-\bsL_{15}-\bsL_{26}\right),\quad{}^1\bsT_2=1/2\left(\bsL_{25}-\bsL_{16}\right),\quad
{}^1\bsT_0=1/2\left(-\bsL_{12}-\bsL_{56}\right),
\end{equation}
\begin{equation}\label{B4}
{}^1\bsS_1=1/2\left(-\bsL_{15}+\bsL_{26}\right),\quad{}^1\bsS_2=1/2\left(-\bsL_{25}-\bsL_{16}\right),\quad
{}^1\bsS_0=1/2\left(\bsL_{12}-\bsL_{56}\right),
\end{equation}
\begin{equation}\label{B5}
{}^1\bsP_1=1/2\left(-\bsL_{35}-\bsL_{46}\right),\quad{}^1\bsP_2=1/2\left(\bsL_{45}-\bsL_{36}\right),\quad
{}^1\bsP_0=1/2\left(-\bsL_{34}-\bsL_{56}\right),
\end{equation}
\begin{equation}\label{B6}
{}^1\bsQ_1=1/2\left(\bsL_{35}-\bsL_{46}\right),\quad{}^1\bsQ_2=1/2\left(\bsL_{45}+\bsL_{36}\right),\quad
{}^1\bsQ_0=1/2\left(\bsL_{34}-\bsL_{56}\right),
\end{equation}
\begin{equation}\label{B7}
{}^2\bsK_1=1/2\left(\bsL_{67}+\bsL_{58}\right),\quad{}^2\bsK_2=1/2\left(-\bsL_{57}+\bsL_{68}\right),\quad
{}^2\bsK_3=1/2\left(\bsL_{56}+\bsL_{78}\right),
\end{equation}
\begin{equation}\label{B8}
{}^2\bsJ_1=1/2\left(\bsL_{67}-\bsL_{58}\right),\quad{}^2\bsJ_2=1/2\left(-\bsL_{57}-\bsL_{68}\right),\quad
{}^2\bsJ_3=1/2\left(\bsL_{56}-\bsL_{78}\right),
\end{equation}
\begin{equation}\label{B9}
{}^2\bsT_1=1/2\left(\bsL_{17}+\bsL_{28}\right),\quad{}^2\bsT_2=1/2\left(-\bsL_{27}+\bsL_{18}\right),\quad
{}^2\bsT_0=1/2\left(\bsL_{12}+\bsL_{78}\right),
\end{equation}
\begin{equation}\label{B10}
{}^2\bsS_1=1/2\left(\bsL_{17}-\bsL_{28}\right),\quad{}^2\bsS_2=1/2\left(\bsL_{27}+\bsL_{18}\right),\quad
{}^2\bsS_0=1/2\left(-\bsL_{12}+\bsL_{78}\right),
\end{equation}
\begin{equation}\label{B11}
{}^2\bsP_1=1/2\left(\bsL_{37}+\bsL_{48}\right),\quad{}^2\bsP_2=1/2\left(-\bsL_{47}+\bsL_{38}\right),\quad
{}^2\bsP_0=1/2\left(\bsL_{34}+\bsL_{78}\right),
\end{equation}
\begin{equation}\label{B12}
{}^2\bsQ_1=1/2\left(\bsL_{37}-\bsL_{48}\right),\quad{}^2\bsQ_2=1/2\left(\bsL_{47}+\bsL_{38}\right),\quad
{}^2\bsQ_0=1/2\left(-\bsL_{34}+\bsL_{78}\right).
\end{equation}
The structure of the basis (\ref{B1})--(\ref{B12}) is determined by the following commutation relations:
\begin{equation}\label{CB1}
\ar\left.
\begin{array}{lll}
\ld{}^1\bsK_1,{}^1\bsK_2\rd=-i{}^1\bsK_2,&\quad\ld{}^1\bsK_2,{}^1\bsK_3\rd=-i{}^1\bsK_1,&\quad
\ld{}^1\bsK_3,{}^1\bsK_1\rd=-i{}^1\bsK_2,\\
\ld{}^1\bsJ_1,{}^1\bsJ_2\rd=-i{}^1\bsJ_2,&\quad\ld{}^1\bsJ_2,{}^1\bsJ_3\rd=-i{}^1\bsJ_1,&\quad
\ld{}^1\bsJ_3,{}^1\bsJ_1\rd=-i{}^1\bsJ_2,\\
&\ld{}^1\bsK_i,{}^1\bsJ_j\rd=0\quad(i,j=1,2,3),&\\
\ld{}^1\bsT_1,{}^1\bsT_2\rd=i{}^1\bsT_2,&\quad\ld{}^1\bsT_2,{}^1\bsT_0\rd=-i{}^1\bsT_1,&\quad
\ld{}^1\bsT_0,{}^1\bsT_1\rd=-i{}^1\bsT_2,\\
\ld{}^1\bsS_1,{}^1\bsS_2\rd=i{}^1\bsS_2,&\quad\ld{}^1\bsS_2,{}^1\bsS_0\rd=-i{}^1\bsS_1,&\quad
\ld{}^1\bsS_0,{}^1\bsS_1\rd=-i{}^1\bsS_2,\\
&\ld{}^1\bsT_i,{}^1\bsS_j\rd=0\quad(i,j=0,1,2),&\\
\ld{}^1\bsP_1,{}^1\bsP_2\rd=i{}^1\bsP_2,&\quad\ld{}^1\bsP_2,{}^1\bsP_0\rd=-i{}^1\bsP_1,&\quad
\ld{}^1\bsP_0,{}^1\bsP_1\rd=-i{}^1\bsP_2,\\
\ld{}^1\bsQ_1,{}^1\bsQ_2\rd=i{}^1\bsQ_2,&\quad\ld{}^1\bsQ_2,{}^1\bsQ_0\rd=-i{}^1\bsQ_1,&\quad
\ld{}^1\bsQ_0,{}^1\bsQ_1\rd=-i{}^1\bsQ_2,\\
&\ld{}^1\bsP_i,{}^1\bsQ_j\rd=0\quad(i,j=0,1,2),&
\end{array}\right\}
\end{equation}
\begin{equation}\label{CB2}
\ar\left.
\begin{array}{lll}
\ld{}^2\bsK_1,{}^2\bsK_2\rd=i{}^2\bsK_2,&\quad\ld{}^2\bsK_2,{}^2\bsK_3\rd=i{}^2\bsK_1,&\quad
\ld{}^2\bsK_3,{}^2\bsK_1\rd=i{}^2\bsK_2,\\
\ld{}^2\bsJ_1,{}^2\bsJ_2\rd=i{}^2\bsJ_2,&\quad\ld{}^2\bsJ_2,{}^2\bsJ_3\rd=i{}^2\bsJ_1,&\quad
\ld{}^2\bsJ_3,{}^2\bsJ_1\rd=i{}^2\bsJ_2,\\
&\ld{}^2\bsK_i,{}^2\bsJ_j\rd=0\quad(i,j=1,2,3),&\\
\ld{}^2\bsT_1,{}^2\bsT_2\rd=-i{}^2\bsT_2,&\quad\ld{}^2\bsT_2,{}^2\bsT_0\rd=i{}^2\bsT_1,&\quad
\ld{}^2\bsT_0,{}^2\bsT_1\rd=i{}^2\bsT_2,\\
\ld{}^2\bsS_1,{}^2\bsS_2\rd=-i{}^2\bsS_2,&\quad\ld{}^2\bsS_2,{}^2\bsS_0\rd=i{}^2\bsS_1,&\quad
\ld{}^2\bsS_0,{}^2\bsS_1\rd=i{}^2\bsS_2,\\
&\ld{}^2\bsT_i,{}^2\bsS_j\rd=0\quad(i,j=0,1,2),&\\
\ld{}^2\bsP_1,{}^2\bsP_2\rd=-i{}^2\bsP_2,&\quad\ld{}^2\bsP_2,{}^2\bsP_0\rd=i{}^2\bsP_1,&\quad
\ld{}^2\bsP_0,{}^2\bsP_1\rd=i{}^2\bsP_2,\\
\ld{}^2\bsQ_1,{}^2\bsQ_2\rd=-i{}^2\bsQ_2,&\quad\ld{}^2\bsQ_2,{}^2\bsQ_0\rd=i{}^2\bsQ_1,&\quad
\ld{}^2\bsQ_0,{}^2\bsQ_1\rd=i{}^2\bsQ_2.\\
&\ld{}^2\bsP_i,{}^2\bsQ_j\rd=0\quad(i,j=0,1,2).&
\end{array}\right\}
\end{equation}

The basis (\ref{B1})--(\ref{B12}) contains 36 generators, which creates a redundant system for the algebra $\mathfrak{so}(4,4)$, since the latter consists of 28 independent generators. The basis of the algebra $\mathfrak{so}(4,4)$ can be obtained from (\ref{B1})--(\ref{B12}) by eliminating eight generators using the relations
\begin{equation}\label{Em1_2}
{}^1\bsJ_3+{}^1\bsK_3={}^1\bsS_0-{}^1\bsT_0={}^2\bsT_0-{}^2\bsS_0=\bsL_{12},
\end{equation}
\begin{equation}\label{Em2_2}
{}^1\bsJ_3-{}^1\bsK_3={}^1\bsP_0-{}^1\bsQ_0={}^2\bsQ_0-{}^2\bsP_0=-\bsL_{34},
\end{equation}
\begin{equation}\label{Em3_2}
{}^1\bsP_0+{}^1\bsQ_0={}^1\bsS_0+{}^1\bsT_0=-{}^2\bsK_3-{}^2\bsJ_3=-\bsL_{56},
\end{equation}
\begin{equation}\label{Em4_2}
{}^2\bsK_3-{}^2\bsJ_3={}^2\bsT_0+{}^2\bsS_0={}^2\bsP_0+{}^2\bsQ_0=\bsL_{78}.
\end{equation}
This reduces the number of generators to 28, as it should be for algebra $\mathfrak{so}(4,4)$. Thus, the set of twelve generators ${}^1\bsK_3$, ${}^1\bsJ_3$, ${}^1\bsT_0$, ${}^1\bsS_0$, ${}^1\bsP_0$, ${}^1\bsQ_0$, ${}^2\bsK_3$, ${}^2\bsJ_3$, ${}^2\bsT_0$, ${}^2\bsS_0$, ${}^2\bsP_0$, ${}^2\bsQ_0$ by virtue of the relations (\ref{Em1_2})--(\ref{Em4_2}) emulates the Cartan subalgebra $\fK=\lf\bsL_{12},\bsL_{34},\bsL_{56},\bsL_{78}\rf$ of the Lie algebra $\mathfrak{so}(4,4)$.

Of the remaining 24 basis generators (\ref{B1})--(\ref{B12}), we form Weyl generators by means of the following linear combinations:
\begin{equation}\label{B14}
{}^1\bsJ_\pm={}^1\bsJ_1\pm i{}^1\bsJ_2,\quad{}^1\bsP_\pm={}^1\bsP_1\pm i{}^1\bsP_2,\quad{}^1\bsS_\pm={}^1\bsS_1\pm i{}^1\bsS_2,
\end{equation}
\begin{equation}\label{B15}
{}^1\bsK_\pm={}^1\bsK_1\pm i{}^1\bsK_2,\quad{}^1\bsQ_\pm={}^1\bsQ_1\pm i{}^1\bsQ_2,\quad{}^1\bsT_\pm={}^1\bsT_1\pm i{}^1\bsT_2.
\end{equation}
\begin{equation}\label{B16}
{}^2\bsJ_\pm={}^2\bsJ_1\pm i{}^2\bsJ_2,\quad{}^2\bsP_\pm={}^2\bsP_1\pm i{}^2\bsP_2,\quad{}^2\bsS_\pm={}^2\bsS_1\pm i{}^2\bsS_2,
\end{equation}
\begin{equation}\label{B17}
{}^2\bsK_\pm={}^2\bsK_1\pm i{}^2\bsK_2,\quad{}^2\bsQ_\pm={}^2\bsQ_1\pm i{}^2\bsQ_2,\quad{}^2\bsT_\pm={}^2\bsT_1\pm i{}^2\bsT_2.
\end{equation}
The 12 generators (\ref{B14})--(\ref{B15}) are Weyl generators of the subalgebra $\mathfrak{so}(4,2)$ (Lie algebra of the conformal group $\SO(4,2)$). A detailed study of the structure of Weyl generators (\ref{B14})--(\ref{B15}) depending on various subalgebras of the algebra $\mathfrak{so}(4,2)$ is contained in the previous paragraphs 5--9.

We form two intermediate bases
\begin{equation}\label{IB1}
\lf{}^1\bsK_3,{}^1\bsJ_3,{}^1\bsT_0,{}^1\bsS_0,{}^1\bsP_0,{}^1\bsQ_0,{}^1\bsK_+,{}^1\bsK_-,{}^1\bsJ_+,{}^1\bsJ_-,
{}^1\bsT_+,{}^1\bsT_-,{}^1\bsS_+,{}^1\bsS_-,{}^1\bsP_+,{}^1\bsP_-,{}^1\bsQ_+,{}^1\bsQ_-\rf,
\end{equation}
\begin{equation}\label{IB2}
\lf{}^2\bsK_3,{}^2\bsJ_3,{}^2\bsT_0,{}^2\bsS_0,{}^2\bsP_0,{}^2\bsQ_0,{}^2\bsK_+,{}^2\bsK_-,{}^2\bsJ_+,{}^2\bsJ_-,
{}^2\bsT_+,{}^2\bsT_-,{}^2\bsS_+,{}^2\bsS_-,{}^2\bsP_+,{}^2\bsP_-,{}^2\bsQ_+,{}^2\bsQ_-\rf.
\end{equation}
The generators of bases (\ref{IB1}) and (\ref{IB2}) satisfy the following relations:
\begin{equation}\label{SYB1}
\ar\phantom{-}\left.
\begin{array}{lll}
\ld{}^1\bsK_3,{}^1\bsK_+\rd={}^1\bsK_+,&\quad\ld{}^1\bsK_3,{}^1\bsK_-\rd=-{}^1\bsK_-,&\quad
\ld{}^1\bsK_+,{}^1\bsK_-\rd=2{}^1\bsK_3,\\
\ld{}^1\bsJ_3,{}^1\bsJ_+\rd={}^1\bsJ_+,&\quad\ld{}^1\bsJ_3,{}^1\bsJ_-\rd=-{}^1\bsJ_-,&\quad
\ld{}^1\bsJ_+,{}^1\bsJ_-\rd=2{}^1\bsJ_3,\\
&\ld{}^1\bsK_i,{}^1\bsJ_j\rd=0\quad(i,j=+,-,3),&\\
\ld{}^1\bsT_0,{}^1\bsT_+\rd=-{}^1\bsT_+,&\quad\ld{}^1\bsT_0,{}^1\bsT_-\rd={}^1\bsT_-,&\quad
\ld{}^1\bsT_+,{}^1\bsT_-\rd=-2{}^1\bsT_0,\\
\ld{}^1\bsS_0,{}^1\bsS_+\rd=-{}^1\bsS_+,&\quad\ld{}^1\bsS_+,{}^1\bsS_-\rd={}^1\bsS_-,&\quad
\ld{}^1\bsS_+,{}^1\bsS_-\rd=-2{}^1\bsS_0,\\
&\ld{}^1\bsT_i,{}^1\bsS_j\rd=0\quad(i,j=+,-,0),&\\
\ld{}^1\bsP_0,{}^1\bsP_+\rd=-{}^1\bsP_+,&\quad\ld{}^1\bsP_0,{}^1\bsP_-\rd={}^1\bsP_-,&\quad
\ld{}^1\bsP_+,{}^1\bsP_-\rd=-2{}^1\bsP_0,\\
\ld{}^1\bsQ_0,{}^1\bsQ_+\rd=-{}^1\bsQ_+,&\quad\ld{}^1\bsQ_0,{}^1\bsQ_-\rd={}^1\bsQ_-,&\quad
\ld{}^1\bsQ_+,{}^1\bsQ_-\rd=-2{}^1\bsQ_0,\\
&\ld{}^1\bsP_i,{}^1\bsQ_j\rd=0\quad(i,j=+,-,0),&
\end{array}\right\}
\end{equation}
\begin{equation}\label{SYB2}
\ar\left.
\begin{array}{lll}
\ld{}^2\bsK_3,{}^2\bsK_+\rd={}^2\bsK_+,&\quad\ld{}^2\bsK_3,{}^2\bsK_-\rd=-{}^2\bsK_-,&\quad
\ld{}^2\bsK_+,{}^2\bsK_-\rd=2{}^2\bsK_3,\\
\ld{}^2\bsJ_3,{}^2\bsJ_+\rd={}^2\bsJ_+,&\quad\ld{}^2\bsJ_3,{}^2\bsJ_-\rd=-{}^2\bsJ_-,&\quad
\ld{}^2\bsJ_+,{}^2\bsJ_-\rd=2{}^2\bsJ_3,\\
&\ld{}^2\bsK_i,{}^2\bsJ_j\rd=0\quad(i,j=+,-,3),&\\
\ld{}^2\bsT_0,{}^2\bsT_+\rd={}^2\bsT_+,&\quad\ld{}^2\bsT_0,{}^2\bsT_-\rd=-{}^2\bsT_-,&\quad
\ld{}^2\bsT_+,{}^2\bsT_-\rd=-2{}^2\bsT_0,\\
\ld{}^2\bsS_0,{}^2\bsS_+\rd={}^2\bsS_+,&\quad\ld{}^2\bsS_0,{}^2\bsS_-\rd=-{}^2\bsS_-,&\quad
\ld{}^2\bsS_+,{}^2\bsS_-\rd=-2{}^2\bsS_0,\\
&\ld{}^2\bsT_i,{}^2\bsS_j\rd=0\quad(i,j=+,-,0),&\\
\ld{}^2\bsP_0,{}^2\bsP_+\rd={}^2\bsP_+,&\quad\ld{}^2\bsP_0,{}^2\bsP_-\rd=-{}^2\bsP_-,&\quad
\ld{}^2\bsP_+,{}^2\bsP_-\rd=-2{}^2\bsP_0,\\
\ld{}^2\bsQ_0,{}^2\bsQ_+\rd={}^2\bsQ_+,&\quad\ld{}^2\bsQ_0,{}^2\bsQ_-\rd=-{}^2\bsQ_-,&\quad
\ld{}^2\bsQ_+,{}^2\bsQ_-\rd=-2{}^2\bsQ_0,\\
&\ld{}^2\bsP_i,{}^2\bsQ_j\rd=0\quad(i,j=+,-,0).&
\end{array}\right\}
\end{equation}
Relations (\ref{SYB1}) and (\ref{SYB2}) show that bases (\ref{IB1}) and (\ref{IB2}) are structurally isomorphic. Moreover, each of the bases (\ref{IB1}) and (\ref{IB2}) is a Yao basis of the Lie algebra $\mathfrak{su}(2,2)$ of the group $\SU(2,2)$, which covers the conformal group $\SO(4,2)$ twice. Taking into account relations (\ref{Em1_2})--(\ref{Em4_2}), which emulate the Cartan subalgebra $\fK=\lf\bsL_{12},\bsL_{34},\bsL_{56},\bsL_{78}\rf$, and combined with \textit{Weyl generators} (\ref{B14})--(\ref{B17}), the two bases (\ref{IB1}) and (\ref{IB2}) form the \textit{Cartan-Weyl basis} for the algebra $\mathfrak{so}(4,4)$:
\begin{multline}
\left\{\bsL_{12},\bsL_{34},\bsL_{56},\bsL_{78},{}^1\bsK_+,{}^1\bsK_-,{}^1\bsJ_+,{}^1\bsJ_-,{}^1\bsT_+,{}^1\bsT_-,
{}^1\bsS_+,{}^1\bsS_-,{}^1\bsP_+,{}^1\bsP_-,{}^1\bsQ_+,{}^1\bsQ_-,\right.\\
\left.{}^2\bsK_+,{}^2\bsK_-,{}^2\bsJ_+,{}^2\bsJ_-,{}^2\bsT_+,{}^2\bsT_-,
{}^2\bsS_+,{}^2\bsS_-,{}^2\bsP_+,{}^2\bsP_-,{}^2\bsQ_+,{}^2\bsQ_-\right\}.\label{CWB}
\end{multline}

By virtue of the relations (\ref{SYB1}), the generators ${}^1\bsK$ and ${}^1\bsJ$ form the Lie algebra $\mathfrak{so}(4)$ of the subgroup $\SO(4)$. At this point, ${}^1\bsK$ and ${}^1\bsJ$ form the bases of two independent algebras $\mathfrak{so}(3)$. Thus, the Lie algebra $\mathfrak{so}(4)$ of the subgroup $\SO(4)$ is isomorphic to the direct sum (\ref{Sum1}.)

Further, by virtue of the relations (\ref{SYB1}), generators ${}^1\bsT$, ${}^1\bsS$ and ${}^1\bsP$, ${}^1\bsQ$ form Lie algebras $\mathfrak{so}(2,2)$ of subgroups $\SO(2,2)$. At this point, generators ${}^1\bsT$, ${}^1\bsS$ (resp. ${}^1\bsP$, ${}^1\bsQ$) form the bases of two independent algebras $\mathfrak{so}(2,1)$. Therefore, by analogy with the algebra $\mathfrak{so}(4)$, we obtain the Lie algebra $\mathfrak{so}(2,2)$, which is locally isomorphic to the direct sum (\ref{sum3}).

The set of generators (\ref{B1})--(\ref{B6}) and the corresponding commutation relations (\ref{CB1}) and (\ref{SYB1}) describe the Yao basis \cite{Yao1} of the  Lie algebra $\mathfrak{su}(2,2)\simeq\mathfrak{so}(4,2)$ of a group of pseudounitary unimodular matrices,
\[
\SU(2,2)\simeq\spin_+(2,4)=\left\{\ar\begin{bmatrix} A & B\\ C & D\end{bmatrix}
\in\C_4:\;\det\begin{bmatrix} A & B \\ C & D\end{bmatrix}=1
\right\},
\]
which is a twofold covering of a conformal group.

It is easy to see that the generators (\ref{B7})--(\ref{B12}) of the second half of the basis (\ref{B1})--(\ref{B12}) by virtue of the commutation relations (\ref{CB2}) and (\ref{SYB2}) describe the basis, structurally identical (with the reverse sign) to the Yao basis. Thus, \textit{the basis (\ref{B1})--(\ref{B12}) of the Lie algebra of the spinor group}
\[
\spin_+(4,4)=\left\{s\in\left.\Mat_8(\R)\oplus\Mat_8(\R)\right|\;N(s)=1
\right\}.
\]
\textit{has a structure split into two halves}.

\section{Spin as the fourth generator of the Cartan subalgebra}
As noted above in paragraph 12, in the systems of Haenzel and Finke, the fourth quantum number $s$, denoted in both systems by two points on the transversals, does not have a clear geometric representation. As a consequence, no three-dimensional system is a complete representation of the periodic law. Spin $s$ is the fourth degree of freedom, therefore, for its adequate representation, a \textit{four-dimensional system} is required. Below we will show that such a system is the weight diagram of the algebra $\mathfrak{so}(4,4)$.

Four generators (\ref{Cartan}) form the basis of the Cartan subalgebra $\fK$ of the algebra $\mathfrak{so}(4,4)$, among which the first three generators form the basis of the Cartan subalgebra of the algebra $\mathfrak{so}(4,2)$ (Lie algebra of the conformal group $\SO(4,2)$). According to the \textit{hydrogen implementation} of the algebra $\mathfrak{so}(4,2)$ (Barut representation \cite{Bar72,ACP82}, see also section 3.1), the first generator $\bsL_{12}$ corresponds to the third component of angular momentum, $\bsL_{12}=\sL_3$, the second generator $\bsL_{34}$ corresponds to the third component of the Laplace-Runge-Lenz vector, $\bsL_{34}=\sA_3$. Finally, the third generator $\bsL_{56}$ defines the third component of the radial part of the wave function of a hydrogen-like system, $\bsL_{56}=\Delta_3$. In turn, the generators $\sL_3$ and $\sA_3$ form the basis of the Cartan subalgebra of the Lie algebra $\mathfrak{so}(4)$ of the group $\SO(4)$, within which Pauli \cite{Pauli} explained the complete degeneracy of the hydrogen spectrum. Three generators $\bsL_{12}$, $\bsL_{34}$ and $\bsL_{56}$ form the basis of a three-dimensional coordinate system in which the weight diagram of the algebra $\mathfrak{so}(4,2)$ is constructed in the form of a double inverted pyramid ($\SO(4,2)$-tower), see Figure \ref{pic7}. The eigenvalues of the generators $\bsL_{56}$, $\bsL_{12}$, $\bsL_{34}$ correspond to the quantum numbers $n$, $l$, $m$. The inclusion of the fourth generator $\bsL_{78}$ leads to an eight-dimensional generalization  $\mathfrak{so}(4,4)$, which makes it possible to identify the eigenvalues of the generator $\bsL_{78}$ with the fourth quantum number $s$. Thus,
\begin{eqnarray}
\bsL_{56}&\longrightarrow & n,\nonumber\\
\bsL_{12}&\longrightarrow & l,\nonumber\\
\bsL_{34}&\longrightarrow & m,\nonumber\\
\bsL_{78}&\longrightarrow & s.\nonumber
\end{eqnarray}

\section{Weight diagram of algebra $\mathfrak{so}(4,4)$}
\begin{figure}[ht] %
\centering
\includegraphics[width=15cm]{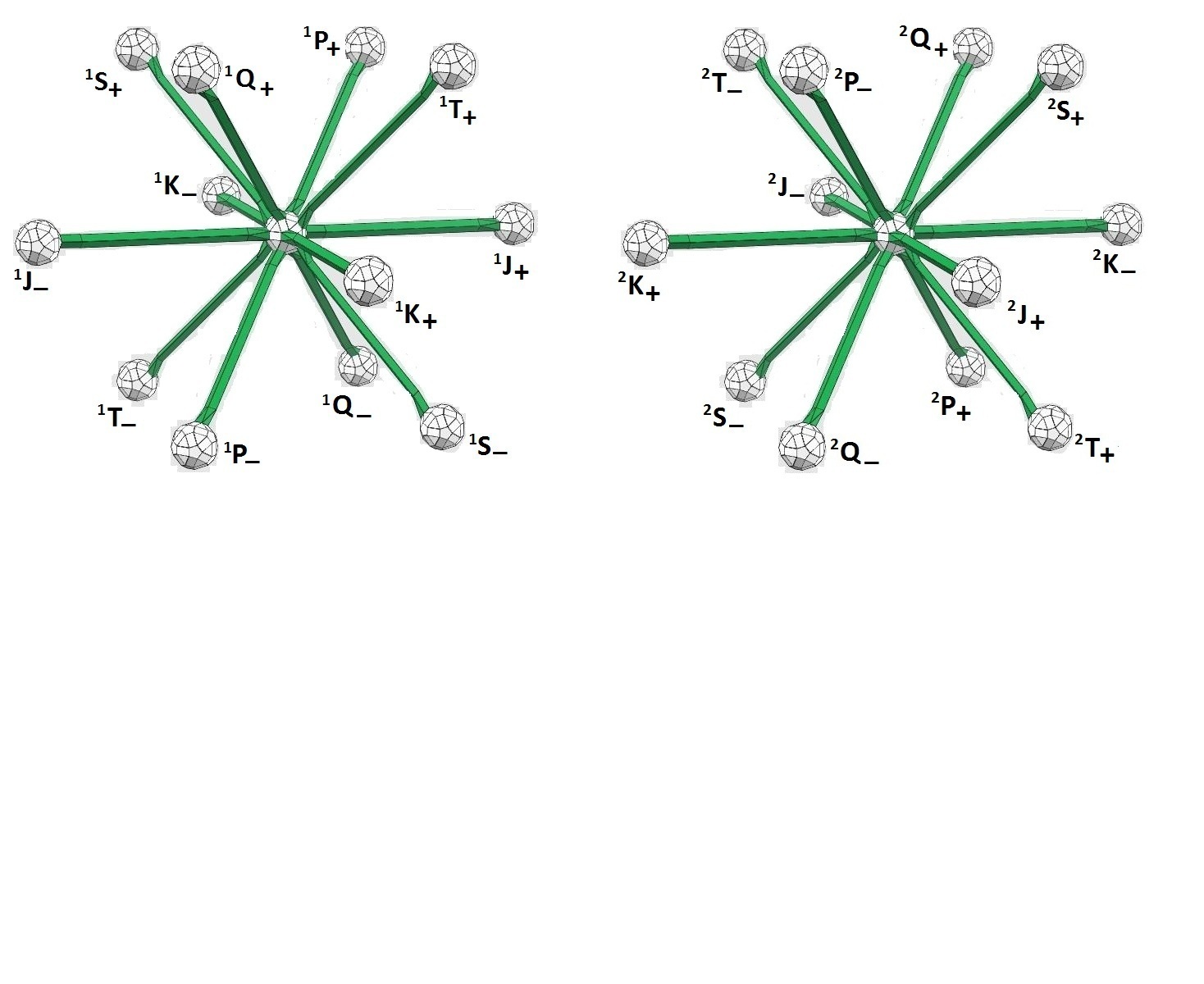}\\
\vspace{-6cm}
\caption{The root diagrams of the split basis (\ref{B1})--(\ref{B12}) of the Lie algebra $\mathfrak{so}(4,4)$.}
\label{pic11}
\end{figure}

Unlike the weight diagram of the Lie algebra $\mathfrak{so}(4,2)$, which allows visualization in three-dimensional space (Figure \ref{pic7}), the weight diagram of the algebra $\mathfrak{so}(4,4)$ is defined in four-dimensional space, which, naturally, significantly complicates its direct visualization. The Dynkin diagram of the algebra $\mathfrak{so}(4,4)\simeq D_4$ has the form
\unitlength=0.35mm
\begin{center}
\begin{picture}(40,45)(0,0)
\put(-4,20){\circle*{6}}
\put(20,20){\circle*{6}}
\put(40,44){\circle*{6}}
\put(40,-4){\circle*{6}}
\put(-4,20){\line(1,0){22}}
\put(20,20){\line(5,6){20}}
\put(20,20){\line(5,-6){20}}
\end{picture}
\end{center}
As noted above, the four generators (\ref{Cartan}) are \textit{Cartan generators} of the subalgebra $\fK\subset\mathfrak{so}(4,4)$. They form the basis of a four-dimensional orthogonal coordinate system and are located at the origin of the root diagram. The structure of the basis (\ref{B1})--(\ref{B12}), determined by the commutation relations (\ref{CB1}) and (\ref{CB2}), as well as the dual Cartan-Weyl bases (\ref{IB1}) and (\ref{IB2}) with the commutation relations (\ref{SYB1}) and (\ref{SYB2}) implies the existence of two root systems
\begin{equation}\label{RS1}
\ar\left.
\begin{array}{ccc}
\boldsymbol{\alpha}({}^1\bsK_+)=(+1,+1,0),&\quad\boldsymbol{\alpha}({}^1\bsT_+)=(+1,0,+1),&
\quad\boldsymbol{\alpha}({}^1\bsP_+)=(0,+1,+1),\\
\boldsymbol{\alpha}({}^1\bsK_-)=(-1,-1,0),&\quad\boldsymbol{\alpha}({}^1\bsT_-)=(-1,0,-1),&
\quad\boldsymbol{\alpha}({}^1\bsP_-)=(0,-1,-1),\\
\boldsymbol{\alpha}({}^1\bsJ_+)=(-1,+1,0),&\quad\boldsymbol{\alpha}({}^1\bsS_+)=(-1,0,+1),&
\quad\boldsymbol{\alpha}({}^1\bsQ_+)=(0,-1,+1),\\
\boldsymbol{\alpha}({}^1\bsJ_-)=(+1,-1,0),&\quad\boldsymbol{\alpha}({}^1\bsS_-)=(+1,0,-1),&
\quad\boldsymbol{\alpha}({}^1\bsQ_-)=(0,+1,-1),\\
\end{array}\right\}
\end{equation}
\begin{equation}\label{RS2}
\ar\left.
\begin{array}{ccc}
\boldsymbol{\alpha}({}^2\bsK_+)=(+1,-1,0),&\quad\boldsymbol{\alpha}({}^2\bsT_+)=(+1,0,-1),&
\quad\boldsymbol{\alpha}({}^2\bsP_+)=(0,+1,-1),\\
\boldsymbol{\alpha}({}^2\bsK_-)=(-1,+1,0),&\quad\boldsymbol{\alpha}({}^2\bsT_-)=(-1,0,+1),&
\quad\boldsymbol{\alpha}({}^2\bsP_-)=(0,-1,+1),\\
\boldsymbol{\alpha}({}^2\bsJ_+)=(+1,+1,0),&\quad\boldsymbol{\alpha}({}^2\bsS_+)=(+1,0,+1),&
\quad\boldsymbol{\alpha}({}^2\bsQ_+)=(0,+1,+1),\\
\boldsymbol{\alpha}({}^2\bsJ_-)=(-1,-1,0),&\quad\boldsymbol{\alpha}({}^2\bsS_-)=(-1,0,-1),&
\quad\boldsymbol{\alpha}({}^2\bsQ_-)=(0,-1,-1),\\
\end{array}\right\}
\end{equation}
\[
\boldsymbol{\alpha}(\sL_3)=(0,0,0),\quad\boldsymbol{\alpha}(\sA_3)=(0,0,0),
\quad\boldsymbol{\alpha}(\Delta_3)=(0,0,0).
\]
Graphically, the root systems (\ref{RS1}) and (\ref{RS2}) can be represented as two cuboctahedra, shown in Figure \ref{pic11}. In this case, the axes of the left cuboctahedron are twisted counterclockwise, and the axes of the right cuboctahedron are clockwise.

Further, using the properties of the basis (\ref{B1})--(\ref{B12}) of the algebra $\mathfrak{so}(4,4)$, which consists in splitting into two structurally identical (isomorphic) Yao bases of the  subalgebra $\mathfrak{so}(4,2)$, we can partially visualize the weight diagram of the algebra $\mathfrak{so}(4,4)$ by means of \textit{two three-dimensional projections}.

The weight space is defined by four Cartan generators $\bsL_{56}=\Delta_3$, $\bsL_{12}=\sL_3$, $\bsL_{34}=\sA_3$ and $\bsL_{78}$, which serve as the basis of a four-dimensional orthogonal coordinate system. Let's define the ket-vector
\begin{equation}\label{Ket}
\left|\nu,\dot{\nu};\lambda,\dot{\lambda};\mu,\dot{\mu};\sigma,\dot{\sigma}\right\rangle,
\end{equation}
where $\nu,\dot{\nu}\in\lf 0,1/2,1,3/2,\ldots\rf$, $\lambda\in\lf-\nu,-\nu+1,\ldots,\nu-1,\nu\rf$, $\dot{\lambda}\in\lf-\dot{\nu},-\dot{\nu}+1,\ldots,\dot{\nu}-1,\dot{\nu}\rf$, $\mu\in\lf-\lambda,-\lambda+1,\ldots,\lambda-1,\lambda\rf$, $\dot{\mu}\in\lf-\dot{\lambda},-\dot{\lambda}+1,\ldots,\dot{\lambda}-1,\dot{\lambda}\rf$, $\sigma,\dot{\sigma}\in\lf-1/2,+1/2\rf$.

The relationship of the numbers included in the ket-vector (\ref{Ket}) with the quantum numbers $n$, $l$, $m$ and $s$ of the ket-vector
\begin{equation}\label{Madelung}
\left|n,l,m,s\right\rangle
\end{equation}
of the Madelung basis is given by the following relations:
\[
n=\left|\nu-\dot{\nu}\right|,\quad l=\left|\lambda-\dot{\lambda}\right|,\quad m=\left|\mu-\dot{\mu}\right|,\quad
s=\left|\sigma-\dot{\sigma}\right|.
\]
In this case, the main quantum number $n$ varies within
\begin{equation}\label{Range}
-\left|\nu-\dot{\nu}\right|,\;-\left|\nu-\dot{\nu}\right|+1,\;-\left|\nu-\dot{\nu}\right|+2,\;\ldots,\;
\left|\nu-\dot{\nu}\right|.
\end{equation}

\begin{figure}[ht] %
\centering
\vspace{-0.5cm}
\includegraphics[width=16cm]{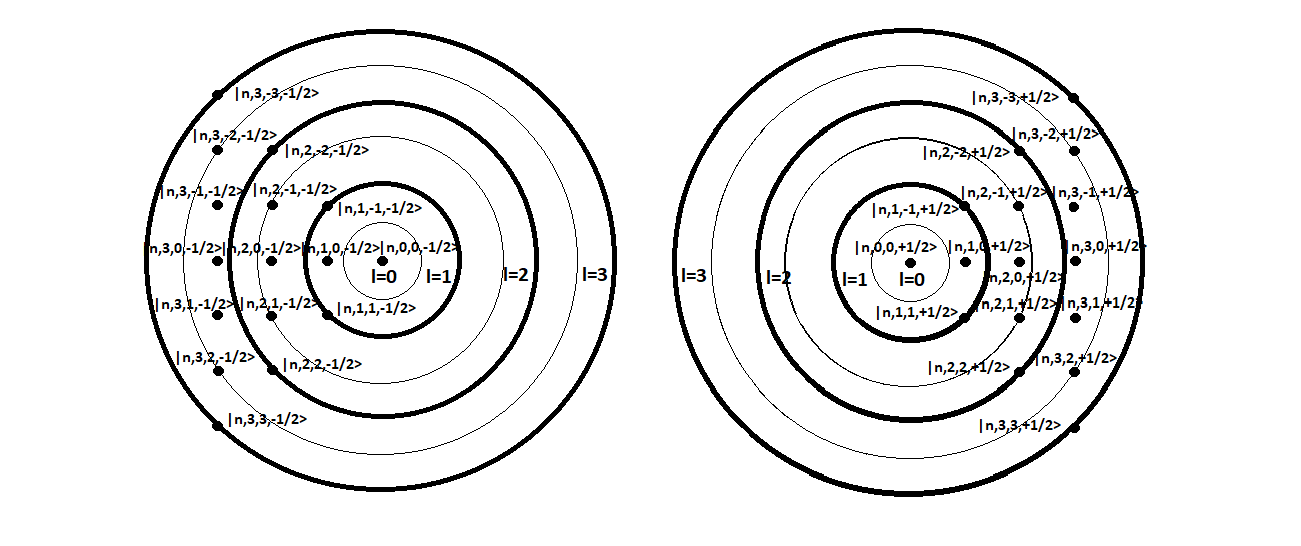}\\
\caption{Two-dimensional sections of three-dimensional projections of the weight diagram of the algebra $\mathfrak{so}(4,4)$ with ket-vectors $\left|n,l,m,-1/2\right\rangle$ and $\left|n,l,m,+1/2\right\rangle$.}
\label{pic12}
\end{figure}

Figure \ref{pic12} shows two-dimensional sections of two three-dimensional projections of the weight diagram of the algebra $\mathfrak{so}(4,4)$.

\begin{figure}[ht] %
\centering
\vspace{-0.5cm}
\includegraphics[width=16cm]{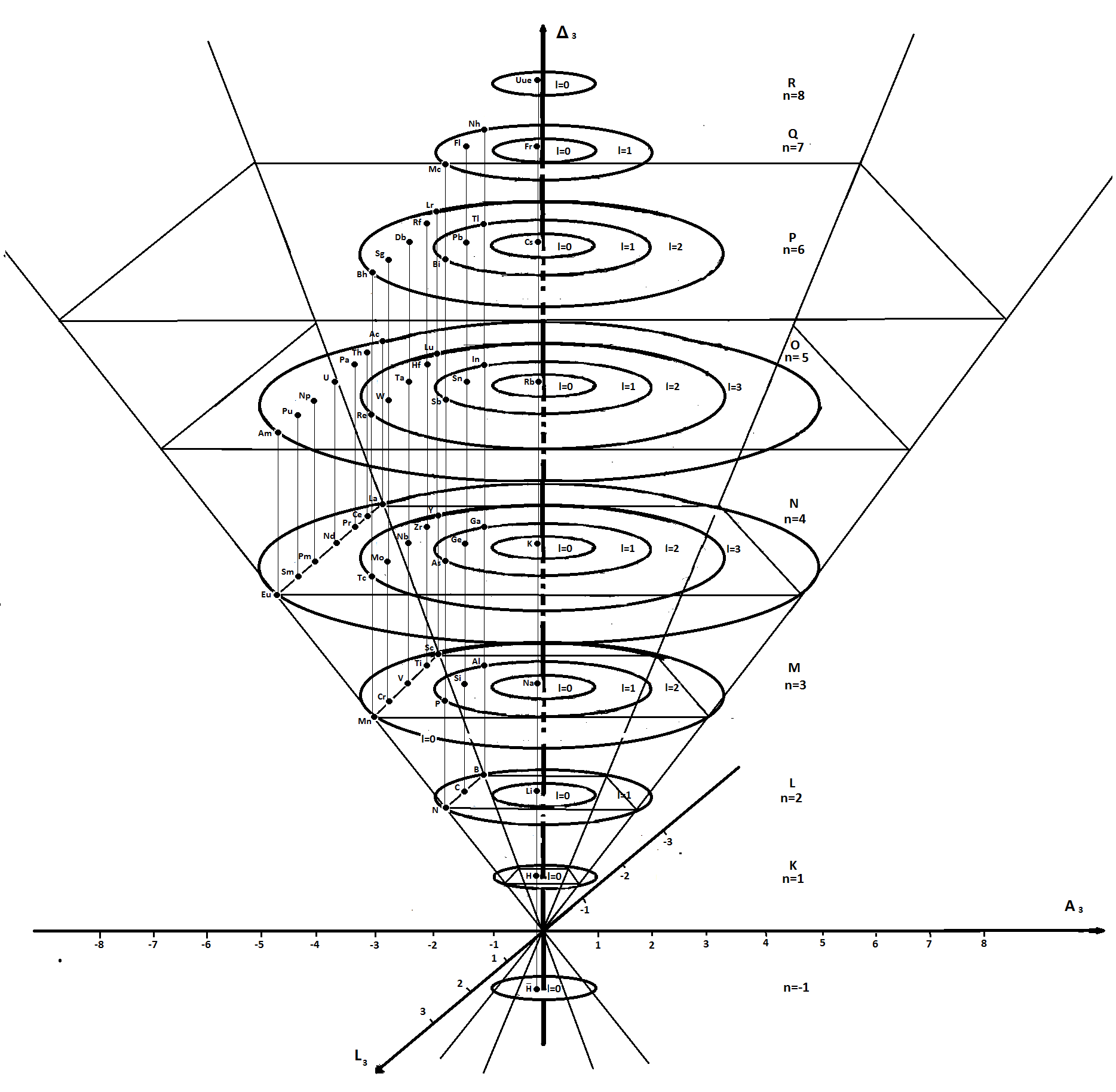}\\
\caption{$\SO(4,2)$-tower for $s=-1/2$ with elements \textbf{H}=$\left|1,0,0,-1/2\right\rangle$, $\ldots$, \textbf{Uue}=$\left|8,0,0,-1/2\right\rangle$.}
\label{pic13}
\end{figure}

\begin{figure}[ht] %
\centering
\vspace{-0.5cm}
\includegraphics[width=16cm]{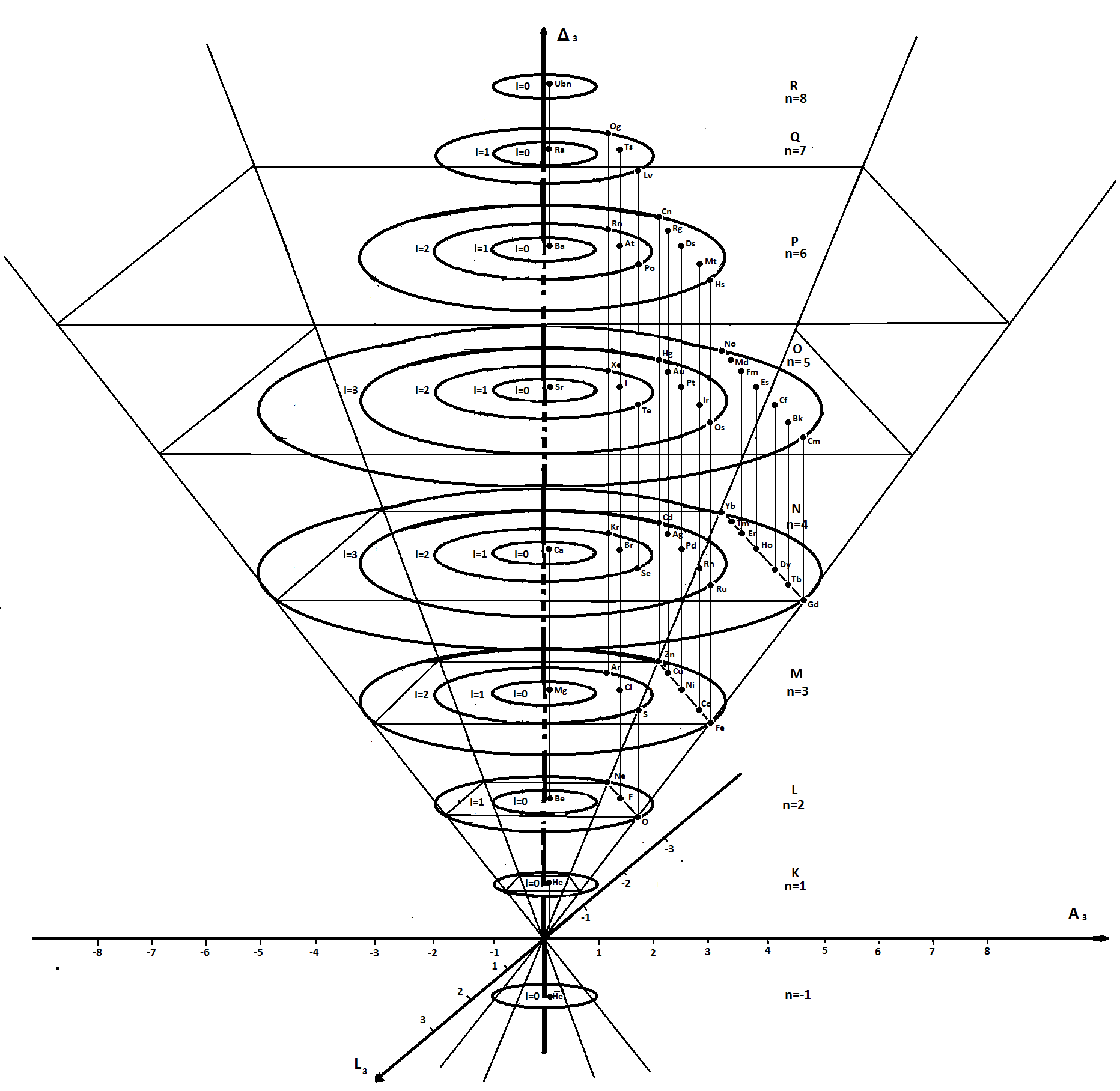}\\
\caption{$\SO(4,2)$-tower for $s=+1/2$ with elements \textbf{He}=$\left|1,0,0,+1/2\right\rangle$, $\ldots$, \textbf{Ubn}=$\left|8,0,0,+1/2\right\rangle$.}
\label{pic14}
\end{figure}

The first three-dimensional projection corresponding to the elements with $s=-1/2$ is shown in Figure \ref{pic13}. The upper pyramid of the $\SO(4,2)$-tower, starting with hydrogen \textbf{H}=$\left|1,0,0,-1/2\right\rangle$ to the hypothetical  element \textbf{Uue}=$\left|8,0,0,-1/2\right\rangle$, forms \textit{a pyramid of matter}. Reflected from the plane $(\sL_3,\sA_3)$, the lower pyramid starts with antihydrogen $\overline{\mathbf{H}}=\left|-1,0,0,-1/2\right\rangle$ and forms \textit{antimatter pyramid}\footnote{The first evidence of the existence of antihydrogen was obtained at accelerators in the 1990s. The capture of antihydrogen atoms into a trap was first demonstrated by the Antihydrogen Laser Physics Apparatus (ALPHA) group at CERN in 2010.}.
The second three-dimensional projection of the weight diagram of the algebra $\mathfrak{so}(4,4)$, corresponding to the elements with $s=+1/2$, is shown in Figure \ref{pic14}. In this case, the pyramid of matter contains elements from helium \textbf{He}=$\left|1,0,0,+1/2\right\rangle$ to oganesson \textbf{Og}=$\left|7,1,1,+1/2\right\rangle$, including the hypothetical element \textbf{Ubn}=$\left|8,0,0,+1/2\right\rangle$. The antimatter pyramid begins with the antihelium $\overline{\mathbf{He}}=\left|-1,0,0,+1/2\right\rangle$.

Homologous elements in the diagrams (Figures \ref{pic13} and \ref{pic14}) are connected by vertical lines (Bailey-Thomsen-Bohr lines). The diagrams show that, starting from the sheet $n=5$ (5th floor of $\SO(4,2)$-tower), the following circular rings (Haenzel rings) remain unfilled: ring $l=4$ for sheet $n=5$, rings $l=3$, $l=4$, $l=5$ for sheet $n=6$, as well as $l=2$, $l=3$, $l=4$, $l=5$ for the sheet $n=7$. At these levels, the hypothetical elements of the Seaborg table and the 10-periodic extension table are located (see \cite{Var1802,Var1901}), which are omitted in the diagrams of Figures \ref{pic13}-\ref{pic14} in order not to clutter the construction. So, for the elements \textbf{Ac}, \textbf{Th}, \textbf{Pa}, \textbf{U}, \textbf{Np}, \textbf{Pu}, \textbf{Am} of the level $(n=5,l=3)$ homologous elements are \textbf{Ute}, \textbf{Uqn}, \textbf{Uqu}, \textbf{Uqb}, \textbf{Uqt}, \textbf{Uqq}, \textbf{Uqp} of level $(n=6,l=3)$ of the Seaborg table, etc. It can be seen from the construction that only the first four floors of the $\SO(4,2)$-towers are fully populated (experimentally discovered elements). In addition, starting from the fourth floor (Haenzel circle) $n=4$, the stability of the elements increases as they approach the radial axis $\boldsymbol{\Delta}_3$.

\section{Conclusion}
This article provides an interpretation of the periodic table of chemical elements within the framework of the weight diagram of the group algebra $\mathfrak{so}(4,4)$ of the rotation group $\SO(4,4)$ of the eight-dimensional pseudo-Euclidean space $\R^{4,4}$. The central place in this generalization is played by the concept of spin. The origin of this concept is inextricably linked to the periodic table. As is known, the concept of spin was introduced by Pauli in 1925, explaining the doublet structure of the spectrum of alkali metals (the anomalous Zeeman effect): "The doublet structure of the alkali spectra, as well as the violation of the Larmor theorem are,
according to this point of view, a result of a classically not describable two-valuedness of the quantum-theoretical properties of the valence electron" \cite{Pauli25}. Later, Van der Waerden noted that this “\textit{\textbf{classically non-describable two-valuedness}}” of electron we call spin \cite[P. 215]{Waerden}.

It is well-known that, starting from the work of Uhlenbeck and Goudsmit \cite{UG},
all the attempts to describe the electron spin classically had failed. In memorials
Yu. B. Rumer \cite{Rum} wrote: “At his time Pauli said to Kronig that the spin theory is
nonsense because a point cannot rotate around itself” \cite[P. 56]{Rum}. Viewing electron as
a point is required in the special relativity; therefore, both electrons in quantum
electrodynamics and quarks in quantum chromodynamics are treated as pointwise
fermions with spin 1/2. As V.A. Fok argued \cite{Fock71}, \textit{\textbf{spin is not a mechanical concept}}. V.A. Fok noted: "The word "spin" literally means "spinning"; the name came about because the operators related to this degree of freedom can be formally interpreted as operators of the intrinsic angular momentum of the electron. This does not mean, however, that an electron can be likened to a spinning top or a rotating ball, etc.; mechanical comparisons are decidedly unsuitable here... spin is not a mechanical concept" \cite[p. 111]{Fock71}.

The first theory providing a
correct mathematical formulation of the “classically non-describable two-valuedness”
of the electron spin was proposed by Pauli in 1927 \cite{Pauli27}. Avoiding any visual
mechanical models, Pauli introduced doubled Hilbert space $\bsH_2\otimes\bsH_\infty$ (vector space of
wave functions), whose vectors are two-component spinors. That was the first advent
of two-component spinors in physics and the first case of doubling. The next doubling that yielded the $\bsH_4\otimes\bsH_\infty$ space of bispinors was accomplished by Dirac in 1928 \cite{Dir28}. In 1933, Gustav Mie introduced 8-component spinors \cite{Mie33}:
\[
\text{\textbf{2-spinors} (Pauli)}\longrightarrow\text{\textbf{4-spinors} (Dirac)}\longrightarrow\text{\textbf{8-spinors} (Mie)}.
\]

Doubling is a universal characteristic of matter. The electron's spin is a manifestation of the duality of matter, not its intrinsic property. Recall that experimentally observed electrons exist with only one direction of spin, and not with two at the same time, duality exists in matter (substance), which gives its accidents (electrons and other states) by one spin value or another, depending on the experimental situation (manifestation).

\end{document}